\nonstopmode
\documentclass[12pt]{article}

\usepackage{amsmath}
\usepackage{amsfonts}
\usepackage{amssymb}
\usepackage{bbm}
\usepackage{a4wide}
\usepackage{graphicx}
\graphicspath{{./}{../graphics/}}
\usepackage[margin=10pt,font=small]{caption}
\usepackage{hyperref}

\newcommand{\R}{\mathbb{R}}

\newcommand{\Z}{\mathbb{Z}}
\newcommand{\uno}{\mathbbm{1}}
\newcommand{\dif}{{\rm d}}
\DeclareMathOperator{\vol}{vol}

\DeclareMathOperator{\diag}{diag}

\newcommand{\arXiv}[2]{\href{http://arxiv.org/abs/#1}
  {\texttt{arXiv:#1 #2}}}

\newcommand{\figIa}{fig1a.eps}
\newcommand{\figIb}{fig1b.eps}
\newcommand{\figII}{fig2.eps}
\newcommand{\figIIIa}{fig3a.eps}
\newcommand{\figIIIb}{fig3b.eps}
\newcommand{\figIIIc}{fig3c.eps}
\newcommand{\figIIId}{fig3d.eps}
\newcommand{\figIVa}{fig4a.eps}
\newcommand{\figIVb}{fig4b.eps}
\newcommand{\figIVc}{fig4c.eps}
\newcommand{\figIVd}{fig4d.eps}
\newcommand{\figVa}{fig5a.eps}
\newcommand{\figVb}{fig5b.eps}
\newcommand{\figVIa}{fig6a.eps}
\newcommand{\figVIb}{fig6b.eps}
\newcommand{\figVII}{fig7.eps}
\newcommand{\figVIII}{fig8.eps}
\newcommand{\figIX}{fig9.eps}
\newcommand{\figXa}{fig10a.eps}
\newcommand{\figXb}{fig10b.eps}
\newcommand{\figXIa}{fig11a.eps}
\newcommand{\figXIb}{fig11b.eps}
\newcommand{\figXIIa}{fig12a.eps}
\newcommand{\figXIIb}{fig12b.eps}
\newcommand{\figXIIc}{fig12c.eps}
\newcommand{\figXIII}{fig13.eps}
\newcommand{\figXIV}{fig14.eps}
\newcommand{\figXV}{fig15.eps}
\newcommand{\figXVIa}{fig16a.eps}
\newcommand{\figXVIb}{fig16b.eps}

\title{%
  Exploring a simple sector of the Einstein-Maxwell landscape%
}

\author{%
  C\'esar Asensio\thanks{\texttt{casencha}\textbf{@}\texttt{unizar.es}}%
  \qquad and \qquad %
  Antonio Segu\'{\i}\thanks{\texttt{segui}\textbf{@}\texttt{unizar.es}}%
  \\
  \\
  \emph{Departamento de F\'{\i}sica Te\'orica, Universidad de Zaragoza}%
}

\begin{document}
\maketitle
\begin{abstract}
  We explore the four dimensional Einstein-Maxwell landscape as a toy
  model in which we can formulate a sphere compactification stabilized
  by an electromagnetic field.  Replacing the compactification sphere
  by $J$ spheres, we obtain a simple sector of the
  $(2J+2)$-dimensional Einstein-Maxwell landscape.  In this toy model,
  we analyze some properties which are very difficult to uncover in
  the string theory landscape, including:  complete moduli
  stabilization, stability conditions, and state counting.  We also
  show how to construct anthropic states in this model.  A detailed
  comparison between the main features of this landscape and the
  Bousso-Polchinski landscape is given.  We finally speculate on the
  impact of these phenomena in the string theory landscape.
\end{abstract}

\section{Introduction}
\label{sec:intro}

As a candidate of a theory-of-everything, string theory has led to
many striking results.  Among them we find the string theory landscape
\cite{BP,Anthr}, a very complicated structure of vacuum states of the
theory which raises its own questions and problems.  All models of
this landscape are rich and complex, and the existence of this
landscape is almost beyond doubt \cite{Is-there}.  When deriving
cosmological models in a landscape we are led to the notion of
multiverse \cite{Multiv-1}, a quantum ensemble of different classical
cosmological models.  The multiverse shows very appealing features,
such as inflation, which is needed to remedy the difficulties of older
cosmological models \cite{EtInf-1,EtInf-2,EtInf-3,EtInf-4,EtInf-5},
and it is generally believed that the cosmological constant
\cite{WW2,B-CC} and the coincidence problems can be solved with
realistic models of the multiverse \cite{Coinc-P-1,Coinc-P-2}.
Nevertheless, the multiverse has its own problems, which will be
described briefly.

The huge amount of possible universes present in the landscape
\cite{Count-1,Count-2,Count-3,Count-4} should be complemented with a
probability distribution which can explain why we are living in this
particular universe.  The standard rule of assigning probabilities as
proportional to $e^{-S}$ for some euclidean action $S$ breaks down in
this context because the classical action of a cosmological model is
divergent due to the infinite spacetime volume.  To extract some
useful information, the action should be regularized, but there are
many different ways of regulating an infinite volume because no
cut-off procedure is invariant under a coordinate change.  Different
regularization procedures can lead to different probability
distributions predicting different universes.  This is known as the
measure problem \cite{Multiv-1,Meas-1}, which consists in giving an
unambiguous definition of the relative probabilities in a given
landscape model.  So far, the measures derived from first principles
need the AdS/CFT correspondence \cite{Maldacena-AdS/CFT} and similar
ideas as a key ingredient \cite{Meas-AdS/CFT-1,Meas-AdS/CFT-2}, which
has an stringy origin, and thus the whole picture is slowly evolving
towards a unified formalism of the multiverse derived from string
theory.

When a multiverse model is used to explain, for example, the smallness
of the observed cosmological constant ($\lambda_{\mathrm{obs}} \approx
10^{-120}$ in Planck units) \cite{SN-1,SN-2}, theory should bring a
probability distribution of $\lambda$ values.  This probability
distribution should take into account the number of states of the
landscape, but also its relative probability computed using a measure
as stated above.  This measure is derived from a mechanism by which
states are populated in the landscape, for example, by some form of
Euclidean Quantum Gravity, with its own difficulties as commented in
the previous paragraph, or by eternal inflation, which populates all
states in the landscape in a stochastic fashion.  When considering the
distribution of $\lambda$ values conditioned on those states which
support some form of observers (like us) \cite{BaTi,WW,Anthr}, one
needs to add an anthropic factor to the probability which should be
provided to complete the prediction.  Therefore, there are three parts
which contribute to the final probability: the distribution of the
states in the landscape (also known as the \emph{prior} part), the
cosmological measure (which needs a mechanism for populating the
landscape), and the anthropic factor (which incorporates the condition
for the existence of observers, such as galaxy formation).

Another problem is the vast complexity of the landscape.  Extracting a
four-dimensional cosmology from string theory requires choosing a
compactification of the remaining dimensions.  Each possible choice of
compactification describes a sector of the entire landscape, and the
number and properties of the states in different sectors can be very
different.  This diversity can be explored by considering different
sectors separately.  Thus, simplified models of some sectors of the
landscape have been built, and they constitute a very important tool
to understand the full implications of the string theory landscape.
The reader can consult \cite{Denef-1,Denef-2} for a review of several
models of the string theory landscape.  For our purposes only the
simplest models are needed.  Among these models we should mention:
\begin{itemize}
\item The Kachru-Kallosh-Linde-Trivedi (KKLT) mechanism
  \cite{GKP,KKLT,KKLMMT} is a supersymmetric string model with many
  stabilized AdS vacua which are lifted to dS by quantum effects.
  This landscape model gives inflationary states represented by
  brane-antibrane pairs evolving in Klebanov-Strassler throats in the
  compactification manifold.  The model has no classical dS vacua, and
  thus its more controversial aspect is the quantum nature of the
  lifting from AdS to dS, which is uncontrollable by its very nature.
\item The Bousso-Polchinski (BP) model \cite{BP} is a simplified
  vacuum sector of M theory compactified from eleven to four
  dimensions with quantized four-form fluxes and M5 branes wrapped
  around three-cycles in the seven-dimensional compactification
  manifold.  The volumes of the three-cycles are the moduli of the
  model, and gives a fundamental charge $q_i$ which determines the
  cosmological constant of the 3+1 cosmological part of the
  eleven-dimensional metric by the formula
  \begin{equation}
    \label{eq:1}
    \lambda = \Lambda + \frac{1}{2}\sum^J_{i=1} q_i^2n_i^2\,.
  \end{equation}
  Here, each $n_i$ is the number of flux units stored in the
  $i^{\mathrm{th}}$ three-cycle, and $\Lambda$ is a bare, negative
  cosmological constant, a parameter of the model.  Equation
  \eqref{eq:1} provides a very elegant mechanism to deal with the
  cosmological constant problem.  By choosing a large number $J$ of
  three-cycles, we can choose the integers $n_i$ to approximately
  cancel the $\Lambda$ contribution, thus obtaining a huge landscape
  of AdS and dS vacua containing states with very low values of the
  effective cosmological constant $\lambda$, also known as the
  \emph{discretuum}.  Calabi-Yau compactifications typically have a
  large amount of three-cycles, so that realistic values of the
  cosmological constant can be obtained without any fine-tuning of the
  model parameters.
\end{itemize}
On one hand, unlike the KKLT mechanism, the BP model does not rely on
quantum effects for producing dS states.  On the other hand, BP moduli
are frozen from the very beginning, and thus no stabilization
mechanism analogous to KKLT is included in the BP model.

While there are some extensions of the KKLT scenario where the
structure of states in the landscape has been elucidated by means of
numerical searches \cite{Frey:2003dm}, a very appealing feature of the
BP landscape is that the closed expression \eqref{eq:1} leads to
analytically tractable counting problems.  Thus, relative
probabilities of different states can be computed based on
combinatorics, but this leads to the conclusion that the number of
states with realistic $\lambda$ values are a very sparse minority,
thus opening the door to anthropic arguments \cite{BaTi,WW,Anthr}.

If we combine the computation of probabilities by naive state counting
with the lack of stability analysis, we find another curious property
of the BP landscape.  Under a large number of three-cycles $J$, flux
space is a very high-dimensional space.  Such spaces have very
counterintuitive geometric features: for example, the opposite
vertexes of a unit hypercube are separated by a distance $\sqrt{J}$,
which can be interpreted (in Planck units) as a huge energy scale.
Another consequence of this elongation of diagonal distances is that
the vast majority of lattice nodes inside a sphere in flux space are
located on hyperplanes, that is, they always have at least one
vanishing component.  In fact, we have seen that as $J$ grows, the
dimension of the most populated hyperplanes is distributed in a narrow
Gaussian window around a typical value $\alpha^* J$ with $\alpha^* <
1$ \cite{Alpha-star}.  But as the KKLT mechanism shows, flux quantum
numbers cannot vanish in stable states.  Of course, the lack of a
stabilization mechanism in the BP model prevents us from directly
excluding those states from the model; if they were excluded, the
states with very low $\lambda$ values could even disappear.  This
``$\alpha^*$-problem'' of the BP model is only present for large $J$,
but a large $J$ is certainly needed to solve the cosmological constant
problem.

Thus, looking for a scenario where we can closely examine this
$\alpha^*$-problem, we are led to consider a model with the following
properties:
\begin{itemize}
\item It should be exactly solvable.
\item It should have many moduli.
\item It should have a stabilization mechanism.
\end{itemize}
The simplest landscape with the first and third requirements is the
Einstein-Maxwell landscape, considered in a plethora of papers as a
toy model landscape \cite{EM6-1,EM6-2,flux-rev,EM6-3}.  This landscape
has compactifications of the form $\mathrm{(A)dS}_2\times\mathrm{S}^4$
or $\mathrm{(A)dS}_4\times\mathrm{S}^2$ having a single modulus,
namely, the radius of the compactification sphere.  We have added many
moduli by considering compactifications of the form
\begin{equation}
  \label{eq:2}
  \mathrm{(A)dS}_2\times\prod_{i=1}^J\mathrm{S}^2\,,
\end{equation}
and thus the moduli are the radii of the $J$ spheres.  We call this
sector \emph{multi-sphere} Einstein-Maxwell compactification.  We will
see that dS states with low quantum numbers are always unstable, and
thus they should be excluded from the landscape.  This legitimates the
$\alpha^*$-problem as an objection against naive counting arguments in
the BP landscape.

The stability of states in multi-sphere compactifications has been
studied previously \cite{Duff:1984sv}, \cite{Berkooz:1998qp},
\cite{DeWolfe:2001nz}.  In particular, in \cite{DeWolfe:2001nz} it is
shown that compactifications of the form $\mathrm{AdS}_p\times
\mathrm{S}^n \times \mathrm{S}^{q-n}$ are unstable for $q < 9$ but
they are stable for $q \geq 9$.  In contrast, the multi-sphere model
(\ref{eq:2}) gives always stable $\mathrm{AdS}_2$ states.  This
example emphasizes the importance of the dimensionality in determining
stability.

We have chosen the cosmological part in the multi-sphere model to be
1+1 instead of 3+1 for simplicity.  On one hand, these cosmologies are
unrealistic, and they have the peculiarities of two-dimensional
dilatonic gravities.  On the other hand, there is no theory predicting
how observers form in a 1+1 universe, and no quantum measure defined
on this multiverse, and therefore the prediction of the cosmological
constant relies on the state counting problem only.  It is also a good
candidate for studying the measure problem in future papers.

The paper is organized as follows.  In section
\ref{sec:one-flux-compactif}, we give a detailed description of the
single-modulus compactification $\mathrm{(A)dS}_2\times\mathrm{S}^2$
in the Einstein-Maxwell theory, together with the stabilization
mechanism and the state counting.  In section \ref{sec:many} we
consider the multi-sphere model and its stabilization mechanism in
detail, and give a very detailed account of the
$\mathrm{(A)dS}_2\times\mathrm{S}^2\times\mathrm{S}^2$ sector of the
six-dimensional Einstein-Maxwell theory.  Section
\ref{sec:state-counting} is devoted to counting the states in the
model.  In section \ref{sec:anthropic} we show how anthropic states
can be constructed in this model.  In section~\ref{sec:EM-vs-BP} we
summarize the differences found between the multi-sphere
Einstein-Maxwell sector and the Bousso-Polchinski landscape, and in
section \ref{sec:implications} we speculate on some consequences that
those phenomena can have on the string theory landscape.  The last
section \ref{sec:conc} summarizes our conclusions.

\section{One-flux  compactification in the four-dimensional Einstein-Maxwell theory}
\label{sec:one-flux-compactif}

\subsection{The one-flux four-dimensional Einstein Maxwell landscape}
\label{sec:one-flux}

By a compactification in four-dimensional gravity we understand a
solution of the Einstein field equations of the form
\begin{equation}\label{eq:3}
  \dif s^2 = e^{2\phi(x,t)}\bigl(-\dif t^2 + \dif x^2\bigr)
  + e^{2\psi(u,v)}\bigl(\dif u^2 + \dif v^2\bigr)\,.
\end{equation}
This is a particular expression of a Kantowski-Sachs cosmology
\cite{Kantowski-Sachs,Linde-300}.  The metric splits in a
$(t,x)$-spacetime part, which we will identify with a two-dimensional
cosmological solution, and a $(u,v)$-surface ${\cal K}$, the compact
part.

The Einstein field equations are, in units with $G=c=1$,
\begin{equation}\label{eq:4}
  R_{\mu\nu} = 8\pi\Bigl(T_{\mu\nu} - \frac{1}{2}Tg_{\mu\nu}\Bigr)\,.
\end{equation}
The stress-energy tensor has two contributions, the first comes from
the electromagnetic field and the second from a vacuum energy density:
\begin{equation}\label{eq:5}
  \begin{split}
    T_{\mu\nu} &= T^{(M)}_{\mu\nu} + T^{(\Lambda)}_{\mu\nu}\,,\\
    T^{(M)}_{\mu\nu} &= \frac{1}{4\pi}\Bigl(F_{\mu\rho}F_{\nu}^{\phantom{\nu}\rho}
    - \frac{1}{4}\,F^2\,g_{\mu\nu}\Bigr)\,\\
    T^{(\Lambda)}_{\mu\nu} &= -\frac{\Lambda}{8\pi}\,g_{\mu\nu}\,.
  \end{split}
\end{equation}
The symbol $F^2=F_{\mu\nu}F^{\mu\nu}$ is the electromagnetic
Lagrangian density, and $\Lambda$ is the cosmological constant of the
four-dimensional theory.

The stress-energy tensor of the Maxwell field is traceless, so we have
\begin{equation}\label{eq:6}
  T = T^{(\Lambda)} = -\frac{\Lambda}{2\pi}\,.
\end{equation}
We will assume a magnetic monopole configuration for the
electromagnetic field:
\begin{equation}\label{eq:7}
  \mathbf{F} = \frac{Q}{V}\,e^{2\psi(u,v)}\dif u\wedge\dif v\,,
\end{equation}
where the boldface is used to denote differential forms.  Here the
constant $Q$ is the magnetic charge of the monopole, and $V$ is the
volume of the compactification manifold, so that
\begin{equation}\label{eq:8}
  V = \vol{\cal K} = \int_{\cal K} e^{2\psi(u,v)}\dif u\wedge\dif v
\end{equation}
and we have
\begin{equation}\label{eq:9}
  \int_{\cal K}\mathbf{F} = Q\,.
\end{equation}
In matrix notation, the Maxwell field is
\begin{equation}\label{eq:10}
  (F_{\mu\nu}) =
  \begin{pmatrix}
    0    &  E_x &  E_u &  E_v \\
    -E_x &  0   & -B_v &  B_u \\
    -E_u &  B_v &  0   & -B_x \\
    -E_v & -B_u &  B_x &  0
  \end{pmatrix}
  =
  \begin{pmatrix}
    0 & 0 & 0   &  0   \\
    0 & 0 & 0   &  0   \\
    0 & 0 & 0   & -B_x \\
    0 & 0 & B_x &  0 
  \end{pmatrix}
  \,,\quad
  B_x = \frac{Q}{V}\,e^{2\psi(u,v)}\,.
\end{equation}
This configuration solves Maxwell equations in curved spacetime
\begin{equation}\label{eq:11}
  \nabla_\nu F^{\mu\nu} = 
  \frac{1}{\sqrt{-g}}\,\partial_\nu\Bigl(\sqrt{-g}\,F^{\mu\nu}\Bigr) = 0
\end{equation}
and the non-vanishing components of the corresponding stress-energy
tensor are
\begin{equation}\label{eq:12}
  \begin{split}
    T^{(M)}_{tt} &= \frac{1}{8\pi}\,B_x^2\, e^{2\phi-4\psi}
    = \frac{1}{8\pi}\biggl(\frac{Q}{V}\biggr)^2e^{2\phi} \,, \\
    T^{(M)}_{xx} &= -\frac{1}{8\pi}\,B_x^2\, e^{2\phi-4\psi}
    = -\frac{1}{8\pi}\biggl(\frac{Q}{V}\biggr)^2e^{2\phi} \,, \\
    T^{(M)}_{uu} &= T^{(M)}_{vv} = \frac{1}{8\pi}\,B_x^2\, e^{-2\psi}
    = \frac{1}{8\pi}\biggl(\frac{Q}{V}\biggr)^2e^{2\psi}\,.
  \end{split}
\end{equation}
The contribution of the cosmological constant is
\begin{equation}\label{eq:13}
  T^{(\Lambda)}_{\mu\nu} -\frac{1}{2}\,T^{(\Lambda)}\,g_{\mu\nu} =
  \frac{\Lambda}{8\pi}\,g_{\mu\nu} \,. 
\end{equation}
Finally, the Ricci tensor has the following nonzero components:
\begin{equation}\label{eq:14}
  \begin{split}
    R_{tt} &= -R_{xx} = -\phi_{tt}+\phi_{xx}\,, \\
    R_{uu} &= -R_{vv} = -\psi_{uu}-\psi_{vv}\,, \\
  \end{split}
\end{equation}
where the subscripts in $\phi_{xx}$ etc.~represent partial
derivatives.  Einstein equations coincide in the $tt$ and $xx$
components, and in the $uu$ and $vv$ components also:
\begin{equation}\label{eq:15}
  \begin{split}
    -\phi_{tt} + \phi_{xx} &= \biggl[-\Lambda +
    \biggl(\frac{Q}{V}\biggr)^2\biggr] e^{2\phi}\,, \\
    -\psi_{uu} - \psi_{vv} &= \biggl[\Lambda +
    \biggl(\frac{Q}{V}\biggr)^2\biggr] e^{2\psi}\,, \\
  \end{split}
\end{equation}
Thus, $\phi$ and $\psi$ are uncoupled and satisfy Liouville equations
of $-+$ and $++$ signatures respectively.

The Liouville equation states that the Gaussian curvature of the
corresponding surface is constant.  We will call these two constants
$\lambda$ and $K$:
\begin{equation}\label{eq:16}
  \begin{split}
    \bigl(\phi_{tt} - \phi_{xx}\bigr)e^{-2\phi} &= \Lambda -
    \biggl(\frac{Q}{V}\biggr)^2 = \lambda \,, \\
    -\bigl(\psi_{uu} + \psi_{vv}\bigr)e^{-2\psi} &= \Lambda +
    \biggl(\frac{Q}{V}\biggr)^2 = K\,. \\
  \end{split}
\end{equation}
Thus, a solution of the $-+$ Liouville equation $\phi(t,x)$ represents
a two-dimensional spacetime of constant curvature $\lambda$, which is
de Sitter ($\text{dS}_2$) if $\lambda>0$, Minkowski ($\text{M}_2$) if
$\lambda=0$ and anti-de Sitter ($\text{AdS}_2$) if $\lambda<0$.
Therefore, $\lambda$ can be interpreted as the cosmological constant
of the dimensionally reduced cosmological model.  On the other hand,
$\psi$ represents a compact surface of constant curvature $K$.  We can
choose $K$ positive, and then the surface will be a sphere of radius
$1/\sqrt{K}$.

The positivity of the constant $K$ is equivalent to zero genus by the
Gauss-Bonnet formula, which also relates volume with curvature:
\begin{equation}\label{eq:17}
  \frac{1}{2\pi}\int_{\mathcal{K}} Ke^{2\psi}\dif u\dif v = 2
  \quad\Rightarrow\quad
  \frac{KV}{2\pi} = 2
  \quad\Rightarrow\quad
  V = \frac{4\pi}{K}\,,
\end{equation}
which leads to an algebraic equation for $K$:
\begin{equation}\label{eq:18}
  K = \Lambda + \biggl(\frac{Q}{V}\biggr)^2 = \Lambda +
  \biggl(\frac{QK}{4\pi}\biggr)^2 \,.
\end{equation}
The previous equation has two solutions:
\begin{equation}\label{eq:19}
  K_{\pm} = 2\Lambda\biggl(\frac{Q_{\text{max}}}{Q}\biggr)^2
  \Biggl[1\pm\sqrt{1 - \biggl(\frac{Q}{Q_{\text{max}}}\biggr)^2}\Biggr]\,,
  \qquad
  Q_{\text{max}} = \frac{2\pi}{\sqrt{\Lambda}}\,.
\end{equation}
The magnetic charge should not exceed $Q_{\text{max}}$.  For greater
charges curvatures become complex and therefore the solution makes no
sense.

The two-dimensional cosmological constant has also two branches
\begin{equation}\label{eq:20}
  \lambda_{\pm} = 2\Lambda - K_{\mp}\,.
\end{equation}
Assuming the usual Dirac quantization condition on the magnetic charge
in terms of the elemental charge $e$ of the particles coupled to the
electromagnetic field,
\begin{equation}\label{eq:21}
  Qe = 2\pi n\,,\quad\text{with $n\in\Z$}\,,
\end{equation}
we have a maximum value of the integer $n$,
\begin{equation}\label{eq:22}
  n_{\text{max}} = \biggl\lfloor\frac{e}{\sqrt{\Lambda}}\biggr\rfloor
\end{equation}
and all integers $n$ satisfying $0<|n|\le n_{\text{max}}$, plus $n=0$,
constitute the one-flux four dimensional Einstein-Maxwell
landscape\footnote{This is the ``pedestrian'' landscape mentioned in
  Footnote 2 of Ref.~\cite{Randall-Jhonson-Carroll}.}.

The case $n=0$ deserves further comment.  In this case we have $Q=0$,
and we have only one branch, $K=\lambda=\Lambda$, which cannot be
supported because the electromagnetic field vanishes.  Thus we should
expect this solution to be unstable, as we will see below.

Thus far, we have considered $\Lambda>0$.  In the case $\Lambda<0$, it
can be seen that the solutions of equation \eqref{eq:18} yield $K_-<0$
and thus only the $K_+$ branch remains as a solution, with a
cosmological constant $\lambda_- = 2\Lambda - K_+$ always negative.
In addition we have no restriction in the quantum number $n$, and
therefore this $\Lambda<0$ infinite landscape is less interesting than
its $\Lambda>0$ counterpart, which we will be considering henceforth.

Equation \eqref{eq:20} can be interpreted as a distribution of the
$\mathrm{dS}_4$ curvature $\Lambda$ between the $\mathrm{(A)dS}_2$ and
$\mathrm{S}^2$ parts.  The solutions obtained show that a positive
curvature $\Lambda$ can be distributed between positive $K$ and
positive or negative $\lambda$ (two possible ways), but a negative
curvature $\Lambda$ should be distributed between a positive $K$ and a
negative $\lambda$ (a unique way), yielding a physically less
interesting landscape.

\begin{figure}
  \centering
  \includegraphics[width=0.5\textwidth]{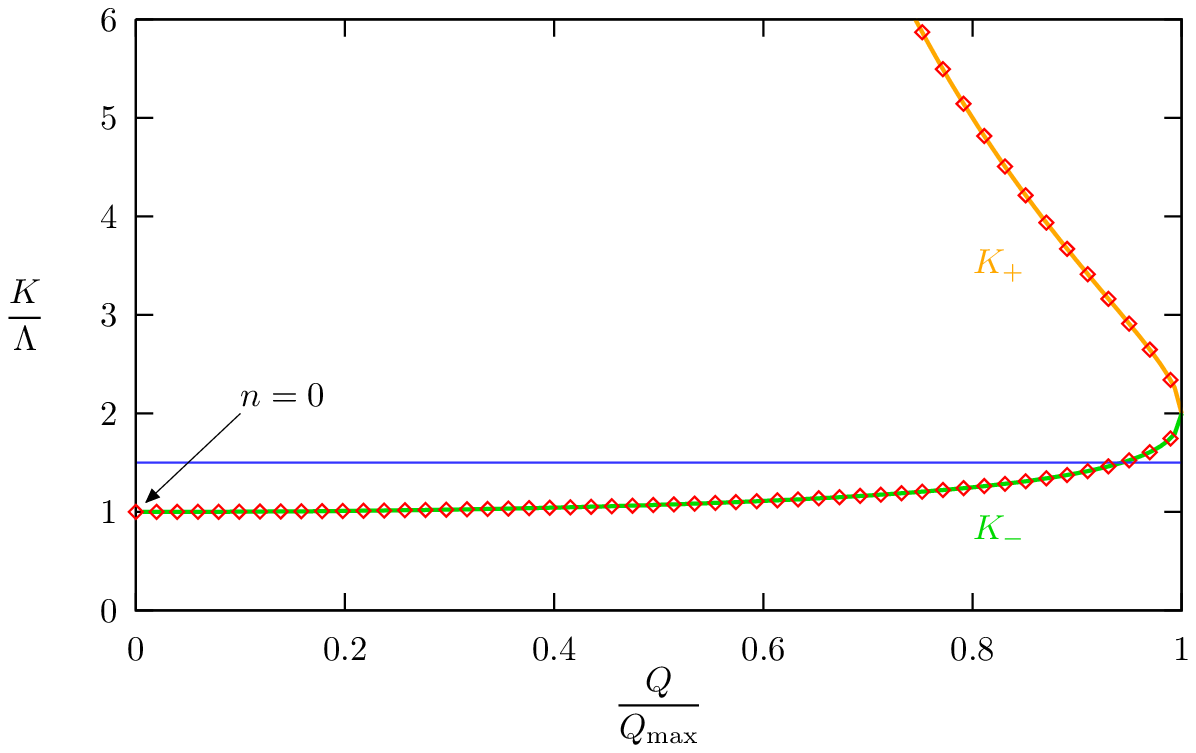}%
  \includegraphics[width=0.5\textwidth]{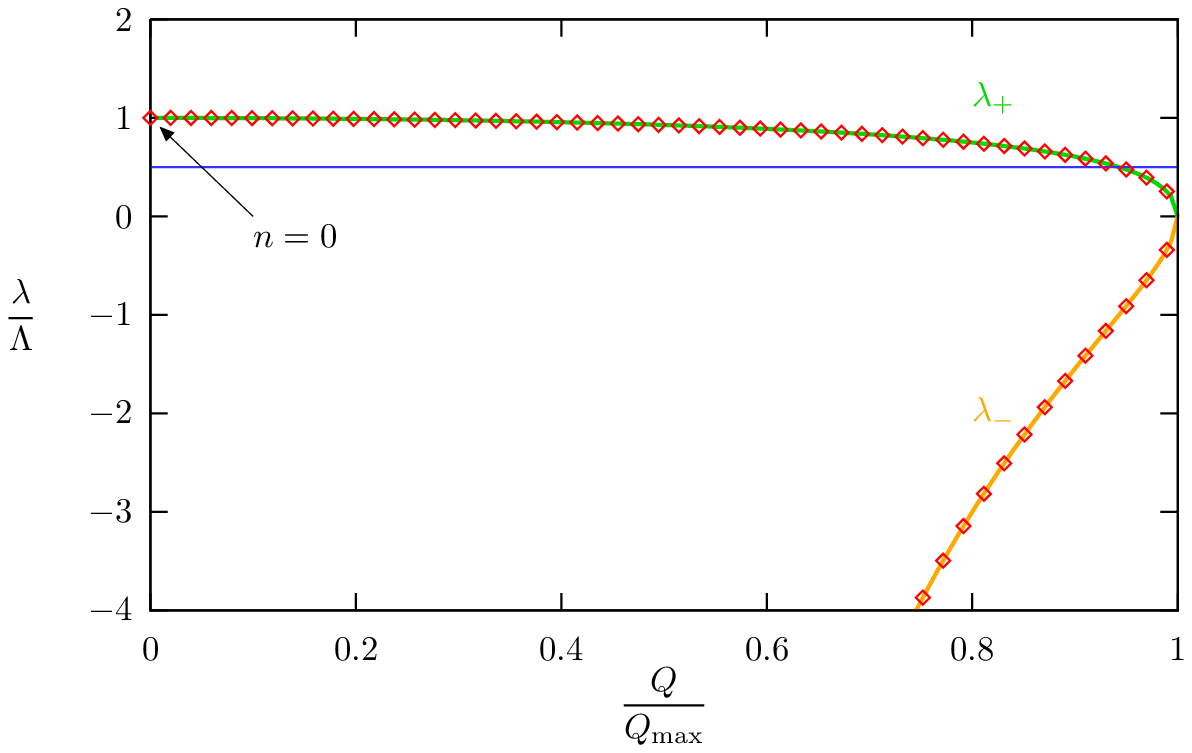}
  \caption{Curvature $K$ (left) and two-dimensional cosmological
    constant $\lambda$ (right) of the two branches of the one-flux
    four-dimensional Einstein-Maxwell landscape are shown with
    $n_{\text{max}}=50$, corresponding to $\Lambda=3.915\times10^{-4}$
    (a random value, expressed in units such that $e=1$).  The state
    $n=0$ is present only in the $K_-$, $\lambda_+$ branch.  The
    branching point is absent because $Q_{\text{max}}$ is
    (generically) not an integer.  The $n=1$ point on the
    $K_+,\lambda_-$ branch produces finite values
    $K_+(1)\approx-\lambda_-(1)\approx10^4\Lambda$.  The horizontal
    line signals the onset of stability, all states above it in the
    $\lambda_+$ branch being unstable.  Thus, this landscape contains
    50 $\text{AdS}_2$ states and 3 $\text{dS}_2$ states.}
\label{fig:one-flux-em4}
\end{figure}

Figure \ref{fig:one-flux-em4} shows curvatures and cosmological
constants for $n_{\text{max}}=50$ for both branches.  Note that this
number controls the number of the states in the landscape, whereas
$\Lambda$ controls the magnitude of the moduli $K$ and $\lambda$.

\subsection{Modulus stabilization}
\label{sec:modulus-stab}

The landscape considered thus far has only one modulus, namely the
Gaussian curvature of the compact sphere or equivalently its radius.
This is a volume modulus, which control the volume (surface area) of
the compact part of the spacetime.  The next question we should ask is
the stability of the solutions encountered.  If a value of $K$ in one
of the two branches for fixed quantum number $n$ is not stable, a small
perturbation will drive the system out of the solution.  If the
subsequent evolution makes the sphere radius to grow unbounded we
speak of \emph{decompactification}.  An unstable state in the
landscape should not be included when counting states in the resulting
two-dimensional multiverse.

The one-flux compactified solution found is
$\text{dS}_2\times\text{S}^2$ or $\text{AdS}_2\times\text{S}^2$ with a
sphere radius which is homogeneous throughout the two-dimensional
spacetime.  The perturbation will alter this situation, assuming that
the compactification radius can be different on different $t,x$
points, yielding an ansatz
\begin{equation}\label{eq:23}
  \dif s^2 =
  e^{2\phi(t,x) - 2\xi(t,x)} \bigl(-\dif t^2 + \dif x^2\bigr) +
  e^{2\psi(u,v) + 2\xi(t,x)} \bigl( \dif u^2 + \dif v^2\bigr)\,.
\end{equation}
The perturbation $\xi(t,x)$ appears also on the non-compact part of
the metric, and in this way the local volume element remains
invariant.  This reflects that the perturbation exchanges locally
volume between the compact and non-compact parts of the metric.
Functions $\phi$ and $\psi$ are solutions of the Liouville equations
encountered before.

We will investigate the dynamics of the perturbation field $\xi(t,x)$
from a two-dimensional perspective.  The four-dimensional action of
the Einstein-Maxwell system is
\begin{equation}\label{eq:24}
  S^{(4)} = \frac{1}{16\pi} \int L^{(4)}
  \,\dif t\,\dif x\,\dif u\,\dif v\,,
  \qquad
  L^{(4)} = \sqrt{-g^{(4)}}
  \bigl(R^{(4)} - 2\Lambda - F^2\bigr)\,,
\end{equation}
where we use superscripts to distinguish between four- and
two-dimensional quantities.  The curvature scalar and the
electromagnetic scalar are
\begin{equation}\label{eq:25}
  \begin{split}
    R^{(4)} &= -2\Bigl[
    \bigl(\phi_{xx} - \phi_{tt}\bigr)e^{-2\phi + 2\xi} +
    \bigl(\psi_{uu} + \psi_{vv}\bigr)e^{-2\psi - 2\xi} +
    \bigl(\xi_{xx} + 3\xi_{x}^2 - \xi_{tt} - 3\xi_{t}^2\bigr)
    e^{2\xi - 2\phi}
    \Bigr]\,,\\
    F^2 &= 2B_x^2\,e^{-4\psi-4\xi}\,.
  \end{split}
\end{equation}
Using $\sqrt{-g^{(4)}} = e^{2\phi+2\psi}$ and integrating by parts the
second derivatives of $\xi$, the four-dimensional Lagrangian is
\begin{multline}\label{eq:26}
  L^{(4)} = -2\Bigl[
    \bigl(\phi_{xx} - \phi_{tt}\bigr)e^{2\psi + 2\xi} +
    \bigl(\psi_{uu} + \psi_{vv}\bigr)e^{2\phi - 2\xi} \\ + 
    \bigl(\xi_{x}^2 - \xi_{t}^2\bigr)
    e^{2\xi + 2\psi}
    + \Lambda\,e^{2\phi+2\psi} + B_x^2\,e^{2\phi-2\psi-4\xi}
    \Bigr]
\end{multline}
Now, we will substitute the sphere ansatz
$-(\psi_{uu}+\psi_{vv})e^{-2\psi}=K$ and the magnetic monopole
configuration $B_x = \frac{Q}{V}e^{2\psi}$ as backgrounds for the
dynamics of the perturbation.  Integrating with respect to $u,v$ we
obtain the two-dimensional action
\begin{equation}\label{eq:27}
  \begin{split}
    S^{(2)} &= \frac{V}{4\pi}\int L^{(2)}\dif t\,\dif x\,, \\
    L^{(2)} &= \sqrt{-g^{(2)}}\Bigl[
    \frac{1}{4}\,e^{2\xi}R^{(2)} -
    \frac{1}{2}\,e^{2\xi}\bigl(g^{(2)}\bigr)^{\alpha\beta}\xi_\alpha\xi_\beta - U(\xi)
    \Bigr]\,,
  \end{split}
\end{equation}
where the summation in $\alpha,\beta\in\{t,x\}$ is implied, and
\begin{equation}
  \label{eq:28}
  \begin{split}
    \bigl(g^{(2)}\bigr)_{\alpha\beta}\,\dif\alpha\,\dif\beta &=
    e^{2\phi}\eta^{\alpha\beta}\,\dif\alpha\,\dif\beta =
    e^{2\phi}\bigl(-\dif t^2 + \dif x^2\bigr)\,,\\
    R^{(2)} &= -2\bigl(\phi_{xx} - \phi_{tt}\bigr)e^{-2\phi}\,, \\
    U(\xi) &= \frac{1}{2}\biggl[
    \Lambda - K\,e^{-2\xi} +
    \biggl(\frac{QK}{4\pi}\biggr)^2 e^{-4\xi}
    \biggr]\,,
  \end{split}
\end{equation}
that is, a model of $1+1$ spacetime with gravity non-minimally coupled
with a scalar (called the \emph{dilaton} or the \emph{radion}) which
experiences self-interaction through a potential.  Note that the
dependence with respect to the dilaton is not relegated to the
potential, and so a direct stability analysis using $U(\xi)$ is not
possible.

The next step is to show that the Euler-Lagrange equations of the
previous two-dimensional model are equivalent to the Einstein
equations of its four-dimensional counterpart, that is, the truncation
of the model is consistent.  Firstly, we recast the Lagrangian
\eqref{eq:27} displaying explicitly all fields and removing the
second derivatives of $\phi$ by integrating by parts, which amounts to
the substitution
\begin{equation}
  \label{eq:29}
  \phi_{\alpha\alpha}\,e^{2\xi} \longrightarrow
  -2\phi_\alpha\xi_\alpha\,e^{2\xi}\,,
\end{equation}
for $\alpha\in\{t,x\}$.  The resulting Lagrangian is
\begin{equation}
  \label{eq:30}
  \begin{split}
  L^{(2)} &= -\frac{1}{2}\,e^{2\xi}\eta^{\alpha\beta}\phi_{\alpha\beta} -
  \frac{1}{2}\,e^{2\xi}\eta^{\alpha\beta}\xi_\alpha\xi_\beta -
  e^{2\phi} U(\xi)\\
  &= e^{2\xi}\eta^{\alpha\beta}\phi_\alpha\xi_\beta -
  \frac{1}{2}\,e^{2\xi}\eta^{\alpha\beta}\xi_\alpha\xi_\beta -
  e^{2\phi} U(\xi)\,.
  \end{split}
\end{equation}
The equations of motion are
\begin{equation}
  \label{eq:31}
  \eta^{\alpha\beta}\bigl(\phi_{\alpha\beta} - \xi_{\alpha\beta} -
  \xi_\alpha\xi_\beta\bigr)e^{2\xi} = -e^{2\phi}U'(\xi)
\end{equation}
with respect to $\xi$, and
\begin{equation}
  \label{eq:32}
  \eta^{\alpha\beta}\bigl(\xi_{\alpha\beta} +
  2\xi_\alpha\xi_\beta\bigr)e^{2\xi} = -2e^{2\phi}U(\xi)
\end{equation}
with respect to $\phi$.  The absence of perturbation $\xi=0$ should be
a solution of the equations, so that equation \eqref{eq:31} with
$\xi=0$ reduces to
\begin{equation}\label{eq:33}
  -e^{-2\phi}\eta^{\alpha\beta}\phi_{\alpha\beta} = 
  \bigl(-\phi_{xx} + \phi_{tt}\bigr)e^{-2\phi} = U'(0) = \lambda\,,
\end{equation}
which is the two-dimensional cosmological solution, and equation
\eqref{eq:32} gives
\begin{equation}\label{eq:34}
  U(0) = 0\quad\Rightarrow\quad
  \Lambda - K +
  \Bigl(\frac{QK}{4\pi}\Bigr)^2 = 0\,,
\end{equation}
which is the same equation previously found for $K$, see
\eqref{eq:18}.  We will now assume that $\xi$ is a small perturbation,
thereby neglecting the backreaction of the perturbation on the
geometry of the cosmological solution.  This allows us to fix $\phi$
as another background field by means of equation \eqref{eq:33}.  This
eliminates $\phi$ as a dynamical variable in the Lagrangian
\eqref{eq:30} and therefore equation \eqref{eq:32} will not be used.
In other words, we assume that equation \eqref{eq:32} is satisfied to
zeroth order, which is the content of \eqref{eq:34}, and we are left
with \eqref{eq:31} as the evolution equation for the perturbation
$\xi$ in the background $\phi$.

Thus, we substitute eq.~\eqref{eq:33} into eq.~\eqref{eq:31},
resulting in a dynamical equation for $\xi$ which is
\begin{equation}
  \label{eq:35}
  -e^{-2\phi}\eta^{\alpha\beta}\bigl(
  \xi_{\alpha\beta} + \xi_\alpha\xi_\beta
  \bigr)
  = \lambda - e^{-2\xi}U'(\xi)
  = - U'_{\text{eff}}(\xi)\,.
\end{equation}
The linear stability analysis of equation~\eqref{eq:35} requires its
linearization (the effect of neglecting the non-linear derivative term
does not spoil linear stability, as discussed in appendix
\ref{sec:deriv-couplings})
\begin{equation}
  \label{eq:36}
  e^{-2\phi}\bigl(\xi_{tt} - \xi_{xx}\bigr) = - U_{\text{eff}}''(0)\xi\,,
\end{equation}
which is a $1+1$ Klein-Gordon equation.  We also consider a small
spacetime region such that $\phi$ can be treated approximately as a
constant.  We thus obtain the linear stability condition of the
solution $\xi=0$ as being a minimum of the effective potential:
\begin{equation}
  \label{eq:37}
  U_{\text{eff}}''(0) = 4(2K-3\Lambda) > 0\,,
\end{equation}
which is $K>\frac{3}{2}\Lambda$.  All points in the
$K_+$ branch satisfy this condition, and therefore all $\text{AdS}_2$
states are stable, but this is not so in the $K_-$ branch.  The
condition $K_- > \frac{3}{2}\Lambda$ is met by all charges satisfying
\begin{equation}
  \label{eq:38}
  Q > Q_{\text{min}} = \frac{2\sqrt{2}}{3}\,Q_{\text{max}}\,.
\end{equation}
Upon quantization, the previous condition is
\begin{equation}
  \label{eq:39}
  n \ge n_{\text{min}} = \Bigl\lceil\frac{Q_{\text{min}}e}{2\pi}\Bigr\rceil\,.
\end{equation}
Thus, all states in the $\text{dS}_2$ branch characterized by a
quantum number $n<n_{\text{min}}$ are unstable.  This includes also
the state $n=0$, as mentioned above.  We conclude that the flux number
labeling true vacuum states should obbey a double inequality, obtained
by combining (\ref{eq:22}) and (\ref{eq:39}):
\begin{equation}
  \label{eq:40}
  n_{\text{min}} \le n \le n_{\text{max}}\,.
\end{equation}

Therefore, the number of stable states in this one-flux landscape is
\begin{equation}
  \label{eq:41}
  \mathcal{N}_1 = \underbrace{n_{\text{max}}}_{\text{AdS}_2} + 
  \underbrace{n_{\text{max}} - n_{\text{min}} + 1}_{\text{dS}_2}
  = 2\biggl\lfloor\frac{e}{\sqrt{\Lambda}}\biggr\rfloor
  -
  \Bigl\lceil\frac{2\sqrt{2}}{3}\,\frac{e}{\sqrt{\Lambda}}\Bigr\rceil
  + 1
  \approx \frac{e}{\sqrt{\Lambda}}\Bigl(2 - \frac{2\sqrt{2}}{3}\Bigr)\,.
\end{equation}
In the example shown in Figure \ref{fig:one-flux-em4} we have
$Q_{\text{max}} = 317.54$ (corresponding to $\Lambda =
3.915\times10^{-4}$) which gives $\mathcal{N}_1=53$ (50 $\text{AdS}_2$
states and only 3 $\text{dS}_2$ states).

The physical description of these states is simple: The
self-gravitating electromagnetic field of the monopole stabilizes a
geometry $\text{(A)dS}_2\times\text{S}^2$ whose natural behaviour is
decompactification\footnote{Note that equation \eqref{eq:36}, when the
  state is unstable, as happens in absence of electromagnetic field,
  predicts an exponential increase with time of the perturbation
  $\xi(t)$.  By inspecting equation \eqref{eq:23}, we see that it
  corresponds to decompactification.}, that is, the curvature of the
sphere part, which tends to vanish, is sustained by the magnetic
field.  The distribution of curvature contributions between the
compact and non-compact parts of the geometry is whatever allowed by
magnetic charge quantization in the $\text{AdS}_2$ case, while in the
$\text{dS}_2$ case only large charges can sustain positive curvature
of de Sitter states.

\section{Adding many fluxes}
\label{sec:many}

Obtaining a non-trivial landscape with many fluxes is not easy.  The
easiest technique is to extend the electromagnetic field $F$ to a
$SO(J)$-invariant $J$-component field, in which all charges are
different \cite{BP,Randall-Jhonson-Carroll}.  This approach is not
convenient to address the problem of stabilization, because the
charges do not come from a transparent compactification scheme, and
therefore nothing is known about the stabilizing potential.

In all known compactifications, the charges come from moduli
describing the shape of the inner manifold.  These moduli are free
geometric parameters, but they are promoted to dynamical scalar fields
in the perturbation analysis.  The charges are considered coupling
\emph{constants}, and therefore their dynamics should be governed by a
potential with at least one minimum.  The stabilization problem
consists of finding this potential.  Different models provide a wide
variety of potentials; if the potential does not possess any minimum,
then the dynamics of the moduli will lead them to grow unbounded; this
phenomenon is known as \emph{decompactification}.

In an ideal model, we should expect that all moduli come from a
compactification manifold which is derived from the equations of
motion.  Nevertheless, these equations are very difficult to solve in
its full generality, and therefore the inner manifold is chosen at the
very beginning of the process, and its validity is confirmed
afterwards, by proving that the chosen ansatz is actually a solution.
Of course, the chosen manifold may not provide a solution, or the
solution may lack some desired properties.

In looking for a many-fluxes landscape, we have tried the following
candidates as compactification manifold:
\begin{itemize}
\item The complex Riemann curves 
  \begin{equation*}
    w^2 = P_k(z)\,,
  \end{equation*}
  where $w$ and $z$ are complex coordinates, are the simplest surfaces
  of known genus $\mathfrak{g}>0$. $P_k(z)$ is a $k$-degree polynomial
  with real coefficients,
  \begin{equation*}
    P_k(z) = z^k + a_{k-2}z^{k-2} + \cdots + a_1z + a_0\,,
  \end{equation*}
  which are the $k-1$ deformation moduli of the surface.  The
  coefficients of $z^k$ and $z^{k-1}$ can be removed by rescaling and
  shifting $z$ respectively.  The genus of the surface is given by
  $k=2\mathfrak{g}+2$ (if $k$ is even) or $k=2\mathfrak{g}+1$ (if $k$ is odd).

  These surfaces are not compact, but they can approximate locally a
  compact surface of the same genus.  Thus, Einstein equations are to
  be solved only locally near the approximation region.  But this
  ansatz turns out to yield no solutions, not even in this approximate
  fashion.
\item The compact hyperbolic manifolds (CHM) are fundamental domains
  of nonabelian lattices in the Lobachevskian plane, in which the
  lattices are generated by discrete subgroups of $SL_2(\R)$.  By
  choosing identification of the sides of the fundamental cell a
  compact surface of genus $\mathfrak{g}\ge2$ is obtained, which has
  constant negative
  curvature\footnote{Refs.~\cite{Hyperb-1,Hyperb-2,Hyperb-3,Hyperb-4,Hyperb-5,Hyperb-6}
    contain more details on hyperbolic compactifications in
    cosmology. }.  These surfaces are indeed solutions of Einstein
  equations, but only if the cosmological constant $\Lambda$ is
  \emph{negative}.  This generates a sector of the Einstein-Maxwell
  landscape with no de Sitter states, and therefore will not be
  considered here.
\end{itemize}
The previous examples show that sometimes the appropiate
compactification can be elusive, maybe because is far more complex
than expected, or because it may not exist.  So we are forced to
consider another simple extension of the four dimensional
Einstein-Maxwell landscape, which is discussed in the following
subsections.

\subsection{Multi-sphere compactification}
\label{sec:multi-sphere}

We consider a $J$-sphere metric ansatz given by
\begin{equation}
  \label{eq:42}
  \dif s^2 = e^{2\phi(x,t)}\bigl(-\dif t^2 + \dif x^2\bigr)
  + \sum^J_{i=1} e^{2\psi_i(u,v)}\bigl(\dif u_i^2 + \dif v_i^2\bigr)\,.
\end{equation}
The metric \eqref{eq:42} represents a manifold of the form
\begin{equation}
  \label{eq:43}
  \mathcal{K} = \text{(A)dS}_2\times
  \overbrace{\text{S}^2\times\cdots\times\text{S}^2}^{\text{$J$
      spheres}}
  = \text{(A)dS}_2\times\bigl[\text{S}^2\bigr]^J
\end{equation}
which is nothing but a sector of the Einstein-Maxwell theory in $2J+2$
dimensions.  The $\phi$ exponent defines the conformal factor of a
two-dimensional cosmological spacetime in $(t,x)$ coordinates.  The
functions $\psi_i$ depend only on the corresponding coordinates
$(u_i,v_i)$ (but not on $(u_j,v_j)$ with $j\ne i$), and they give a
conformal representation of the $i$-th sphere metric.

The Ricci tensor of the metric \eqref{eq:42} is
\begin{equation}
  \label{eq:44}
  \begin{split}
    R_{xx} &= - R_{tt} = \phi_{tt} - \phi_{xx}\,,\\
    R_{u_iu_i} &= R_{v_iv_i} = -\Delta_i\psi_i\,,
  \end{split}
\end{equation}
all remaining components being zero.  The $i$-th Laplacian operator is
$\Delta_i = \partial_{u_i}^2 + \partial_{v_i}^2$.

The metric \eqref{eq:42} should be complemented with the
electromagnetic field
\begin{equation}
  \label{eq:45}
  \mathbf{F} = \sum_{i=1}^J B_i(u_i,v_i)\,\dif u_i\wedge\dif v_i
\end{equation}
where the magnetic $u_i,v_i$-component $B_i$ depends only on the
corresponding coordinates $(u_i,v_i)$ (but not on $(u_j,v_j)$ with
$j\ne i$).  Thus, the electromagnetic tensor $F_{\mu\nu}$ is analogous
to \eqref{eq:10}, with $J$ $2\times2$ blocks along the diagonal.
Maxwell equations \eqref{eq:11} reduce to
\begin{equation}
  \label{eq:46}
  \partial_{u_i}\sqrt{-g}\,F^{v_i u_i} = 0\,,
  \qquad
  \partial_{v_i}\sqrt{-g}\,F^{u_i v_i} = 0\,.
\end{equation}
Using the volume element prefactor $\sqrt{-g}=e^{2\phi + 2\sum_{i=1}^J
\psi_i}$ and the relation $F^{u_iv_i} = e^{-4\psi_i}F_{u_iv_i}$, we
find that a solution of the equations is
\begin{equation}
  \label{eq:47}
  F_{u_iv_i} = B_i = \frac{Q_i}{V_i}\,e^{2\psi_i}\,,
\end{equation}
where $V_i$ is the volume of the $i$-th sphere,
\begin{equation}
  \label{eq:48}
  V_i = \int_{\text{S}^2} e^{2\psi_i}\,\dif u_i\wedge\dif v_i\,,
\end{equation}
and $Q_i$ is an integration constant.  When integrating the two-form
we obtain
\begin{equation}
  \label{eq:49}
  \int \mathbf{F}
  = \sum_{i=1}^J \frac{Q_i}{V_i}\int_{\text{S}^2} e^{2\psi_i}
  \,\dif u_i\wedge\dif v_i
  = \sum_{i=1}^J Q_i = Q\,,
\end{equation}
so that $Q$ is the total magnetic charge of the configuration, and
each constant $Q_i$ can be interpreted as the magnetic charge
contribution of the corresponding magnetic field component.

The most convenient way of obtaining the field equations is writing
the action
\begin{equation}
  \label{eq:50}
  S = \frac{1}{16\pi}\int \sqrt{-g}\bigl(R - 2\Lambda - F^2\bigr)\,
  \dif t\,\dif x\prod_{i=1}^J\dif u_i\,\dif v_i
\end{equation}
in terms of the fields $\phi$, $\psi_i$ and $B_i$ and then derive the
equations from it.  The curvature scalar of the metric ansatz
\eqref{eq:42} is
\begin{equation}
  \label{eq:51}
  R = 2(\phi_{tt} - \phi_{xx})e^{-2\phi} - 2\sum_{i=1}^J
  \Delta_i\psi_i\,e^{-2\psi_i} \,.
\end{equation}
The Lagrangian density of the electromagnetic field \eqref{eq:45} is
\begin{equation}
  \label{eq:52}
  F^2 = 2\sum_{i=1}^J B_i^2\,e^{-4\psi_i}\,.
\end{equation}
Thus, the action specialized to our ansatz is
\begin{equation}
  \label{eq:53}
  S = \frac{1}{8\pi}\int e^{2\phi + 2\sum_{i=1}^J \psi_i}
  \Bigl[
  (\phi_{tt} - \phi_{xx})e^{-2\phi} - \sum_{i=1}^J
  \Delta_i\psi_i\,e^{-2\psi_i}
  - \Lambda
  - \sum_{i=1}^J B_i^2\,e^{-4\psi_i}
  \Bigr]\,
  \dif t\,\dif x\prod_{i=1}^J\dif u_i\,\dif v_i
\end{equation}
The variation of the action \eqref{eq:53} with respect to the vector
potential $A_\mu$ (which determines the magnetic field $B_i
= \partial_{u_i} A_{v_i} - \partial_{u_i} A_{v_i}$) gives the Maxwell
equations \eqref{eq:46}.  Varying with respect to $\phi$ and
$\psi_j$ gives
\begin{equation}
  \label{eq:54}
  \begin{split}
    \Lambda &= -\sum_{i=1}^J \Bigl\{\Delta_i\psi_i\,e^{-2\psi_i} +
    B_i^2\,e^{-4\psi_i}\Bigr\}\,,\\
    \bigl(\phi_{tt} - \phi_{xx}\bigr)e^{-2\phi} &=
    \Lambda + \sum_{i=1}^J \Bigl\{\Delta_i\psi_i\,e^{-2\psi_i} +
    B_i^2\,e^{-4\psi_i}\Bigr\}
    - \Delta_j\psi_j\,e^{-2\psi_j}
    - 2 B_j^2\,e^{-4\psi_j}\,.
  \end{split}
\end{equation}
Note that the first equation in \eqref{eq:54} cancels some terms in
the second.  Now, we recognize the Gaussian curvatures of conformally
flat metrics with signatures $-+$ and $++$; so we substitute the
constant curvature ansatz implied in the geometry of $\mathcal{K}$
\eqref{eq:43} as we did previously \eqref{eq:16}, introducing the
constants
\begin{equation}
  \label{eq:55}
  \lambda = \bigl(\phi_{tt} - \phi_{xx}\bigr)e^{-2\phi}\,,\qquad
  K_i = -\Delta_i\psi_i\,e^{-2\psi_i}\,,
\end{equation}
where $\lambda$ is the curvature of the non-compact part (the
cosmological constant of the two-dimensional spacetime) and $K_i$ is
the curvature of the $i$-th sphere in the product $[\text{S}^2]^J$.
If we finally substitute $B_i$ by the solution \eqref{eq:47}, we
obtain
\begin{align}
  \label{eq:56}
  \Lambda &= \sum_{i=1}^J \Bigl\{
  K_i - \biggl(\frac{Q_i}{V_i}\biggr)^2
  \Bigr\}\,, \\
   \label{eq:57}
  \lambda &= K_j - 2\biggl(\frac{Q_j}{V_j}\biggr)^2\,.
\end{align}
We can substitute equation \eqref{eq:57} in \eqref{eq:56}, obtaining
\begin{equation}
  \label{eq:58}
  \Lambda = \frac{1}{2}\biggl(J\lambda + \sum_{i=1}^J K_i\biggr)\,.
\end{equation}
The relation \eqref{eq:17} between volume and curvature is valid, and
transforms equation \eqref{eq:57} in an algebraic equation for $K_j$:
\begin{equation}
  \label{eq:59}
  2\biggl(\frac{Q_jK_j}{4\pi}\biggr)^2 - K_j + \lambda = 0\,,
\end{equation}
which has two solutions
\begin{equation}
  \label{eq:60}
  K_j^{(\pm)} = \frac{4\pi^2}{Q_j^2}
  \left[1 \pm \sqrt{1 - 2\lambda\,\frac{Q_j^2}{4\pi^2}}\ \right]\,.
\end{equation}
Substituting \eqref{eq:60} in equation \eqref{eq:58}, we obtain a
single equation for $\lambda$, whose solution can be substituted back
in equation \eqref{eq:60}, determining the curvatures.

It should be noted that if $Q_j=0$, then equation \eqref{eq:59} has a
single solution, namely $K_j = \lambda$, which is the limit of the
$K^{(-)}_j$ solution when $Q_j\to0$.

Now, the usual Dirac quantization condition is
\begin{equation}
  \label{eq:61}
  Q_je = 2\pi n_j\quad\text{with $n_j\in\Z$}\,,
\end{equation}
which can be justified in the following way.  We will use the
conformal representation of the sphere metric:
\begin{equation}
  \label{eq:62}
  \bigl(\dif s^2\bigr)_{\text{S}^2_j}
  = \frac{\dif u_j^2 + \dif v_j^2}{K_j\cosh^2 u_j}
\end{equation}
such that the $u_j$ coordinates separate hemispheres ($u_j > 0$ is the
northern hemisphere, $u_j=0$ is the equator, etc.) and $v_j$ are
angles mod $2\pi$.  A quantum wavefunction $\Psi$ defined on the
manifold $\mathcal{K}$ depends on the coordinates $t$, $x$,
$\{u_j,v_j\}_{j=1,\cdots,J}$.  The loops $\gamma_j$ in which $v_j$
varies along $[0,2\pi]$ and the remaining coordinates are fixed can be
used to define holonomies $h_j$ acting on the wavefunction:
\begin{equation}
  \label{eq:63}
  h_j\Psi = \exp\Bigl(ie\int_{\gamma_j}A\Bigr)\Psi\,,
\end{equation}
where the electromagnetic potential is used as the connection to
parallel transport the wavefunction values along the loop.  It is well
known that the potential of a magnetic monopole can be defined in two
patches on the sphere which overlap at the equator:
\begin{equation}
  \label{eq:64}
  A = \sum_{i=1}^J A_j\,,\qquad
  A_j =
  \begin{cases}
    A^{(+)}_j = \frac{Q_j}{K_jV_j}\,\bigl[\tanh u_j - 1\bigr]\dif v_j 
    & \text{if $u_j \ge 0$,}\\
    A^{(-)}_j = \frac{Q_j}{K_jV_j}\,\bigl[\tanh u_j + 1\bigr]\dif v_j
    & \text{if $u_j \le 0$.}\\
  \end{cases}
\end{equation}
Here, $V_j=\frac{4\pi}{K_j}$, as usual, and the magnetic field is
\begin{equation}
  \label{eq:65}
  B_j = \dif A_j^{(\pm)} = 
  \frac{Q_j}{K_jV_j}\,\frac{\dif u_j\wedge\dif v_j}{\cosh^2u_j}
  = \frac{Q_j}{V_j}\,e^{2\psi_j}\dif u_j\wedge\dif v_j\,.
\end{equation}
Thus, the potential is discontinuous at the equator, but the
discontinuity is given by a gauge transformation $\chi_j$, namely
\begin{equation}
  \label{eq:66}
  A^{(+)}_j - A^{(-)}_j = -\frac{2Q_j}{K_jV_j}\,\dif v_j = \dif \chi_j\,.
\end{equation}
If we move the loop slightly upwards or downwards from the equator,
the discontinuity in $A$ will leave a different phase on $\Psi$, thus
violating gauge invariance, unless the phase difference is an integer
times $2\pi$, that is, equation \eqref{eq:61}.

Therefore, we can substitute \eqref{eq:61} in the equation satisfied
by $\lambda$ \eqref{eq:58}, obtaining
\begin{equation}
  \label{eq:67}
  \Lambda = \frac{1}{2}\Bigl[J\lambda + \sum_{j=1}^J \frac{e^2}{n_j^2}
  \left(1 \pm \sqrt{1 - 2\frac{\lambda}{e^2}\,n_j^2}\ \right)
  \Bigr]\,.
\end{equation}
Note that $e^2$ is a scale which can be used to measure $\Lambda$ and
$\lambda$.  In order to simplify the formulae, we will assume
henceforth that the substitutions $\frac{\lambda}{e^2}\to\lambda$ and
$\frac{\Lambda}{e^2}\to\Lambda$ have been made.  Equation
\eqref{eq:67} takes the form
\begin{equation}
  \label{eq:68}
  L_{\{s_j\},\{n_j\}}(\lambda) = \Lambda\,,\quad\text{with}\quad
  L_{\{s_j\},\{n_j\}}(\lambda) = 
  \frac{1}{2}\Bigl[J\lambda + \sum_{j=1}^J \frac{1}{n_j^2}
  \left(1 + s_j \sqrt{1 - 2\lambda\,n_j^2}\ \right)
  \Bigr]\,.
\end{equation}
The function $L_{\{s_j\},\{n_j\}}(\lambda)$ depends on the signs
$\{s_j\}$ of the curvatures involved and on the winding numbers
$\{n_j\}$ of the magnetic field.  When some $n_j$ vanishes, the
curvature should be taken as $K_j=K_j^{(-)}=\lambda$, the $K^{(+)}_j$
branch is absent.  The solutions of equation \eqref{eq:68} constitute
the states of the multi-sphere Einstein-Maxwell landscape.

\begin{figure}[htbp]
  \centering
  \includegraphics[width=0.75\textwidth]{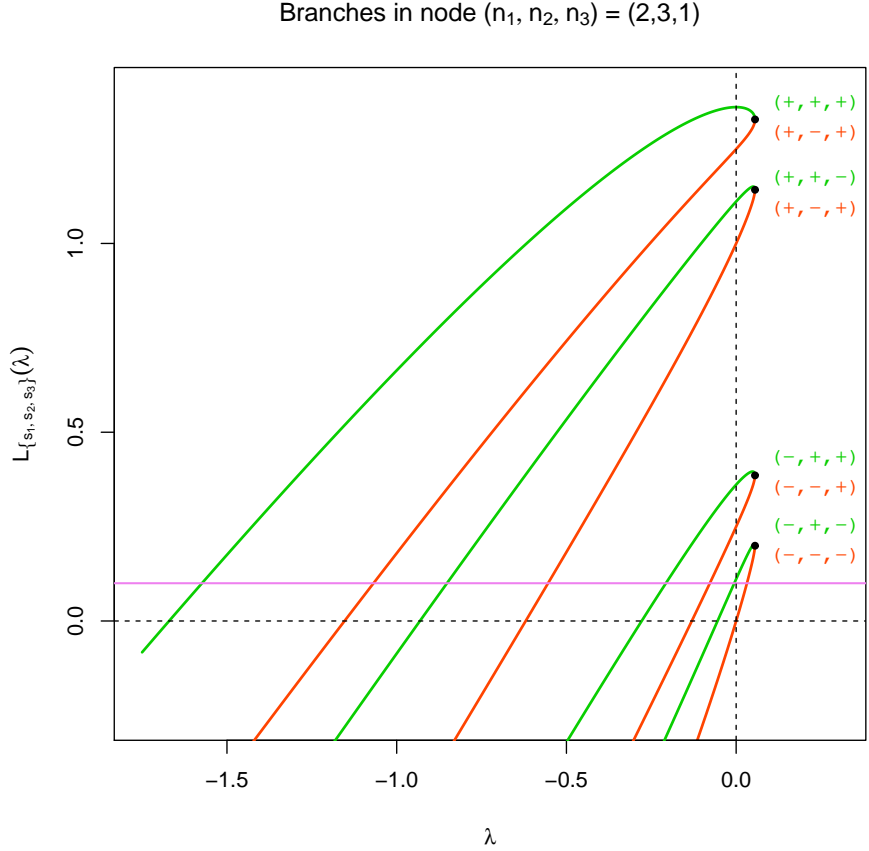}
  \caption{The eight branches of the $J=3$ multi-sphere
    Einstein-Maxwell landscape at the node $(n_1,n_2,n_3) = (2,3,1)$
    are shown.  Note that each pair of branches meet at a branching
    point, whose horizontal position is the same in all branches, see
    text.  The solid horizontal line corresponds to a random value of
    the cosmological constant $\Lambda$; with this choice, there are
    eight solutions of the equation $L(\lambda)=\Lambda$, yielding a
    single de Sitter state and seven anti-de Sitter states.  Note that
    if $\Lambda$ were at 0.5 height, there would be only four
    solutions, and if it were at 1.5 height, there would be no
    solutions at all. All statistic plots in this paper were done
    using R \cite{R-proj}.}
\label{fig:J3-branches}
\end{figure}

The set of integers $\{n_j\}$ will be called a \emph{node}, while the
set of signs $\{s_j\}$ will be called a \emph{branch} of the
$L_{\{s_j\},\{n_j\}}$ function.  Fixing a node with all $n_j$ nonzero,
we have $2^J$ branches, one for each possible choice of the signs
$\{s_j\}$.  The number of solutions of the equation is different for
each $\Lambda$ value; in figure \ref{fig:J3-branches} it is shown an
example with $J=3$, where the eight branches of the node
$(n_1,n_2,n_3) = (2,3,1)$ are shown.  The displayed value of $\Lambda$
yields eight solutions, but is it obvious from figure
\ref{fig:J3-branches} that the number of solutions vary when $\Lambda$
is moved upwards, and becomes zero when the $\{+,+,+\}$ branch is
surpassed.  The highest branch will be always the all-$+$ branch, and
will be called the \emph{principal} branch.

Not all solutions of equation \eqref{eq:68} give valid states; for
example, curvatures \eqref{eq:60} can become complex if $\lambda$ is
too large.  The smallest value of $\lambda$ at which some of the pairs
of curvatures $K_j^{(\pm)}$ meet is the \emph{branching point} of the
node, and is given by
\begin{equation}
  \label{eq:69}
  \lambda_{\text{b}} = \frac{1}{2\max_{1\le j\le J}\{n_j^2\}}\,.
\end{equation}
This is the maximum positive value that $\lambda$ can achieve when all
$n_j\ne0$.  Curvatures can also become negative; this happens when
$\lambda<0$ in all branches except the principal one.  Those states
are not well defined, but we might define them in detail by replacing
the corresponding sphere by a CHM.  Nevertheless, to keep things
simple, this sector of the landscape will be deliberately left out,
because all its states are AdS.  This leaves the principal branch as
the only source of AdS states with positive curvature in all places of
the compact part.

The problem of the stability of the states just found is addressed in
subsection \ref{sec:moduli-stab}, and the problem of counting them is
the subject of section \ref{sec:state-counting}.

\subsection{Moduli stabilization}
\label{sec:moduli-stab}

The next step in the analysis is to determine if the states of the
multi-sphere Einstein-Maxwell landscape are stable or not.  Our
approach will follow closely that of subsection
\ref{sec:modulus-stab}.  We begin by introducing perturbations
$\xi_j(t,x)$ which represent deviations of the curvatures found as
solutions of the Einstein equations; the perturbed metric ansatz in
the Einstein frame is
\begin{equation}
  \label{eq:70}
  \dif s^2 =
  e^{2\phi(t,x) - 2\sum_{j=1}^J \xi_j(t,x)}
  \bigl(-\dif t^2 + \dif x^2\bigr) +
  \sum_{j=1}^J
  e^{2\psi_j(u_j,v_j) + 2\xi_j(t,x)} \bigl( \dif u_j^2 + \dif v_j^2\bigr)\,.  
\end{equation}
As before, $\phi(t,x)$ represents a $1+1$ cosmological solution, and
$\psi_j(u_j,v_j)$ corresponds to the metric of the internal spheres in
conformal coordinates.  The perturbations $\xi_j(t,x)$ describe
changes in the radii of the internal spheres, and thus they will be
referred to as the \emph{multi-radion} fields.

We will proceed by writing the action \eqref{eq:50} specialized for
the metric \eqref{eq:70}.  The Ricci scalar is
\begin{multline}
  \label{eq:71}
  R = 2\,e^{-2\phi + 2\sum_{j=1}^J \psi_j}
  \biggl\{
  \phi_{tt} - \phi_{xx}
  + \sum_{j=1}^J
  \biggl[
  \bigl(\xi_j\bigr)_{tt} - \bigl(\xi_j\bigr)_{xx}
  + 3\bigl(\xi_j\bigr)_{t}^2 - 3\bigl(\xi_j\bigr)_{x}^2
  \\
  + 2\bigl(\xi_j\bigr)_{t}\Bigl(\sum_{k\ne j} \xi_k \Bigr)_t
  - 2\bigl(\xi_j\bigr)_{x}\Bigl(\sum_{k\ne j} \xi_k \Bigr)_x
  - e^{2\phi - 2\psi_j - 2\xi_j - 2\sum_{k=1}^J \xi_k}\Delta_j\psi_j
  \biggr]
  \biggr\}
\end{multline}
Note that the previous expression involves second derivatives of the
multi-radion fields.  We can replace those terms by first derivatives
by integrating by parts in the action.  This amounts to the following
replacement rule:
\begin{equation}
  \label{eq:72}
  \bigl(\xi_j\bigr)_{tt} e^{2\sum_{k=1}^J \xi_k} \longrightarrow
  -2 \bigl(\xi_j\bigr)_{t}\Bigl(\sum_{k=1}^J \xi_k\Bigr)_t
  e^{2\sum_{k=1}^J \xi_k}\,,
\end{equation}
and another analogous equation with the $x$ derivatives.

The next step is to note that the expression \eqref{eq:47} for the
magnetic field remains unchanged, because the perturbations appear in
Maxwell equations \eqref{eq:46} only as factors depending on variables
$(t,x)$, and thus they can be factored out of the equations.  Thus,
the contribution of the magnetic field to the action is not exactly
\eqref{eq:52} but
\begin{equation}
  \label{eq:73}
  F^2 = 2\sum_{j=1}^J B_j^2\,e^{-4\psi_j-4\xi_j}\,,
\end{equation}
with the same magnetic field $B_j = \frac{Q_j}{V_j}\,e^{2\psi_j}$.

Finally, we will insert in the action the unperturbed sphere metric
ansatz $-e^{2\psi_j}\Delta_j\psi_j = K_j$.

Gathering all these ingredients, we obtain the following Lagrangian
for the perturbed metric:
\begin{equation}
  \label{eq:74}
  \begin{split}
    L &= \sqrt{-g}\bigl(R - 2\Lambda - F^2\bigr) \\
    &= 2\,e^{2\phi + 2\sum_{k=1}^J \psi_k}
    \biggl\{
    \bigl(\phi_{tt} - \phi_{xx}\bigr)e^{-2\phi + 2\sum_{k=1}^J \xi_k}
    - \Lambda
    \\
    &+ \sum_{j=1}^J
    \biggl[
    \Bigl(\bigl(\xi_j\bigr)_{t}^2 - \bigl(\xi_j\bigr)_{x}^2\Bigr)
    e^{-2\phi + 2\sum_{k=1}^J \xi_k}
    + K_j\,e^{-2\xi_j}
    - \Bigl(\frac{Q_j}{V_j}\Bigr)^2 e^{-4\xi_j}
    \biggr]
    \biggr\}\,.
  \end{split}
\end{equation}
Now, we can perform the integration in the internal variables
$\{u_j,v_j\}$ and thus obtain a dimensionally reduced model for the
cosmological part together with the multi-radion fields:
\begin{equation}
  \label{eq:75}
  S = \frac{1}{16\pi}\int L\,\dif t\,\dif x\prod_{j=1}^J
  \dif{u_j}\,\dif{v_j}
  = \frac{1}{4\pi}\biggl(\prod_{j=1}^J V_j\biggr) \int
  L^{(2)}\,\dif{t}\,\dif{x}\,,   
\end{equation}
with a $1+1$ Lagrangian
\begin{equation}
  \label{eq:76}
  L^{(2)} = \sqrt{-g^{(2)}}
  \left\{
    \frac{1}{4}\,e^{2\sum_{k=1}^J \xi_k} R^{(2)} -
    \frac{1}{2}\,e^{2\sum_{k=1}^J \xi_k}\sum_{j=1}^J
    \bigl(g^{(2)}\bigr)^{\alpha\beta}\bigl(\xi_j\bigr)_{\alpha}\bigl(\xi_j\bigr)_{\beta}
    - \sum_{j=1}^J U_j(\xi_j)
  \right\}
\end{equation}
where the summation in $\alpha,\beta\in\{t,x\}$ is implied, and we
have used (see \eqref{eq:28})
\begin{equation}
  \label{eq:77}
  \begin{split}
    \bigl(g^{(2)}\bigr)_{\alpha\beta}\,\dif\alpha\,\dif\beta
    &=    e^{2\phi}\eta^{\alpha\beta}\,\dif\alpha\,\dif\beta
    = e^{2\phi}\bigl(-\dif t^2 + \dif x^2\bigr)\,,\\
    R^{(2)} &= -2\bigl(\phi_{xx} - \phi_{tt}\bigr)e^{-2\phi}\,, \\
    U_j(\xi_j) &= \frac{1}{2}\biggl[
    \frac{\Lambda}{J} - K_j\,e^{-2\xi_j} +
    \biggl(\frac{Q_jK_j}{4\pi}\biggr)^2 e^{-4\xi_j}
    \biggr]\,.
  \end{split}
\end{equation}
It is apparent from \eqref{eq:76} that all radions are coupled by the
dilatonic factors.  When exhibiting all fields explicitly we obtain
\begin{equation}
  \label{eq:78}
  \begin{split}
    L^{(2)} &= -\frac{1}{2}\,e^{2\sum_k \xi_k}\,
    \eta^{\alpha\beta}\phi_{\alpha\beta} - \frac{1}{2}\,e^{2\sum_k
      \xi_k}\, \sum_{j=1}^J \eta^{\alpha\beta}
    \bigl(\xi_j\bigr)_{\alpha}\bigl(\xi_j\bigr)_{\beta} -
    e^{2\phi}\sum_{j=1}^J U_j(\xi_j)\,,\\
    &= \,e^{2\sum_k \xi_k}\,
    \eta^{\alpha\beta}\phi_{\alpha} \sum_{j=1}^J
    \bigl(\xi_j\bigr)_{\beta} - \frac{1}{2}\,e^{2\sum_k \xi_k}\,
    \sum_{j=1}^J \eta^{\alpha\beta}
    \bigl(\xi_j\bigr)_{\alpha}\bigl(\xi_j\bigr)_{\beta} -
    e^{2\phi}\sum_{j=1}^J U_j(\xi_j)\,,
  \end{split}
\end{equation}
which is the generalization of the Lagrangian previously found for
$J=1$, see eq.~\eqref{eq:30}.  Note that a substitution rule
analogous to \eqref{eq:29}
\begin{equation}
  \label{eq:79}
  \phi_{\alpha\alpha}\,e^{2\sum_k\xi_k} \longrightarrow
  -2\phi_\alpha\sum_{j=1}^J \bigl(\xi_j\bigr)_\alpha\,e^{2\sum_k\xi_k}\,,
\end{equation}
has been used in passing from the first line to the second in
\eqref{eq:78}.  The equations of motion are as follows;
$\partial_\alpha\frac{\partial L^{(2)}}{\partial\phi_\alpha} =
\frac{\partial L^{(2)}}{\partial\phi}$ is
\begin{equation}
  \label{eq:80}
  e^{2\sum_k\xi_k}\eta^{\alpha\beta}
  \biggl[
  2\Bigl(\sum_k(\xi_k)_\alpha\Bigr)\Bigl(\sum_k(\xi_k)_\beta\Bigr)
  + \sum_k(\xi_k)_{\alpha\beta}
  \biggr]
  = -2e^{2\phi}\sum_k U_k(\xi_k)\,,
\end{equation}
and $\partial_\alpha\frac{\partial L^{(2)}}{\partial(\xi_j)_\alpha} =
\frac{\partial L^{(2)}}{\partial\xi_j}$ are
\begin{equation}
  \label{eq:81}
  e^{2\sum_k\xi_k}\eta^{\alpha\beta}
  \biggl[
  \phi_{\alpha\beta} - (\xi_j)_{\alpha\beta}
  - 2(\xi_j)_\alpha\Bigl(\sum_k(\xi_k)_\beta\Bigr)
  + \sum_k(\xi_k)_\alpha (\xi_k)_\beta
  \biggr]
  = -e^{2\phi} U'_j(\xi_j)\,.
\end{equation}
The absence of perturbations should be a solution of the previous
equations.  We can verify this requirement by substituting $\xi_j=0$
in \eqref{eq:81}:
\begin{equation}
  \label{eq:82}
  -\eta^{\alpha\beta}\phi_{\alpha\beta}\,e^{-2\phi} = U'_j(0) = \lambda\,,
\end{equation}
which is the cosmological solution previously obtained
\eqref{eq:16},\eqref{eq:55}.  Using eq.~\eqref{eq:77},
eq.~\eqref{eq:82} is reduced to
\begin{equation}
  \label{eq:83}
  U'_j(0) = K_j - 2\biggl(\frac{Q_jK_j}{4\pi}\biggr)^2 = \lambda\,,
\end{equation}
that is, exactly equation \eqref{eq:59} determining the curvatures.
By substituting $\xi_j=0$ in \eqref{eq:80} we obtain
\begin{equation}
  \label{eq:84}
  \sum_j U_j(0) = 0
  \quad\Rightarrow\quad
  \frac{1}{2}\sum_j \biggl[\frac{\Lambda}{J} - \frac{K_j}{2} -
  \frac{\lambda}{2}\biggr] = 0\,,
\end{equation}
which is exactly equation \eqref{eq:58}.  Thus, both equations
\eqref{eq:83},\eqref{eq:84} are the correct, unperturbed ones.

At this point, we proceed as in subsection \ref{sec:modulus-stab} by
fixing the cosmological background $\phi$ by means of \eqref{eq:82}.
This fixing is equivalent to neglecting the backreaction of the
perturbations $\xi_j$ on the cosmology $\phi$.  This being done,
$\phi$ is not a degree of freedom anymore, and the variation of the
Lagrangian $L^{(2)}$ in \eqref{eq:78} with respect to $\phi$ is
meaningless.  This background fixing step can also be viewed as
solving equation \eqref{eq:80} to zeroth order.  In this
approximation, the multi-radion field moves in a fixed cosmological
background and its evolution is determined by equation~\eqref{eq:81}.

We now turn to the analysis of equation~\eqref{eq:81}.
It is a strongly coupled, nonlinear system of equations with only
one known solution, $\xi_j=0$, which gives rise to the landscape under
study.  There is no hope of finding a nontrivial solution to this
system; nevertheless, we are only interested in a description of the
stability of the trivial solution.

To this end, we substitute \eqref{eq:82} back to equation
\eqref{eq:81}, obtaining
\begin{equation}
  \label{eq:85}
  -e^{-2\phi}\eta^{\alpha\beta}
  \biggl[
  (\xi_j)_{\alpha\beta}
  + 2(\xi_j)_\alpha\Bigl(\sum_k(\xi_k)_\beta\Bigr)
  - \sum_k(\xi_k)_\alpha (\xi_k)_\beta
  \biggr]
  = 
  \lambda - e^{-2\sum_k\xi_k} U'_j(\xi_j)\,.
\end{equation}
The derivative couplings appear in a quadratic form.  The linear
stability analysis allows us to approximate the system of equations by
Taylor-expanding to first order the right-hand side of equation
\eqref{eq:85} and neglecting the quadratic derivative couplings, thus
considering the much simpler linear system (written in matrix form)
\begin{equation}
  \label{eq:86}
  -e^{-2\phi}\eta^{\alpha\beta}
  \boldsymbol{\xi}_{\alpha\beta}
  = 
  -H\boldsymbol{\xi}\,.
\end{equation}
We have used the symbol $\boldsymbol{\xi}$ to denote the $J$-component
column vector of the perturbations $\xi_j$, and the frequency matrix
$H$ is given by
\begin{equation}
  \label{eq:87}
  H_{jk} = \left.
    \frac{\partial }{\partial\xi_k} e^{-2\sum_\ell\xi_\ell} U'_j(\xi_j)
  \right|_{\boldsymbol{\xi}=\boldsymbol{0}}
  = 2\bigl[(K_j-2\lambda)\delta_{jk} - \lambda\bigr]\,,
\end{equation}
that is,
\begin{equation}
  \label{eq:88}
  H = 2
  \begin{pmatrix}
    K_1 -3\lambda &      -\lambda & \cdots & -\lambda \\
    -\lambda      & K_2 -3\lambda & \cdots & -\lambda \\
    \vdots        &       \vdots  & \ddots & \vdots   \\
    -\lambda      &      -\lambda & \cdots & K_J - 3\lambda
  \end{pmatrix}
  \,.
\end{equation}
The linear stability condition is therefore that all eigenvalues of
the matrix $H$ should be positive.  We will call $\kappa$ the minimum
eigenvalue of $H$, so that the stability criterion is simply
\begin{equation}
  \label{eq:89}
  \kappa > 0\,.
\end{equation}
In order to complete the linear stability analysis we should justify
the neglecting of the derivative couplings.  These non-linear terms
have a generically negative sign\footnote{Note that the derivative
  couplings appear in a quadratic form having all negative eigenvalues
  except for one, see the paragraph following equation \eqref{eq:154}
  in appendix \ref{sec:deriv-couplings}.}, as opposed to a positive
sign characteristic of a dissipative force, and thus they could be
interpreted as an ``anti-dissipative'' force.  It is legitimate to ask
if these non-linear terms can spoil the linear stability of the
solution.

It turns out that these nonlinear terms do not spoil the linear
stability criterion \eqref{eq:89} as long as the amplitude of the
perturbation is sufficiently small.  The magnitude of the threshold and
the corresponding heuristic argument leading to these conclusions, not
being central to this discussion, have been placed in appendix
\ref{sec:deriv-couplings}.

The exact computation of the spectrum of $H$ is not possible in the
general case.  Therefore, we cannot derive a formula $\kappa(\lambda)$
to quickly establish the stability of a state.  As a result, the
computation of $\kappa$ should be done numerically on each individual
state.  Nevertheless, we can obtain some general stability results by
computing the determinant of $H$, a task which can be done as follows.

Firstly we note that if all $K_j$ are equal, the determinant of $H$
would be (we drop from now on the unimportant factor 2 of $H$) the
characteristic polynomial of the matrix
\begin{equation}
  \label{eq:90}
  B =
  \begin{pmatrix}
    3\lambda &  \lambda & \cdots &  \lambda \\
    \lambda  & 3\lambda & \cdots &  \lambda \\
    \vdots   &  \vdots  & \ddots &  \vdots  \\
    \lambda  &  \lambda & \cdots & 3\lambda
  \end{pmatrix}
  \,.
\end{equation}
This matrix has an eigenvalue $2\lambda$ with degeneracy $(J-1)$ and
a simple eigenvalue $(J+2)\lambda$.  Its characteristic polynomial is
\begin{equation}
  \label{eq:91}
  \det(K\uno - B) = \bigl[K - (J+2)\lambda\bigr]
  \prod_{j=1}^{J-1}\bigl(K-2\lambda\bigr)\,.
\end{equation}
Specializing the variable $K$ at a single curvature value $K_j$ would
give the determinant of $H$ if all curvatures were equal to $K_j$.
This is not the value of the determinant we are seeking; but we can
form a permutation-invariant superposition of all those expressions:
\begin{equation}
  \label{eq:92}
  \det H = \frac{1}{J}\sum_{i=1}^J
  \bigl[K_i - (J+2)\lambda\bigr]
  \prod_{\substack{j=1\\j\ne i}}^{J}\bigl(K_j - 2\lambda\bigr)
  \,.
\end{equation}
The factor $1/J$ comes from a normalization condition.  This turns out
to be the correct expression for the determinant of $H$, and it is
straightforwardly transformed in the characteristic polynomial of $H$.
\begin{equation}
  \label{eq:93}
  \det(H-\mu\uno) = \frac{1}{J}\sum_{i=1}^J
  \bigl[K_i - \mu - (J+2)\lambda\bigr]
  \prod_{\substack{j=1\\j\ne i}}^{J}\bigl(K_j - \mu - 2\lambda\bigr)
  \,.
\end{equation}
Nevertheless, the computation of its roots is not possible in general.

Based on expression \eqref{eq:93}, it follows that whenever
$\lambda<0$ the determinant $\det H$ is positive, and furthermore
$\det(H-\mu\uno)$ cannot vanish at a negative value of $\mu$, and thus
the stability eigenvalue should be positive.  As a result, \emph{all
  AdS states of the model are stable}.

Another case worth investigating is those states which have at least a
vanishing quantum number $n_j=0$.  The corresponding $K_j^{(+)}$
curvature is not defined in this case, because equation \eqref{eq:59}
is linear and it has only the solution $K_j^{(-)} = \lambda$.  Thus,
we can substitute $K_j = \lambda + \delta_j$ ($\delta_j>0$) in $\det
H$.  Assuming we can vary independently $\lambda$ and $\delta_j$, we
can expand $\det H$ for small $\lambda$ values:
\begin{equation}
  \label{eq:94}
  \det H \xrightarrow{\lambda\to0} \prod_{i=1}^J \delta_i
  - 2\lambda \sum_{i=1}\prod_{\substack{j=1\\j\ne i}}^{J} \delta_j
  + \mathcal{O}(\lambda^2)\,.
\end{equation}
The previous expression shows again that a negative value value of
$\lambda$ cannot make this determinant to change sign.  A positive
value indeed can change the sign in the determinant, and this
indicates that de Sitter states are likely to be unstable.  Of course
no general statement of this sort can be formulated, because this
depends on the magnitude of $\lambda$ as well as on all $\delta_j$:
for sufficiently small $\lambda>0$ and fixed $\delta_j$, the
determinant can be positive.  But if a \emph{single} $\delta_k=0$,
then the determinant reduces to
\begin{equation}
  \label{eq:95}
  \det H \xrightarrow{\lambda\to0}
  - 2\lambda \prod_{\substack{j=1\\j\ne k}}^{J} \delta_j
  + \mathcal{O}(\lambda^2)\,,
\end{equation}
which is certainly negative for $\lambda>0$!  Thus, we conclude that
an odd number of eigenvalues of $H$ have changed their signs and the
state is unstable.  We can suspect that in this case a single
eigenvalue has reversed sign, because if we would take two vanishing
$\delta_k$ the sign of the determinant would again be positive.  We
can see that if all $\delta_j=0$, $H$ has a completely negative
spectrum.  Thus we can expect that the vanishing of each $\delta_j$
changes sign of an eigenvalue, and thus \emph{all states with some
  $n_j=0$ are unstable}.

The previous reasoning is heuristic, because we cannot assume that
$\lambda$ and $\delta_j$ vary independently.  They depend on the
discrete numbers $\{n_1,\cdots,n_J\}$ and thus the stability criterion
should be validated numerically.  Nevertheless, heuristics works in
this case.  As it is shown in the following subsection, all states
with a single $n_j=0$ are unstable, at least in all searches we have
carried out.

We will close this subsection by giving a perturbative argument
showing that all low-$\lambda$ states lying in all non-principal
branches are unstable.  This way, the principal branch remains as the
only source of AdS and stable dS states.  We begin by splitting the
$H$ matrix \eqref{eq:88} as follows:
\begin{equation}
  \label{eq:96}
  H = 2\diag\{K_1-2\lambda,\cdots,K_J-2\lambda\}
  - 2\lambda U\,,
\end{equation}
where $U$ is a $J\times J$ matrix filled with ones.  If $\lambda$ is
small, we can consider the diagonal matrix as a ``dominant'' term and
the off-diagonal terms as a perturbation.  The perturbation is
permutation-invariant, and thus all eigenvalues of the diagonal matrix
are shifted the same amount.  The perturbative stability eigenvalue is
\begin{equation}
  \label{eq:97}
  \kappa = 2\bigl[\min_j K_j - 3\lambda\bigr]\,.
\end{equation}
If the minimum curvature is taken from the $(+)$ branch, then
$K_j^{(+)}\xrightarrow{\lambda\to0}\frac{2}{n_j^2}$ and the state has
a chance of being stable.  But if some curvature is taken from a $(-)$
branch, then we have $K_j^{(-)}\xrightarrow{\lambda\to0}\lambda$ and
the corresponding eigenvalue is negative.  But then $\kappa$ should be
negative also, showing that no matter how small $\lambda$ might be, if
the state comes from a non-principal branch, it will be unstable.

The previous argument does not rule out the existence of higher
$\lambda$ stable dS states in non-principal branches, but our
numerical searches have not found them.

\subsection{State searching in concrete examples}
\label{sec:state-searching}

Once the analysis of the model is reasonably complete, we should ask
for concrete examples where we can exhibit some states and their
associated magnitudes such as reduced cosmological constant $\lambda$,
curvatures $K_i$ and the stability eigenvalue $\kappa$.

We repeat here the relevant equations for the reader's convenience.
Given a $J$-tuple of integers $\{n_1,\cdots,n_J\}$ we compute the
solutions of equation \eqref{eq:68}, which is
\begin{equation}
  \label{eq:98}
  L_{\{s_j\},\{n_j\}}(\lambda) = \Lambda\,,\quad\text{with}\quad
  L_{\{s_j\},\{n_j\}}(\lambda) = 
  \frac{1}{2}\Bigl[J\lambda + \sum_{j=1}^J \frac{1}{n_j^2}
  \left(1 + s_j \sqrt{1 - 2\lambda\,n_j^2}\ \right)
  \Bigr]\,.
\end{equation}
All solutions for each branch $\{s_1,\cdots,s_J\}$, where the $s_j$
are signs $\pm1$, should be computed.  The corresponding solutions must
be real, and all its curvatures must be real and positive:
\begin{equation}
  \label{eq:99}
  K_j^{(s_j)} = \frac{1}{n_j^2}
  \left(1 + s_j \sqrt{1 - 2\lambda\,n_j^2}\ \right)\,.
\end{equation}
If a single curvature turns out to be real and negative or complex,
then the state should be discarded.  If all curvatures are positive,
we form the frequency matrix $H$ (dropping the unimportant factor 2
which appears in \eqref{eq:88}) and compute its minimum eigenvalue
$\kappa$, called the \emph{stability eigenvalue} of the state:
\begin{equation}
  \label{eq:100}
  H = 
  \begin{pmatrix}
    K_1 -3\lambda &      -\lambda & \cdots & -\lambda \\
    -\lambda      & K_2 -3\lambda & \cdots & -\lambda \\
    \vdots        &       \vdots  & \ddots & \vdots   \\
    -\lambda      &      -\lambda & \cdots & K_J - 3\lambda
  \end{pmatrix}
  \,,\quad
  \kappa = \min_{\mu\in\mathrm{spec}(H)}\mu
  \,.
\end{equation}
The stability condition is simply $\kappa>0$.

Thus, the searching method has the following steps:
\begin{enumerate}
\item Choose $J$ and $\Lambda$, the parameters of the model.
\item Choose a set of integers $\{n_1,\cdots,n_J\}$.
\item Solve equation \eqref{eq:98} for $\lambda$.
\item Compute the curvatures \eqref{eq:99} associated with the
  solutions found and accept the state if all curvatures are real and
  positive.
\item Compute the stability eigenvalue and flag the state as stable or
  unstable.
\item Go to step 2 until some bounding search criterion is met.
\end{enumerate}
The choosing of the integers can be done in various ways.  In the
simplest models with $J=2$\footnote{The two-sphere Einstein-Maxwell
  lansdcape is a sector of the six-dimensional Einstein-Maxwell not
  considered in Refs.~\cite{EM6-2,EM6-3}.}  we can scan a large square
in the $(n_1,n_2)$ plane in a brute-force search.  In this way we
cannot miss any state.  The symmetries $n_1\leftrightarrow n_2$ and
$n_j \to -n_j$ allow us to restrict to $n_2 \ge n_1 \ge 0$.  This
brute-force method cannot be used for higher values of $J$.  In those
cases, we should generate states randomly in an efficient manner; but
prior to the discussion on how this is done, we present some results
in the $J=2$ case.

\begin{figure}
  \centering
  \includegraphics[width=0.5\textwidth]{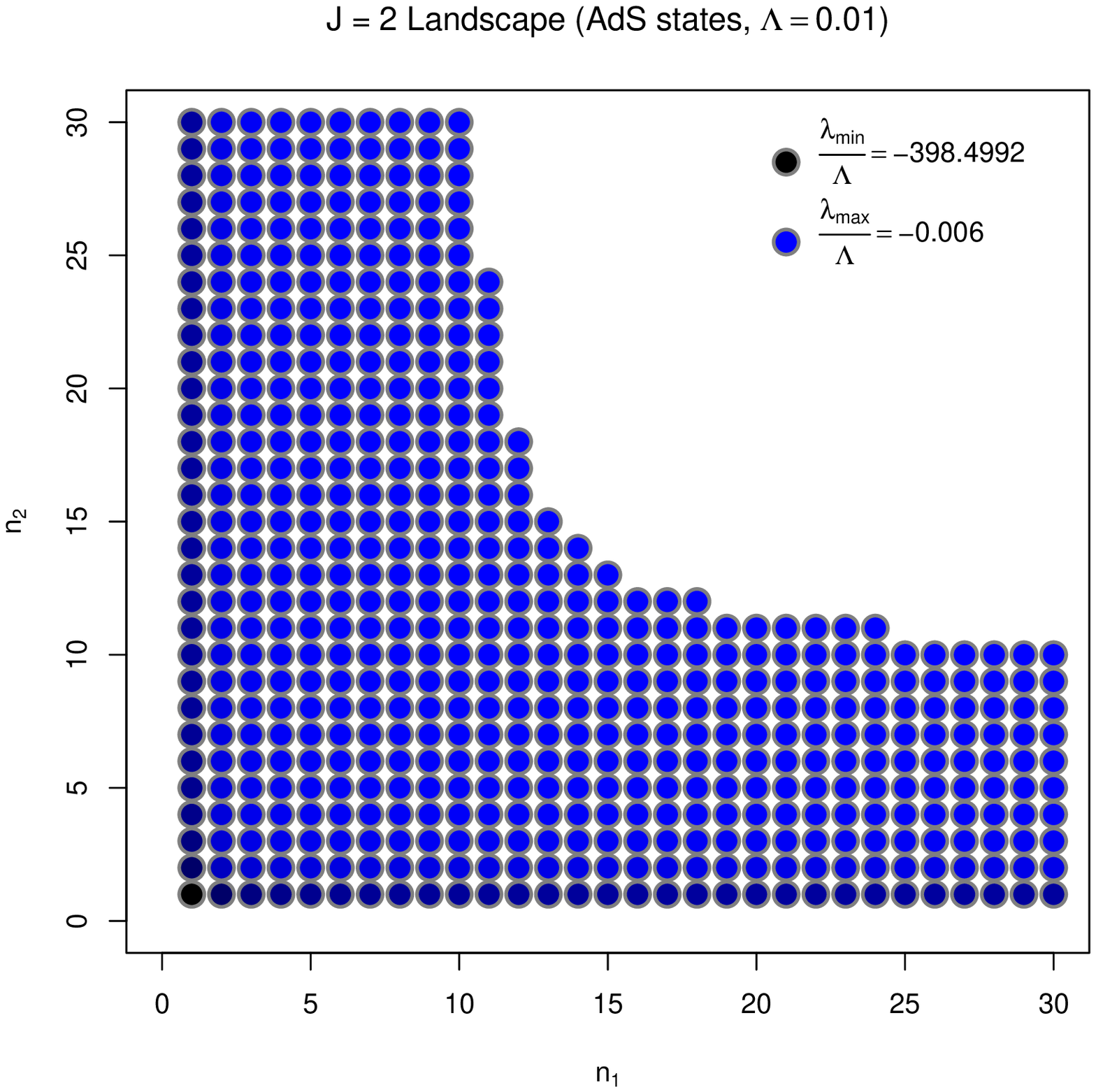}%
  \includegraphics[width=0.5\textwidth]{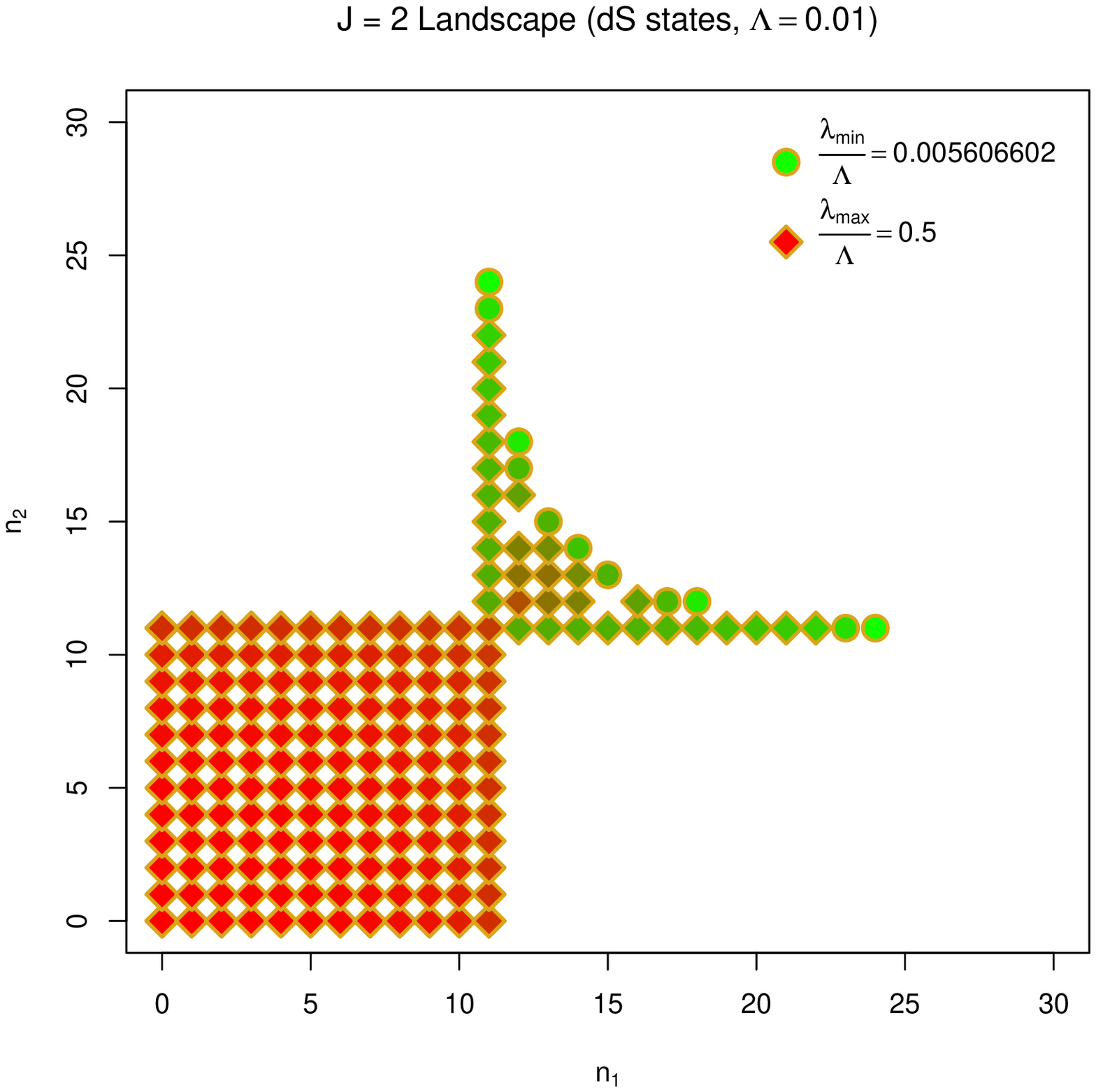}
  \includegraphics[width=0.5\textwidth]{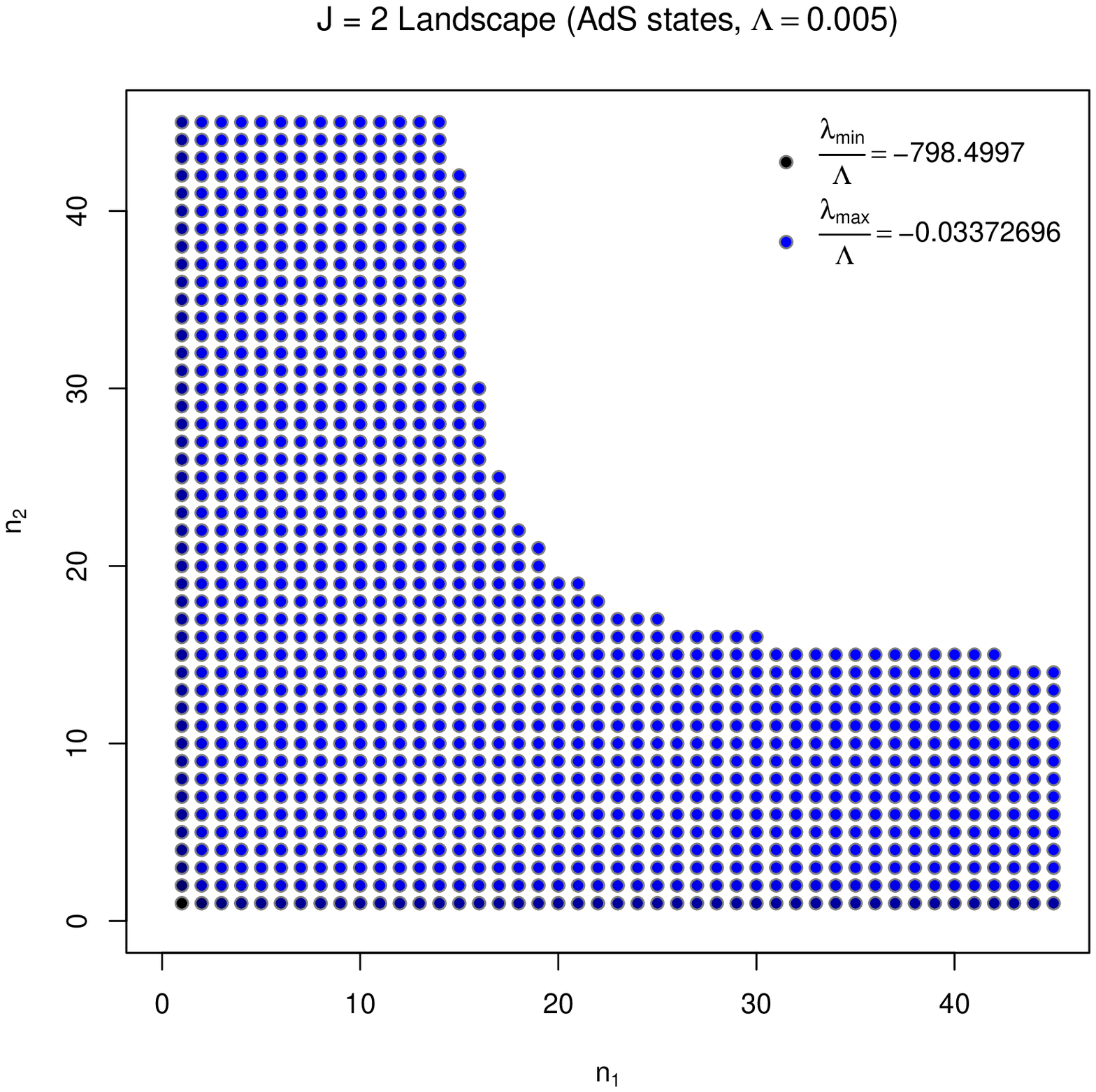}%
  \includegraphics[width=0.5\textwidth]{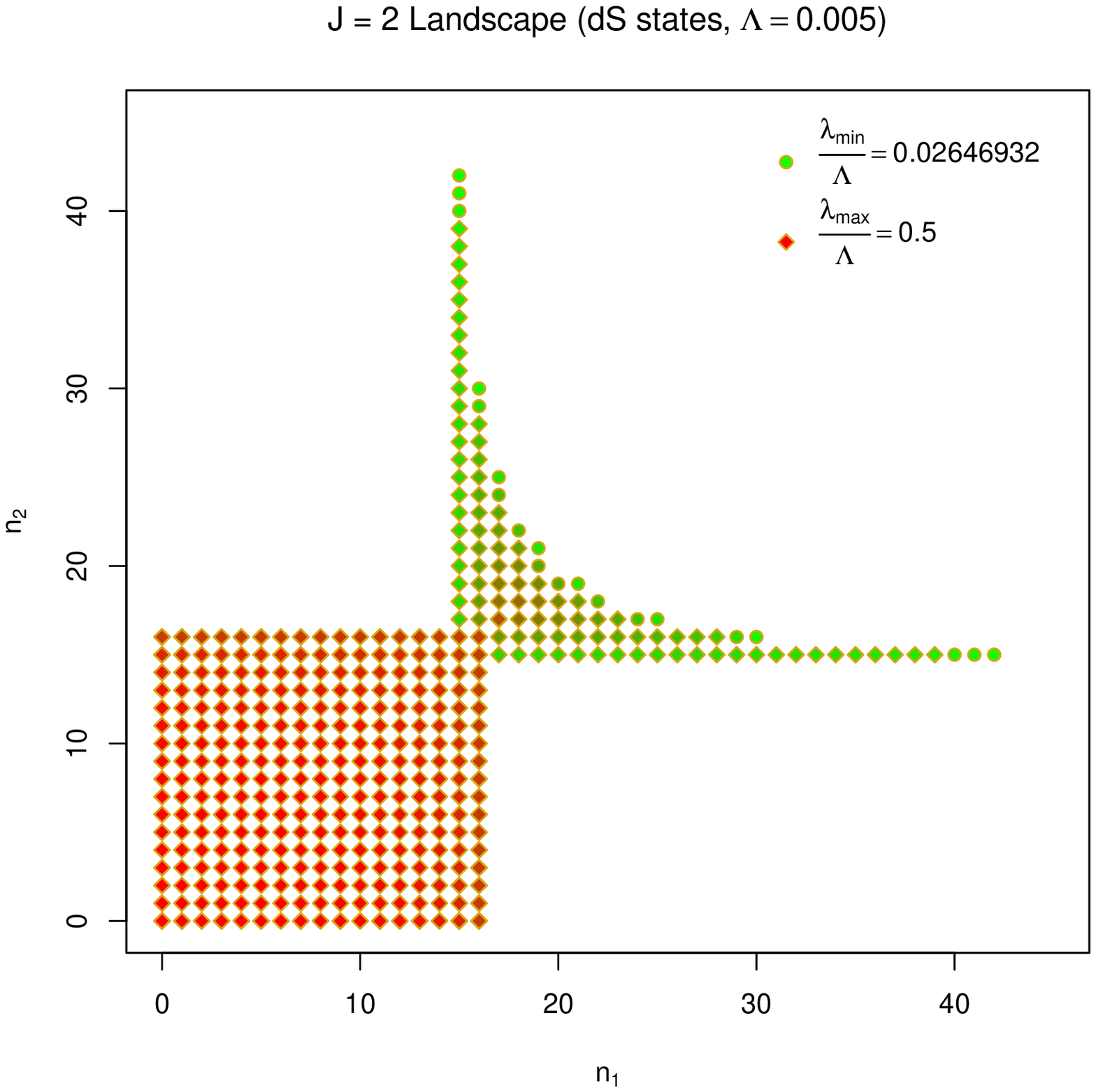}
  \caption{Landscape examples with $\Lambda=0.01$ (top panels) and
    $\Lambda=0.005$ (bottom panels).  AdS (left) and dS (right) states
    are shown, using a circle for a stable state and a diamond for a
    unstable one.  A state is drawn if a solution has been found for
    $\lambda$ with real and positive curvatures.  The magnitude of
    $\lambda$ is reflected in the color of each symbol.  Finally, all
    dS states lie above AdS ones, and meet at the branching curve,
    beyond which the landscape has no states.}
  \label{fig:landscapes-J2}
\end{figure}

We have chosen two values of the 4D cosmological constant $\Lambda$,
0.01 and 0.005, in order to exhibit how the lowering of $\Lambda$
causes the proliferation of states.  In figure \ref{fig:landscapes-J2}
we plot a point in each place of the $(n_1,n_2)$ plane where a state
has been found; of course, we have four branches for searching, so
some states overlap here.  States with negative and positive 2D
cosmological constant have been separated, so that we can see AdS
states at left panels and dS states at right panels.  Stable states
have been drawn using circles, and diamonds for unstable states.  Note
that all AdS states are stable, while most dS states are unstable.
The colors are related with the magnitude of the cosmological constant
as shown in the legend of each graphic.

Features of these models which can be seen in figure
\ref{fig:landscapes-J2} include:
\begin{itemize}
\item All AdS states are stable.
\item There is no AdS states with $n_1=0$ or $n_2=0$ because in those
  cases there is no curvature associated with this branch.
\item All dS states with $n_1=0$ or $n_2=0$ are unstable, as expected
  by the heuristic argument exhibited at the end of the previous
  subsection.
\item There is a curve which limits the existence both of dS and AdS
  states.  The form of this curve is easily computed by substituting
  $\lambda=0$ in equation \eqref{eq:98} with all positive $s_j$:
  \begin{equation}
    \label{eq:101}
    \sum_{j=1}^J \frac{1}{n_j^2} = \Lambda\,.
  \end{equation}
  Minkowski states, which ideally may be present, lie on this
  hypersurface which will be called \emph{branching hypersurface}
  (curve in the $J=2$ case).  In practice, Minkowski states require
  fine-tuning of $\Lambda$ and thus they are generically absent.
\item All dS states are located \emph{above} AdS ones, and they are
  located near the branching curve.  The discrete states are located
  on a multi-branch surface whose branches meet at the branching
  curve, hence its name.  Thus, states of $\lambda$ near zero from
  either side are located near this curve, which would contain, if
  present, the $\lambda=0$ states.
\item All stable dS states are located near the branching curve, and
  only there, but closeness is not enough for a state to be stable, as
  will be seen below.
\end{itemize}

A feature of these models which cannot be seen in figure
\ref{fig:landscapes-J2} is that all AdS states come from the
$\{+,+\}$ branch (which will be called the \emph{principal branch}).
This is so because the solutions in the remaining branches have the
$K_j^{(-)}$ curvatures negative, as can be seen in formula
\eqref{eq:99}.  Thus, only the principal branch is a source of AdS
states.

\begin{figure}
  \centering
  \includegraphics[width=0.5\textwidth]{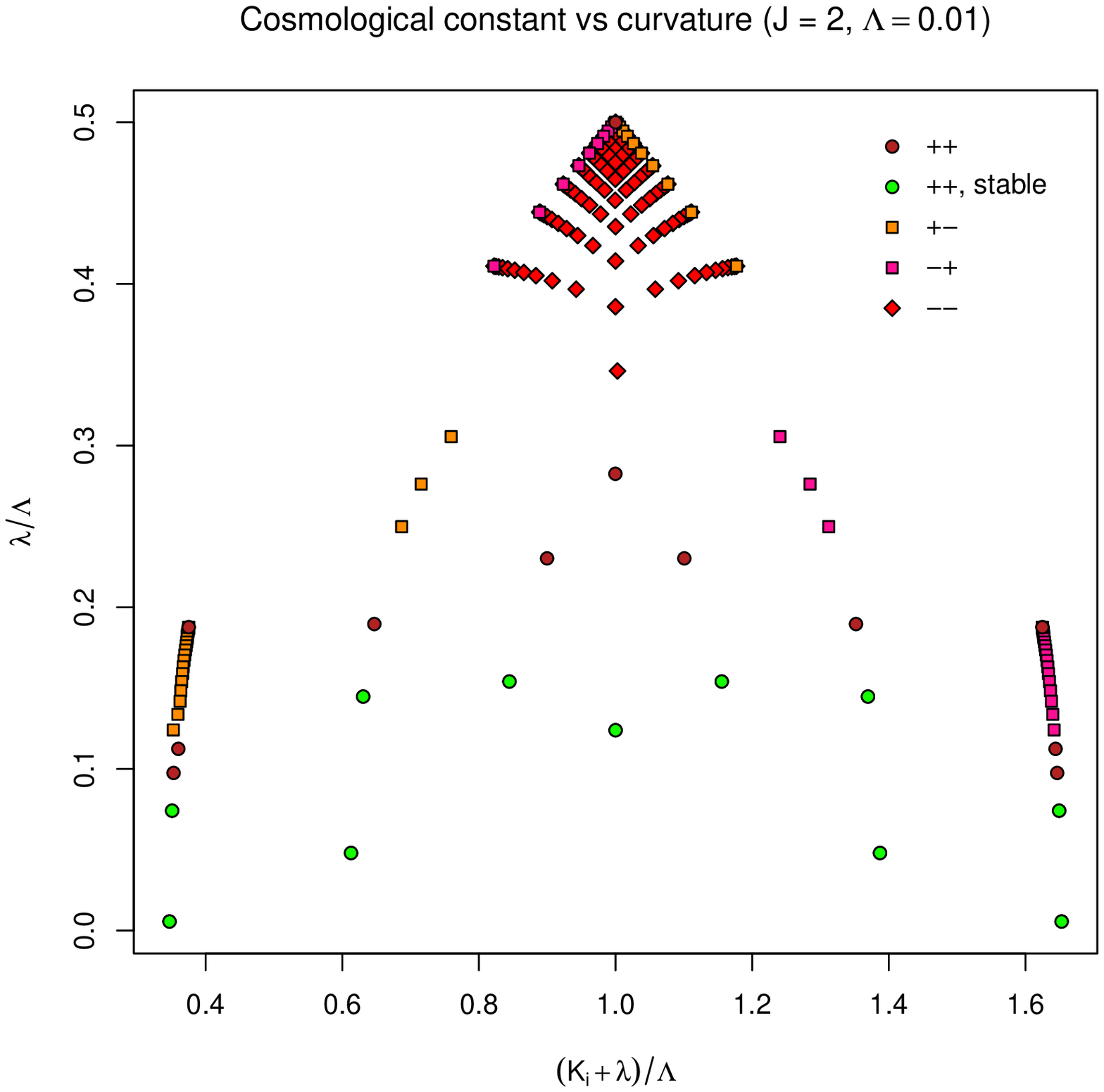}%
  \includegraphics[width=0.5\textwidth]{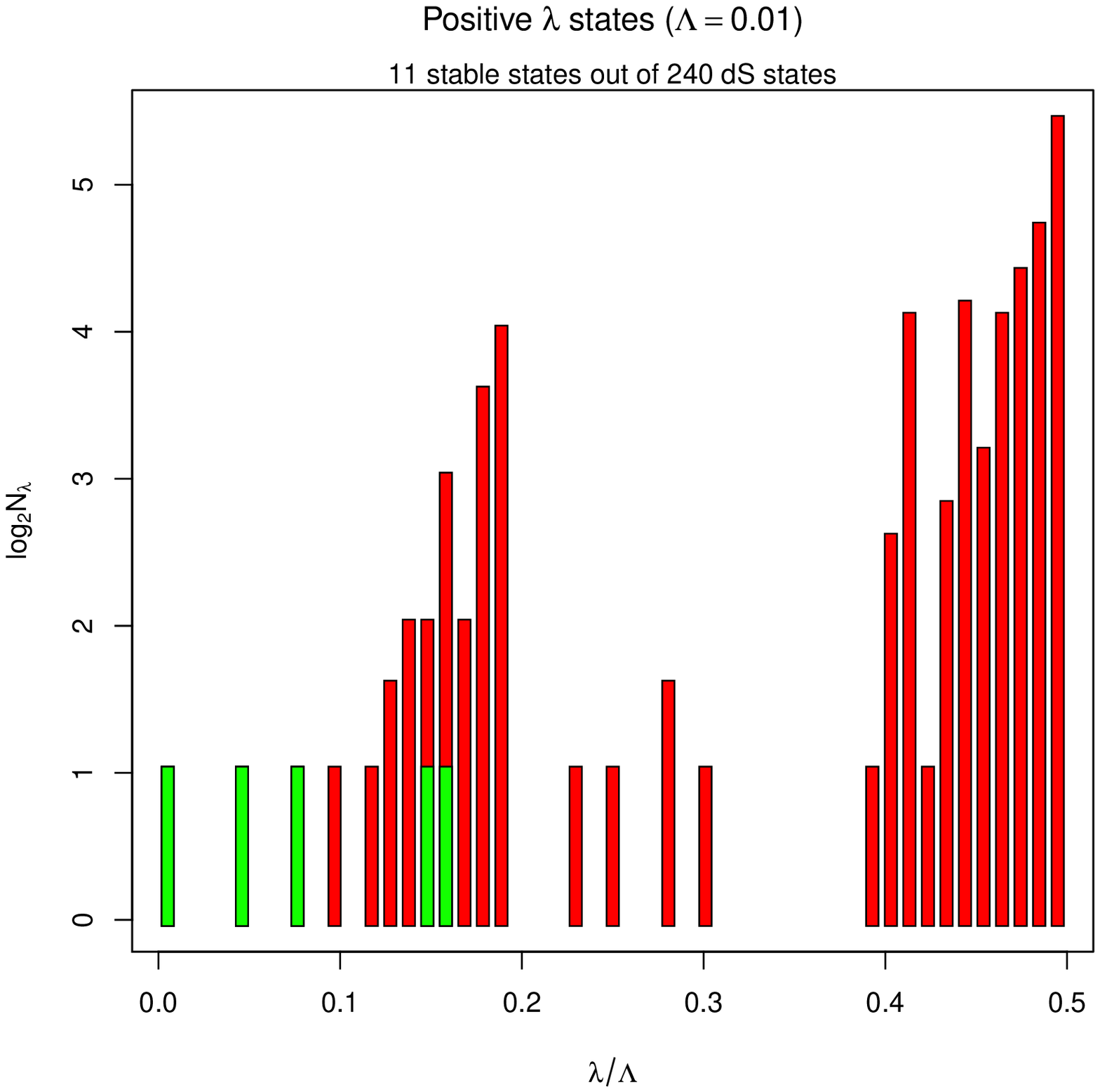}
  \includegraphics[width=0.5\textwidth]{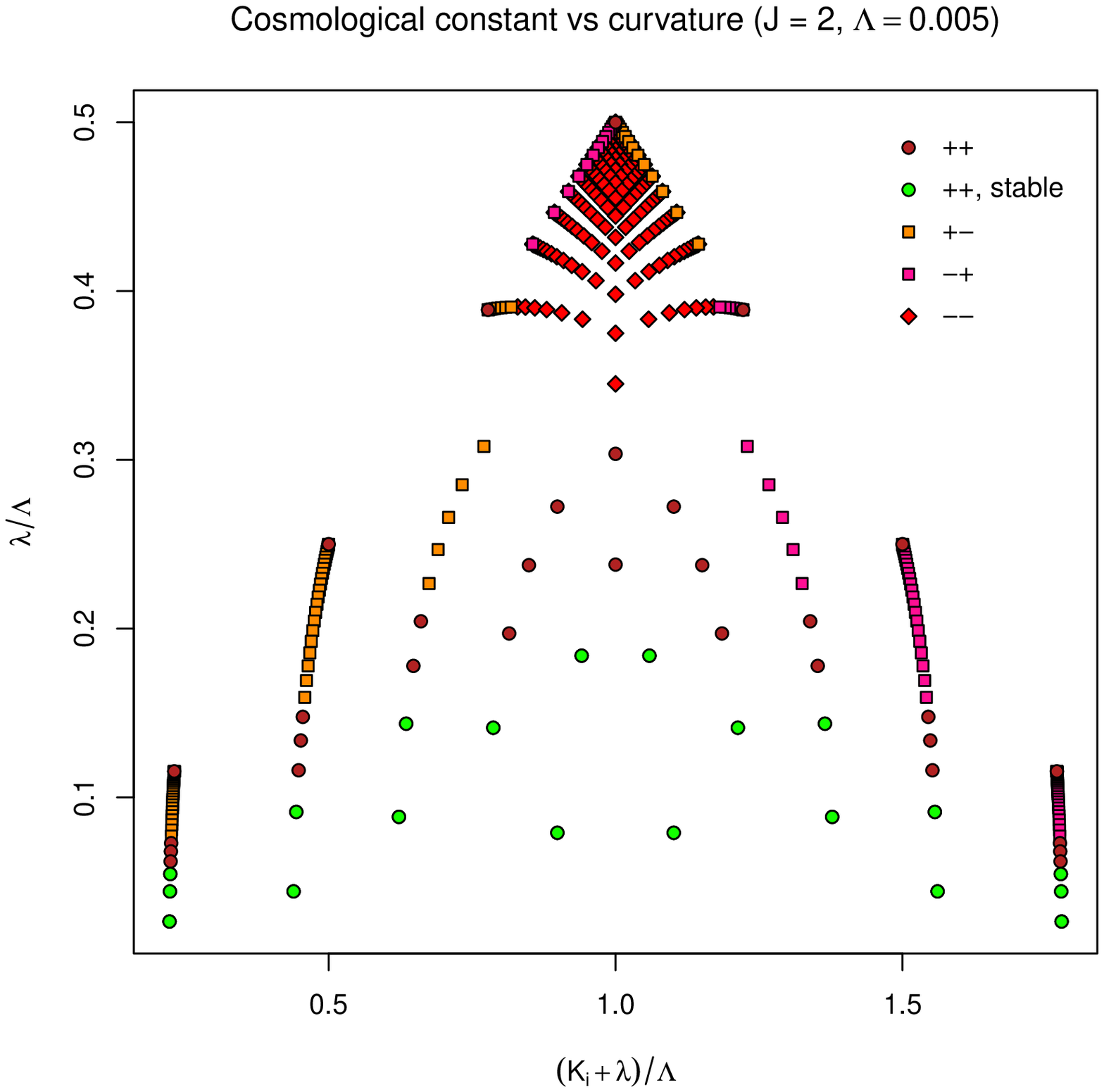}%
  \includegraphics[width=0.5\textwidth]{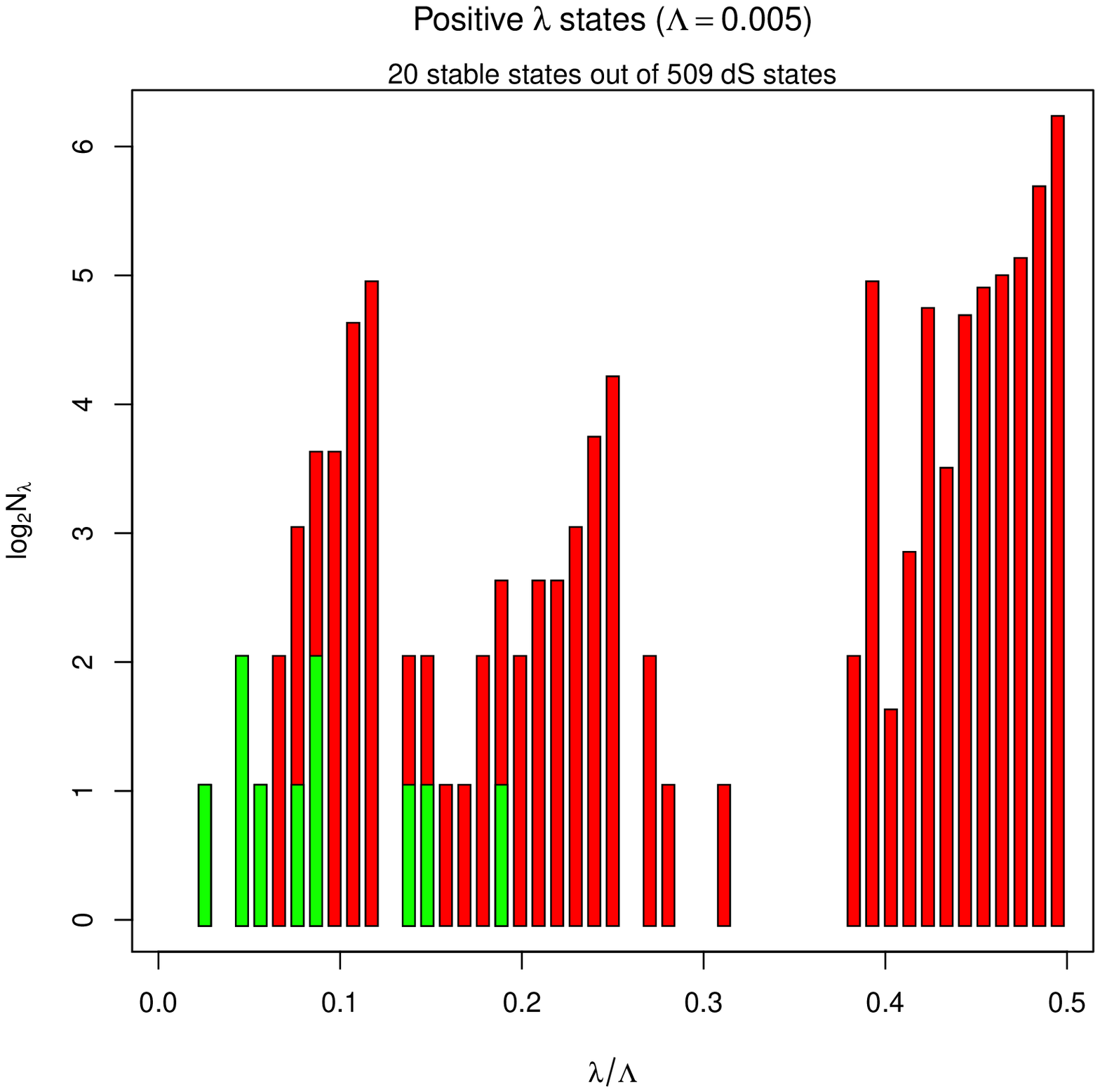}
  \caption{Landscape examples with $\Lambda=0.01$ (top panels) and
    $\Lambda=0.005$ (bottom panels).  Left panels show the variation
    of the cosmological constant $\lambda$ with curvature.  These
    triangle-shaped graphics show structures which the states seem to
    follow, and clearly separates the different branches by the
    $\lambda$ values they provide.  Right panels show the $\lambda$
    distributions of dS states, whose peaks come from the different
    branches.  The only branch that has not a peak associated to it is
    the principal branch, which provides all stable dS states.}
  \label{fig:magnitudes-J2}
\end{figure}

\begin{figure}
  \centering
  \includegraphics[width=0.5\textwidth]{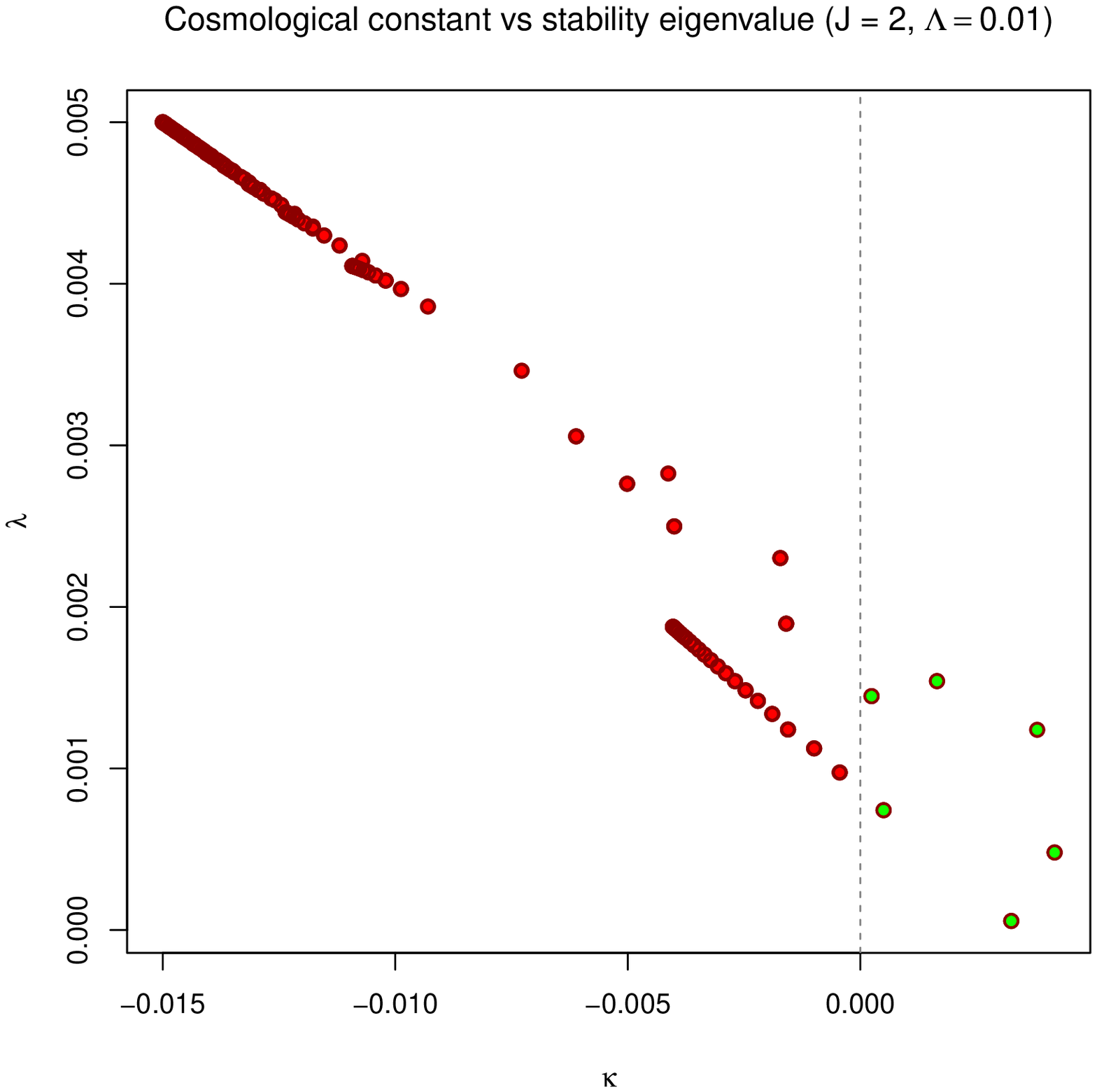}%
  \includegraphics[width=0.5\textwidth]{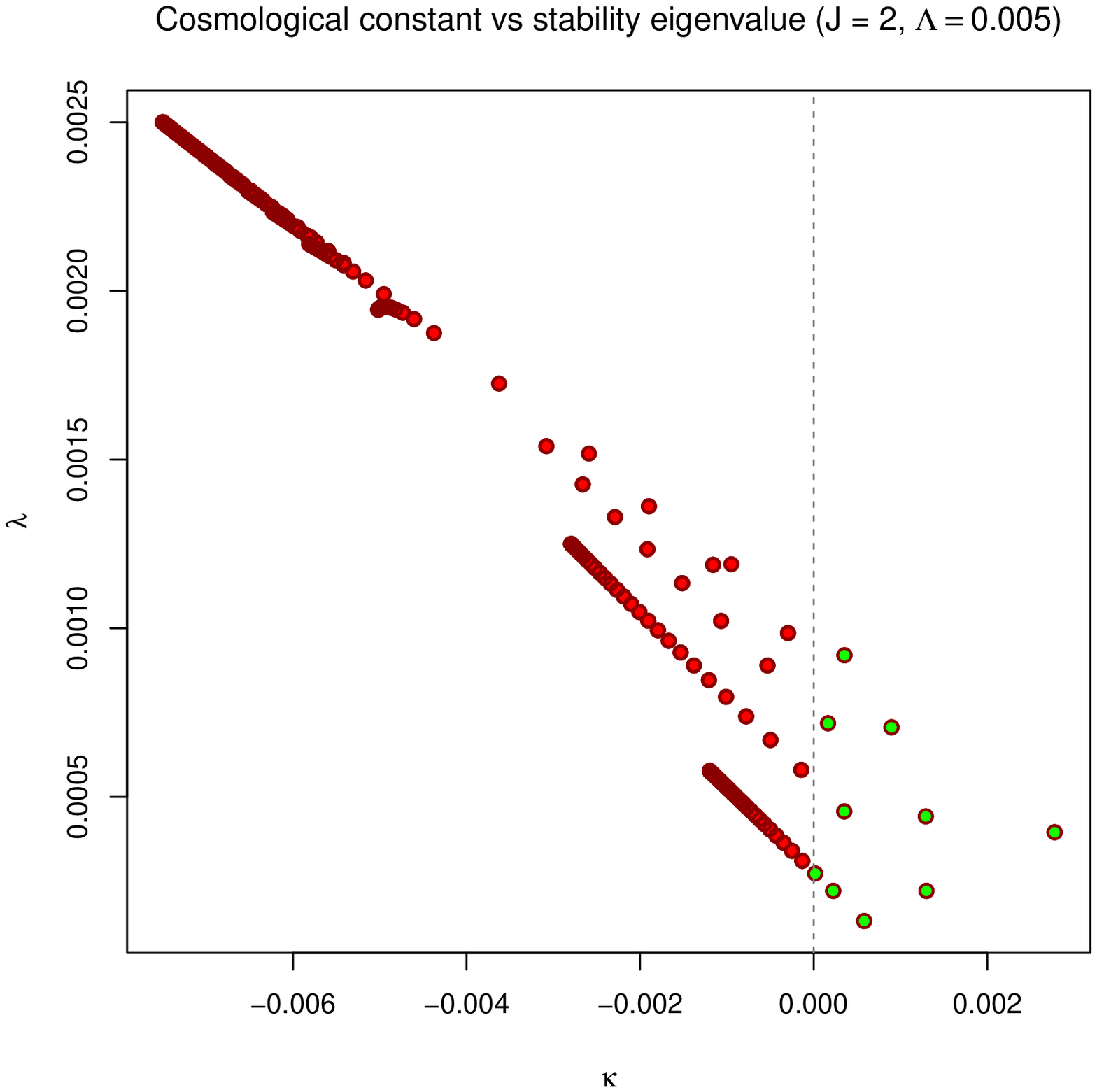}
  \caption{Landscape examples with $\Lambda=0.01$ (left panel) and
    $\Lambda=0.005$ (right panel).  Dependence of the cosmological
    constant versus the stability eigenvalue is shown, with an almost
    linear behaviour.  Stable states are located to the right of the
    vertical dashed line, which is thus the onset of stability.}
  \label{fig:lambda-vs-kappa-J2}
\end{figure}

In contrast, dS states can come from each of the four branches, but
each branch brings in states with very different properties.  For
example, all dS states near the branching curve come from the
principal branch.  In particular, all stable dS states come from this
branch.  The bunch of dS states lying in the reddish square in the
right panels of figure \ref{fig:landscapes-J2} come from the
$\{-,-\}$ branch, and they are ``most'' unstable in the sense that
they have the greater (in absolute value) negative stability
eigenvalue.  All these features can be seen in the left panels in
figure \ref{fig:magnitudes-J2}.  These left panels show cosmological
constant versus curvature, and they exhibit clearly the different
nature of the states coming from different branches:
\begin{itemize}
\item The states coming from the principal branch (bullets) form the
  core of the figure, and at the bottom of this figure are located the
  stable states, all of them coming from the principal branch.  Note
  that stability does not mean lower cosmological constant!  In the
  right panels of figure \ref{fig:magnitudes-J2} we can see that
  stable states are mixed with unstable ones in the cosmological
  constant value distribution.  Nevertheless, all lowest-lying states
  are stable.
\item The states coming from the $\{+,-\}$ and $\{-,+\}$ branches are
  at both sides of the triangle-shaped graphic shown at the left
  panels of figure \ref{fig:magnitudes-J2}.  All of them are unstable,
  but the values of the cosmological constant in this subset range
  from lowest to highest.
\item The states coming from the $\{-,-\}$ branch are located at the
  cusp of the triangle (figure \ref{fig:magnitudes-J2}, left panels)
  and all of them have the highest values of the cosmological constant
  and also the highest (in absolute value) negative stability
  exponents (see figure \ref{fig:lambda-vs-kappa-J2}).  In particular,
  the $n_1=0,n_2=0$ state, which is the cusp of the triangle, belongs
  to the $\{-,-\}$ branch. 
\end{itemize}

In the right panels of figure \ref{fig:magnitudes-J2} we can see the
cosmological constant distribution of dS states.  The contribution of
the branches can be seen as different peaks; while the principal
branch contributes with the stable states and other unstable
distributed in the lowest range, the $\{+,-\}$ and $\{-,+\}$ show two
peaks in the low and middle range, and the bulk of the $\{-,-\}$
states are relegated to the high range.

Finally, in figure \ref{fig:lambda-vs-kappa-J2} is shown the behaviour
of the cosmological constant versus the stability eigenvalue.  Its
almost linear relation can be seen to be dispersed in branches, which
are the same structures showing up in left panels of figure
\ref{fig:magnitudes-J2}.  Here, we can see that the states with lower
cosmological constant have a ``less negative'' stability eigenvalue
than the states with higher $\lambda$, which include the $\{-,-\}$
states, as said above.

Thus, the different branches provide very different states.  Among
them, the most interesting seem to be those coming from the principal
branch, because they include AdS as well as stable and unstable dS
states, which are the ingredients we need to construct a toy model of
a multiverse.

For $J>2$, we can exploit the fact that stable dS states are near the
branching hypersurface (it is a curve only for $J=2$) and design a
sampling method which looks for states in $(n_1,\cdots,n_J)$ space
whose Voronoi cell (which is the cube with its center at the point in
question) intersects the branching hypersurface.  Those states are
called \emph{secant} states \cite{RHM,Jul}, and all states near the
branching surface belong to this category (but the reciprocal is
false, that is, a secant state may not be near the branching surface
if $J$ is large enough!).  Thus, we can sample the principal branch by
sampling the secant states.  In this way, we always find a state with
a fair chance of being a true state of the model in the principal
branch, that is, we have an efficient sampling method, much better
than brute-force node enumeration or completely random node sampling.

The sampling of secant states is simple: we choose a uniformly random
direction in $J$-space and find the point of intersection between the
ray having the chosen direction starting from the origin and the
branching hypersurface.  This intersection point belongs to the
Voronoi cell of a single secant state, which we find by rounding the
coordinates of the intersection point.  Once we have the state, we
solve the equation for $\lambda$ in the principal branch and follow
the steps detailed above.

In the following section we use this sampling method to obtain a
sample of the cosmological constant distribution which can be compared
with an approximate formula to be obtained below.

\section{State counting}
\label{sec:state-counting}

As we have seen in the previous section, the states of the
multi-sphere Einstein-Maxwell landscape can come from different
branches, and the richest of those branches is the principal one.  In
this section we turn to the problem of counting states on this
branch.  Our main aim is to compute the distribution of cosmological
constant values in this branch.

As discussed above, only the principal branch can have both AdS and dS
states, both stable and unstable, and moreover the large-$J$ sampling
method is especially adapted to the principal branch.  Therefore, we
can obtain samples to compare with the approximate formula to be
obtained in subsection \ref{sec:small-cc-distribution} below.

\subsection{Counting states in the principal branch}
\label{sec:exact-counting}

Given a node $n=(n_1,\cdots,n_J)$ in $J$-space, there exists a state
in the principal branch of the multi-sphere Einstein-Maxwell landscape
with cosmological constant $\Lambda$ if the equation
\begin{equation}
  \label{eq:102}
  L_n(\lambda) = \Lambda
\end{equation}
has a solution in $\lambda$.  The principal branch $L_n$ function is
given in \eqref{eq:98} with all positive signs $s_j=+$.  In the
principal branch there are no states with some $n_j=0$, and thus the
function $L_n(\lambda)$ has a maximum at $\lambda=0$ and it is
monotonically decreasing from $\lambda=0$ to
$\lambda=\lambda_{\mathrm{b}}$, where $\lambda_{\mathrm{b}}$, see
equation \eqref{eq:69}, is the branching point of the node $n$.  A dS
state exists therefore if $\Lambda$ is between the two extremal values
of the $L_n(\lambda)$ function:
\begin{equation}
  \label{eq:103}
  L_n(\lambda_\mathrm{b}) \le \Lambda \le L_n(0)\,.
\end{equation}
Equation \eqref{eq:103} is the existence condition for a dS state at
node $n$ in the principal branch.  The corresponding equalities define
two surfaces in node space:  The \emph{branching surface}
\begin{equation}
  \label{eq:104}
  L_n(0) = \sum_{j=1}^J \frac{1}{n_j^2} = \Lambda\,,
\end{equation}
whose integer points, if any, have vanishing $\lambda$, and the
\emph{limiting surface}
\begin{equation}
  \label{eq:105}
  L_n(\lambda_\mathrm{b}) = 
  \frac{J}{4n_{\mathrm{max}}^2} + \frac{1}{2}\sum_{j=1}^J \frac{1}{n_j^2}
  \left(1 + \sqrt{1 - \frac{n_j^2}{n_{\mathrm{max}}^2}}\ \right)
   = \Lambda\,,
\end{equation}
which signals the end of the principal branch.  In terms of a
characteristic function, the existence condition is
\begin{equation}
  \label{eq:106}
  \chi_{[L_n(\lambda_\mathrm{b}),L_n(0)]}(\Lambda) =
  \begin{cases}
    1 & \text{if a state exists at $n$,}\\
    0 & \text{if a state does not exist at $n$.}
  \end{cases}
\end{equation}
The previous existence condition should be supplemented with the
stability condition \eqref{eq:89}, that is, $\kappa(n)>0$, where
$\kappa(n)$ is the minimum eigenvalue of the stability matrix $H$ at
node $n$ if a state exists there.  This stability condition can also
be ascribed to a surface (the \emph{stability surface}) signaling the
stability threshold.  The stability surface should be comprised
between the branching and limiting surfaces; unfortunately, the
analytic expression of it cannot be found for general $J>1$.  As a
consequence, we will represent this condition by adding the factor
$\theta(\kappa(n))$ to the existence condition.  Therefore, the exact
number of stable states with given $\Lambda$ in the principal branch
of the multi-sphere EM landscape is
\begin{equation}
  \label{eq:107}
  \mathcal{N}_J(\Lambda) = \sum_{\substack{n\in\Z^J\\ n_j\ne0}}
  \chi_{[L_n(\lambda_{\mathrm{b}}),L_n(0)]}(\Lambda)\,
  \theta(\kappa(n))\,.
\end{equation}
The exact evaluation of the previous expression is possible only for
$J=1$ as is showed in subsection \ref{sec:modulus-stab}, equation
\eqref{eq:41}.  We repeat it here for the reader's convenience,
omitting from it AdS states and setting $e=1$:
\begin{equation}
  \label{eq:108}
  \mathcal{N}_1(\Lambda) = \biggl\lfloor\frac{1}{\sqrt{\Lambda}}\biggr\rfloor
  -
  \biggl\lceil\frac{2\sqrt{2}}{3\sqrt{\Lambda}}\biggr\rceil
  + 1
  \approx  \frac{1}{\sqrt{\Lambda}}\Bigl(1 - \frac{2\sqrt{2}}{3}\Bigr)\,.
\end{equation}
We will shortly turn into the approximate evaluation of
\eqref{eq:107}.  But prior to that, we need to grasp some general
ideas on the structure of the stable state set which we are willing to
count.

We will emphasize two main aspects:  asymptotic hyperplanes and state
chains.
\begin{description}
\item[Asymptotic hyperplanes] Both equations \eqref{eq:104} and
  \eqref{eq:105} corresponding to the branching and limiting surfaces,
  and likewise the stability surface, have asymptotic hyperplanes
  located at
  \begin{equation}
    \label{eq:109}
    |n_j| = \frac{1}{\sqrt{\Lambda}}
    \quad(\text{for fixed $j$}),\quad
    |n_{k\ne j}| \to\infty\,,
  \end{equation}
  and thus all states are restricted to the region
  \begin{equation}
    \label{eq:110}
    |n_j| > \frac{1}{\sqrt{\Lambda}} = \nu_0
    \quad(1\le j \le J)\,.
  \end{equation}
  Therefore, all dS states should have a charge greater than $\nu_0$
  as a necessary condition.

  Note that no integer $n_j$ can be equal to $\nu_0$ while preserving
  the existence condition unless all the remaining integers $n_{i\ne
    j}$ are infinite, hence the name ``asymptotic''.  It is easy to
  see that the corresponding states would have $\lambda=0$ and all
  curvatures vanishing except for one, and thus they would not
  represent compactified states.  Moreover, the associated stability
  matrix has $J-1$ zero eigenvalues (see eq.~\eqref{eq:100}), and thus
  these states are only marginally stable.  These properties suggest
  that they should be excluded from the landscape.

\item[State chains] There is also a natural upper bound on the charge,
  which can be obtained by considering the following straight line in
  flux space:
  \begin{equation}
    \label{eq:111}
    n_1 = \cdots = n_{J-1} = \sqrt{\frac{J-1}{\Lambda}}\,,
    \quad n_J \in\R\quad\text{(free parameter)}\,.
  \end{equation}
  The previous line is asymptotic to the branching surface, in the
  sense that it satisfies equation \eqref{eq:104} when $n_J\to\infty$.
  This line do not contain nodes because the quotient
  $\sqrt{\frac{J-1}{\Lambda}}$ is generically not an integer.  But we
  can slightly modify the previous line:
  \begin{equation}
    \label{eq:112}
    n_1 = \cdots = n_{J-1} =
    \left\lceil\sqrt{\frac{J-1}{\Lambda}}\right\rceil = \nu_1\,,
    \quad n_J \in\R\quad\text{(free parameter)}\,.
  \end{equation}
  This modified line can contain valid states if $n_J$ lies between
  $\lceil\nu_0\rceil$ and $\nu_2$, where $\nu_2$ is the intersection
  height with the branching surface:
  \begin{equation}
    \label{eq:113}
    \nu_2 = \frac{\nu_1}{
      \sqrt{\Lambda\nu_1^2 - (J-1)}
    }\,.
  \end{equation}
  All states on the line above $\nu_2$ are beyond the branching
  surface.  Therefore, we have the bound $\nu_0 < n_j < \nu_2$ ($1\le
  j\le J$).  All states which might happen to lie on this line are
  said to form a \emph{state chain}.

  It should be noted that when $\sqrt{\frac{J-1}{\Lambda}}$ coincides
  with the integer $\nu_1$, then $\nu_2$ diverges, which at first
  sight would be interpreted as an infinite dS state chain of ever
  decreasing $\lambda$.  But those nodes in the chains have no states,
  as can be seen by explicitly writing equation \eqref{eq:102} for the
  nodes in the line \eqref{eq:112}, and look for solutions with small
  $\lambda$ and large $n_J$:
  \begin{equation}
    \label{eq:114}
    2\Lambda = J\lambda + (J-1)\frac{1 + \sqrt{1 - 2\lambda \nu_1^2}}{\nu_1^2}
    + \frac{1 + \sqrt{1 - 2\lambda n_J^2}}{n_J^2}\,,
  \end{equation}
  We can consider $\lambda \ll \frac{1}{\nu_1^2}$, but $\lambda \ll
  \frac{1}{2n_J^2}$ is not true, because $n_J$ is large.  Therefore,
  equation \eqref{eq:114} can be rewritten as
  \begin{equation}
    \label{eq:115}
    2\Lambda - \frac{2(J-1)}{\nu_1^2} - \lambda =
    \frac{1 + \sqrt{1 - 2\lambda n_J^2}}{n_J^2} \,.
  \end{equation}
  The right hand side of \eqref{eq:115} is positive, thus a solution
  to equation \eqref{eq:115} can exist only if $\Lambda -
  \frac{J-1}{\nu_1^2}$ is strictly positive, which leads to
  \begin{equation}
    \label{eq:116}
    \nu_1 = 
    \left\lceil \sqrt{\frac{J-1}{\Lambda}}\right\rceil
    > \sqrt{\frac{J-1}{\Lambda}}\,,
  \end{equation}
  that is, if $\sqrt{\frac{J-1}{\Lambda}}$ is an integer then there is
  no solution to equation \eqref{eq:115}.  As a consequence, $\nu_2$
  can be made as large as we want by fine-tuning $\Lambda$ but it is
  never infinite.  This argument shows that all state chains are
  finite.

  This discussion on the state chains can be generalized to other
  asymptotic affine manifolds that the branching surface can have.
  For instance, asymptotic hyperplanes \eqref{eq:109}, as we have seen
  above, are likewise devoid of states, but there are close
  hyperplanes (having nodes) each one containing a replica of a
  $(J-1)$-dimensional landscape.
\end{description}

We will now consider the evaluation of the number of stable states
$\mathcal{N}_J(\Lambda)$.  Firstly, we get rid of the sign degeneracy
$2^J$, which is always trivially present.  Secondly, we invoke the
permutation symmetry, which allows us to arrange the integers $n_j$ in
decreasing order.  The corresponding permutation degeneracy is $J!$
except on those nodes having repeated components.  This difference
will be ignored for simplicity; we will see below that it will be of
little importance in the small-$\lambda$ region.  Thirdly, we consider
a node $n=\{n_1,\cdots,n_J\}$ with $n_1 > n_2 > \cdots > n_J$ and the
corresponding equation for the existence of a state \eqref{eq:102}:
\begin{equation}
  \label{eq:117}
  L_n(\lambda) = 
  \frac{1}{2}\biggl[J\lambda + \sum_{j=1}^J K_j(\lambda)\biggr] =
  \Lambda \,.
\end{equation}
The $L_n(\lambda)$ curve has a branching point $\lambda_\mathrm{b}$
given by \eqref{eq:69}, that is,
$\lambda_\mathrm{b}=\frac{1}{2n^2_1}$.  Any approximation method we
might wish to apply on the $L_n(\lambda)$ curve should respect this
branching point in order to accurately represent the existence
condition.  In particular, we cannot assume $\lambda \ll
\lambda_\mathrm{b}$.  But we do have $n_1 > n_J=\min_j\{n_j\}$, and in
the case $n_J \ll n_1$, we can assume $\lambda\ll\frac{1}{2n_J^2}$ and
write
\begin{equation}
  \label{eq:118}
  K_J(\lambda) \approx \frac{2}{n_J^2} - \lambda\,,
\end{equation}
which leaves equation \eqref{eq:117} as
\begin{equation}
  \label{eq:119}
  \frac{1}{2}\biggl[(J-1)\lambda + \sum_{j=1}^{J-1} K_j(\lambda)\biggr] =
  \Lambda - \frac{1}{n_J^2}\,.
\end{equation}
Equation \eqref{eq:119} represents the solutions of a landscape in
which the $J^{\mathrm{th}}$ curvature has been removed, and the
cosmological constant $\Lambda$ has been replaced with $\Lambda -
\frac{1}{n_J^2}$.  We can now let $n_J$ run from
$\bigl\lfloor\frac{1}{\sqrt{\Lambda}}\bigr\rfloor+1$ through the diagonal node
having $n_J=\bigl\lfloor\sqrt{\frac{J}{\Lambda}}\bigr\rfloor$, thus obtaining the
recurrence law
\begin{equation}
  \label{eq:120}
  \mathcal{N}_J(\Lambda) \approx J!
  \sum_{m = \bigl\lfloor\frac{1}{\sqrt{\Lambda}}\bigl\rfloor+1}
  ^{\bigl\lfloor\sqrt{\frac{J}{\Lambda}}\bigr\rfloor}
  \mathcal{N}_{J-1}\Bigl(\Lambda - \frac{1}{m^2}\Bigr)\,.
\end{equation}
The previous formula is valid under the following conditions:
\begin{itemize}
\item The fraction of states with repeated components is small.
\item The cosmological constant $\lambda$ of the states included is
  small, so that equation \eqref{eq:118} can be valid.
\end{itemize}
The states near the asymptotic hyperplanes will satisfy the previous
conditions more accurately, so that the first terms in the sum
\eqref{eq:120} will be more precise than the terms near the diagonal.
The latter states will fail to satisfy the strong inequality $n_J \ll
n_1$.  This means that the low-lying (that is, small-$\lambda$) states
will be taken into account, but the formula can miss or overcount some
high-lying (high-$\lambda$) states.

Equation \eqref{eq:108} triggers the recurrence relation, the first
consequence being
\begin{equation}
  \label{eq:121}
  \mathcal{N}_2(\Lambda) \approx 2
  \sum_{m = \bigl\lfloor\frac{1}{\sqrt{\Lambda}}\bigl\rfloor+1}
  ^{\bigl\lfloor\sqrt{\frac{2}{\Lambda}}\bigr\rfloor}
  \Biggl\{
  \biggl\lfloor\frac{1}{\sqrt{\Lambda - \frac{1}{m^2}}}\biggr\rfloor
  -
  \biggl\lfloor\frac{2\sqrt{2}}{3\sqrt{\Lambda - \frac{1}{m^2}}}\biggr\rfloor
  + 1\Biggr\}\,.
\end{equation}
The previous equation can be approximated by a smoother version by
removing the floor-ceiling functions inside the sum:
\begin{equation}
  \label{eq:122}
  \mathcal{N}_2(\Lambda) \approx 2\biggl(1 - \frac{2\sqrt{2}}{3}\biggr)
  \sum_{m = \bigl\lfloor\frac{1}{\sqrt{\Lambda}}\bigl\rfloor+1}
  ^{\bigl\lfloor\sqrt{\frac{2}{\Lambda}}\bigr\rfloor}
  \frac{1}{\sqrt{\Lambda - \frac{1}{m^2}}} 
  \,.
\end{equation}
The previous formula can be refined by cutting off the chains whose lower ends pass the
diagonal and counting the states on the diagonal accurately.  This ``diagonal
corrected'' formula is to be used in figures \ref{fig:N2-vs-Lambda-J2}
and \ref{fig:N2-vs-Lambda-J2-half} below.

Further simplification can be achieved by isolating the first term
(which carries the discontinuities) and estimating the remaining sum
by means of an integral:
\begin{equation}
  \label{eq:123}
  \mathcal{N}_2(\Lambda)
  \approx 2\biggl(1 - \frac{2\sqrt{2}}{3}\biggr)
  \Biggl\{
  \frac{\bigl\lfloor\frac{1}{\sqrt{\Lambda}}\bigl\rfloor+1}
  {\sqrt{\Lambda\bigl(\bigl\lfloor\frac{1}{\sqrt{\Lambda}}\bigl\rfloor+1\bigr)^2
      - 1}}
  + \int_{\frac{1}{\sqrt{\Lambda}} + 1}^{\sqrt{\frac{2}{\Lambda}}}
    \frac{\dif x}{\sqrt{\Lambda - \frac{1}{x^2}}}
  \Biggr\}
  \,.
\end{equation}
Formulae \eqref{eq:122} and \eqref{eq:123} show clearly the effect of
state chains as discontinuities at integer values of
$\frac{1}{\sqrt{\Lambda}}$.  When $\frac{1}{\sqrt{\Lambda}}$
approaches an integer from below, a very long state chain develops
which increases dramatically the number of states.  When $\Lambda$ is
reduced, the number of ``bulk'' states, that is, those not in the
chains, increases as reflected by the well-behaved integral
contribution, which for $\Lambda\ll1$ is
\begin{equation}
  \label{eq:124}
  \int_{\frac{1}{\sqrt{\Lambda}} + 1}^{\sqrt{\frac{2}{\Lambda}}}
  \frac{\dif x}{\sqrt{\Lambda - \frac{1}{x^2}}}
  \xrightarrow{\quad\Lambda\ll1\quad}
  \frac{1}{\Lambda} - \frac{\sqrt{2}}{\Lambda^{\frac{3}{4}}}
  \,.
\end{equation}

Figure \ref{fig:N2-vs-Lambda-J2} provides a good example of the
performing of equations \eqref{eq:121}, \eqref{eq:122} and
\eqref{eq:123}.  These formulae are to be compared with brute-force
determination of the number of stable states in the corresponding
models.  The discreteness of the lattice induces strong fluctuations
in the actual number of states, which is well represented by formula
\eqref{eq:123}, provided we interpret it as an average behaviour.

\begin{figure}[htbp]
  \centering
  \includegraphics[width=0.5\textwidth]{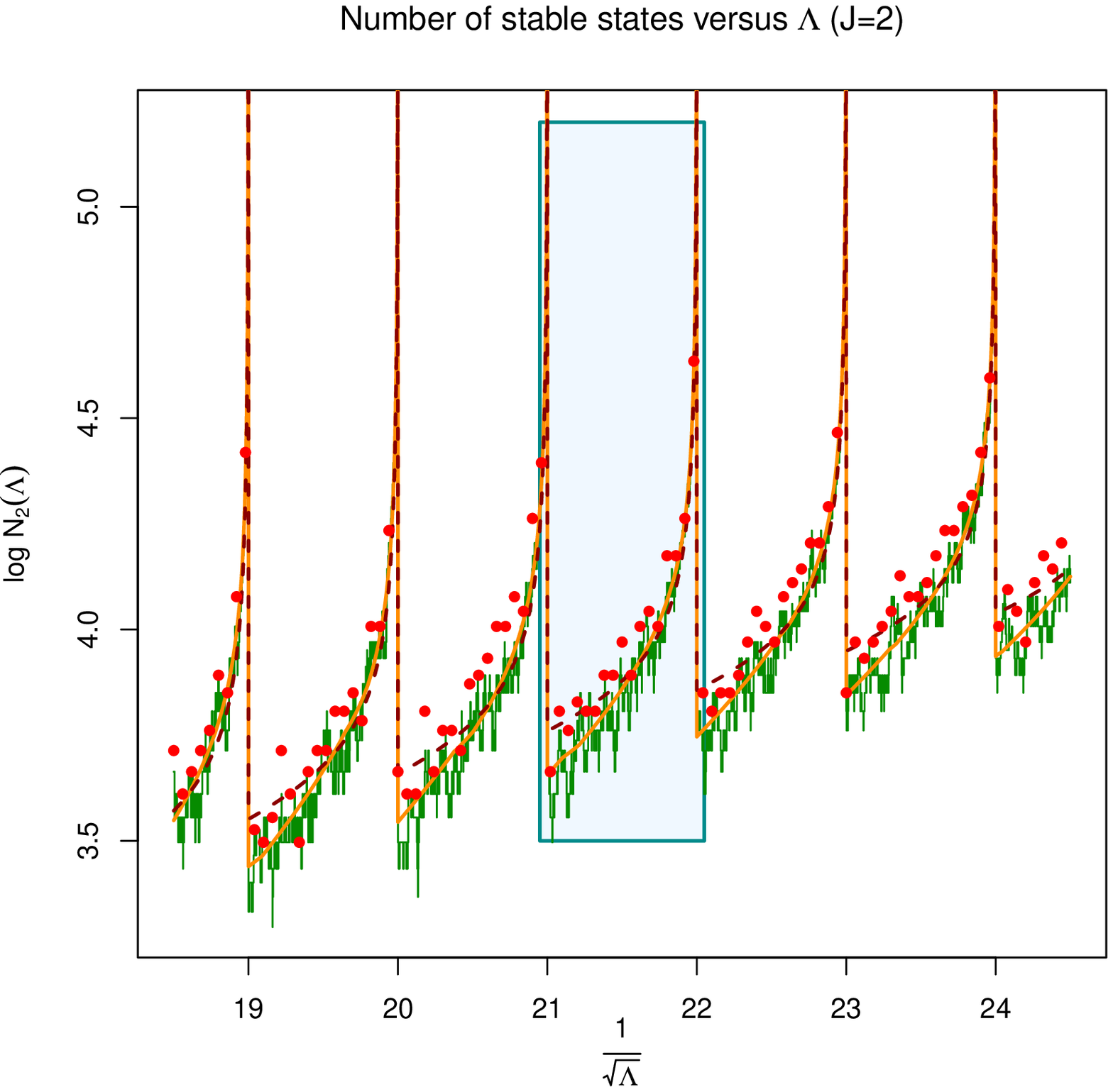}%
  \includegraphics[width=0.5\textwidth]{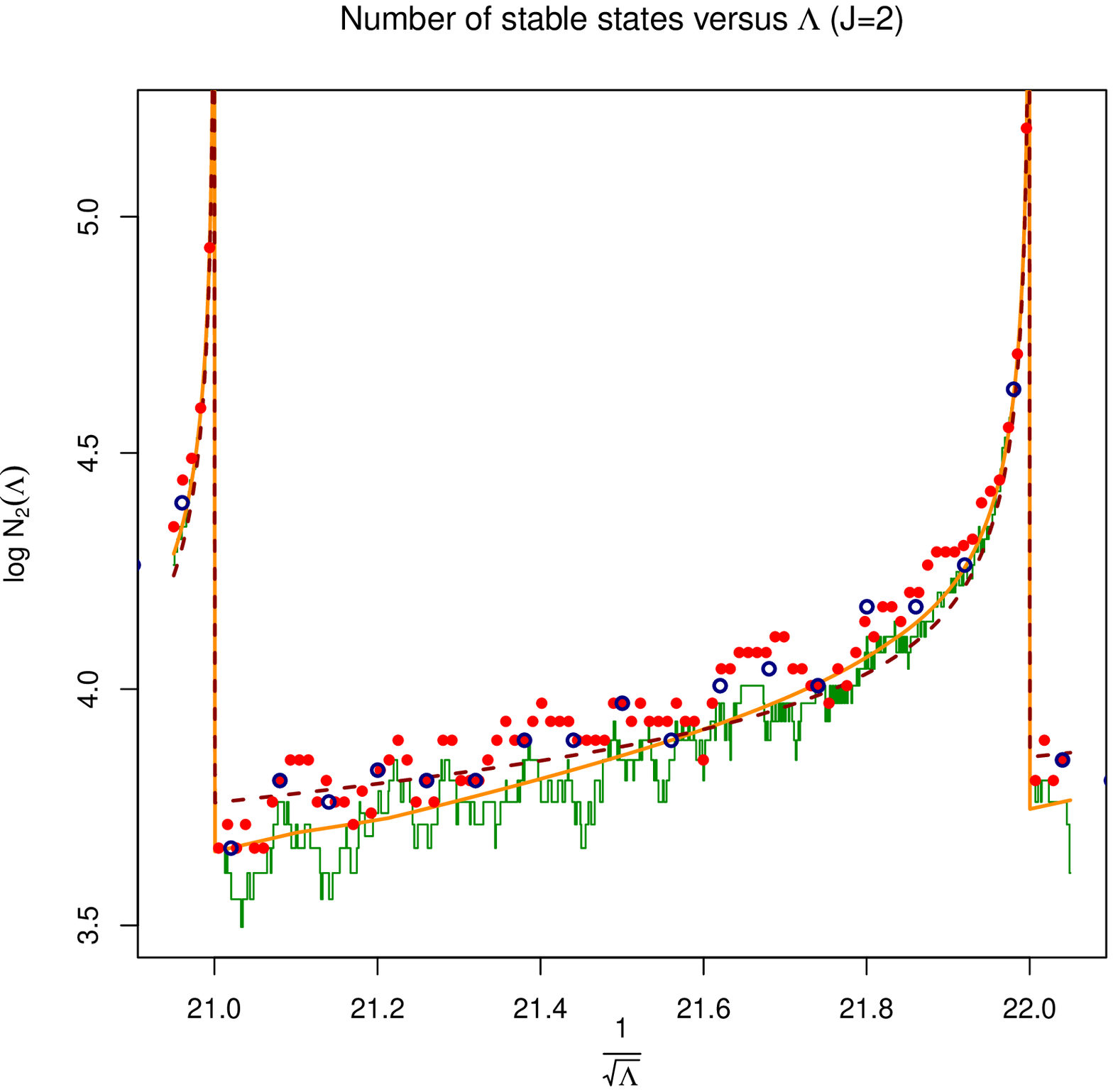}
  \caption{Number of stable dS states as a function of $\Lambda$ for
    the two-sphere Einstein-Maxwell landscape.  Left panel shows the
    strongly discontinuous variation of the state number when
    $\frac{1}{\sqrt{\Lambda}}$ crosses several integer values.
    Bullets are brute-force computed state numbers, thin green line is
    the outcome of formula \eqref{eq:121}, thick solid line is formula \eqref{eq:122} (with
    diagonal corrections) and thick dashed line is formula
    \eqref{eq:123}.  Right panel shows an amplification of the small
    rectangle shown in left panel.  Data are to be interpreted as
    before, with the addition of hollow bullets, which mark the data
    displayed in left panel.  Simplified formulae seem to have an
    averaging effect on the lattice details, which are reflected in
    the fast-varying nature of the discrete formula and brute-force
    data.}
  \label{fig:N2-vs-Lambda-J2}
\end{figure}

While figure \ref{fig:N2-vs-Lambda-J2} emphasizes the strongly
discontinuous nature of the state number, we can also show the steady
increase in the state number by avoiding the discontinuities.  This
can be done, for example, by sampling landscape models with
half-integer values of $\frac{1}{\sqrt{\Lambda}}$.  These samples
never encounter large state chains and thus a regular, well-behaved
curve emerges, very well described by the formulae just obtained.
This smooth component of the state number is illustrated in figure
\ref{fig:N2-vs-Lambda-J2-half}.

\begin{figure}[htbp]
  \centering
  \includegraphics[width=0.75\textwidth]{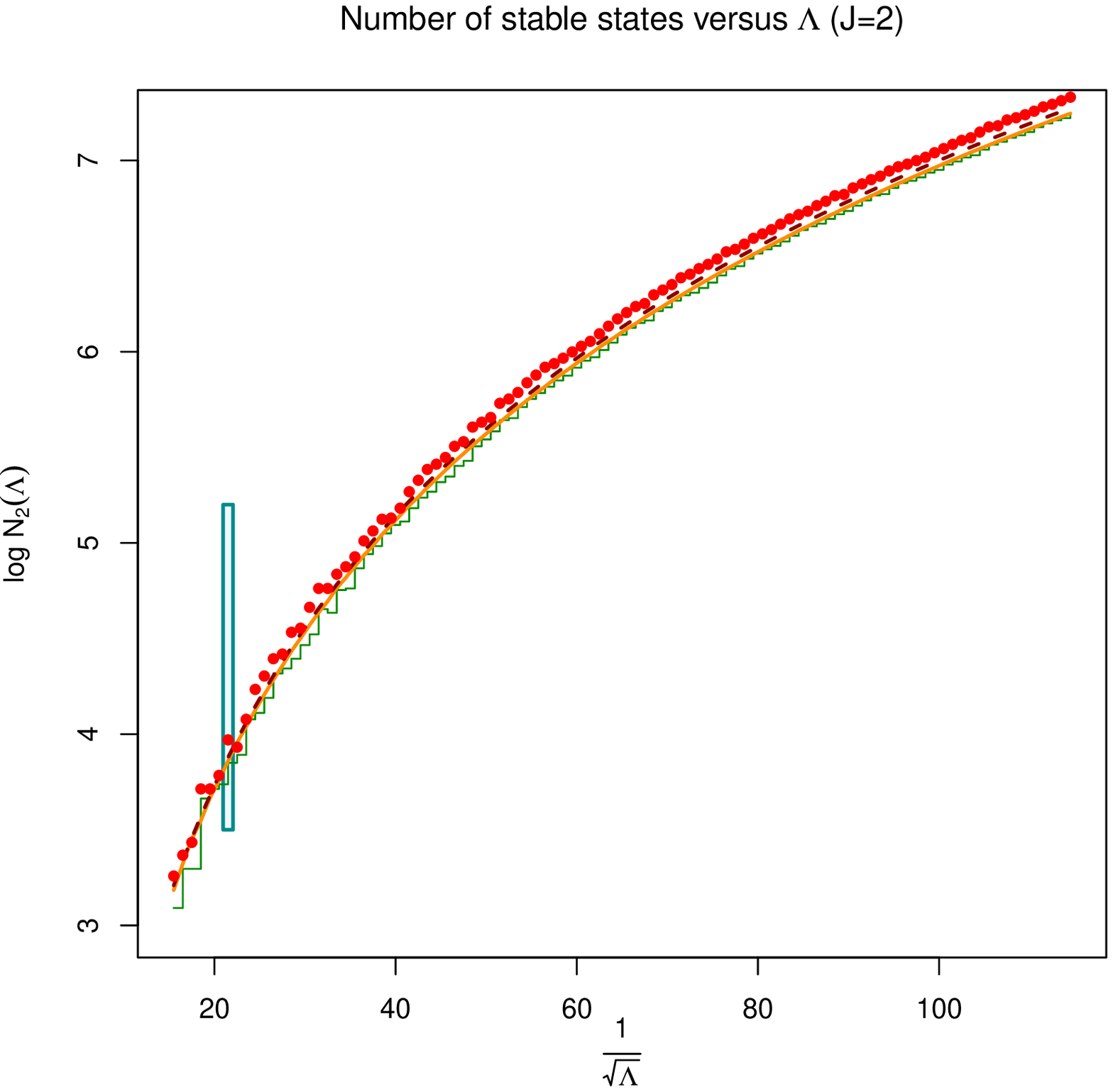}
  \caption{Number of stable dS states as a function of $\Lambda$ for
    the two-sphere Einstein-Maxwell landscape.  Only half-integer
    values of $\frac{1}{\sqrt{\Lambda}}$ have been considered in this
    plot, in order to avoid the spikes shown in figure
    \ref{fig:N2-vs-Lambda-J2}.  Bullets are brute-force computed state
    numbers, thin green line is the outcome of formula \eqref{eq:121},
    thick solid line is formula \eqref{eq:122} (with diagonal
    corrections) and thick dashed line is formula \eqref{eq:123}.  A
    smooth behaviour is observed, showing a very good agreement
    between approximate formulae and numerical searches.  The narrow
    vertical rectangle is at the same position as the rectangle shown
    in figure \ref{fig:N2-vs-Lambda-J2} (left panel).}
  \label{fig:N2-vs-Lambda-J2-half}
\end{figure}

We will close this subsection by summarizing the properties of formula
\eqref{eq:123} as follows:
\begin{itemize}
\item It accurately captures the spikes in the state number when
  $\frac{1}{\sqrt{\Lambda}}$ approaches integer values from below.
  These spikes come from the presence of very long state chains in
  this regime.
\item It correctly represents the main behaviour of the state number
  in a generic sense, that is, when $\frac{1}{\sqrt{\Lambda}}$ is not
  near integer values.
\item We can interpret formula \eqref{eq:123} as an average behaviour
  which turn the fine details of the lattice into a smooth profile
  while taking into account the main discontinuities.
\item Finally, the approximation formulae seem to be missing some
  states.  The reason for this is the approximation we are using to
  count stable states: Equation \eqref{eq:120} implies using the $J-1$
  stability criterion to count stable states in the $J$ model, which
  introduces the error.  In the following subsection we will see that
  the missing states are located near the discontinuities of the
  density of states, and thus they correspond to relatively high
  values of $\lambda$.  This is precisely the condition which makes
  \eqref{eq:119} to break down, so this behaviour was to be expected.
\end{itemize}

\subsection{Small cosmological constant distribution}
\label{sec:small-cc-distribution}

In this subsection, we will call $\lambda(n)$ the 1+1 cosmological
constant of a stable dS state at node $n$ (assuming the state exists),
and we will denote by $\rho$ a fixed value to be compared with
$\lambda(n)$.  With this in mind, we define the distribution function
of $\rho$ in a given multi-sphere EM landscape as the number of stable
dS states whose $\lambda(n)$ value does not exceed $\rho$:
\begin{equation}
  \label{eq:125}
  \Omega_J(\rho,\Lambda) = \sum_{\substack{n\in\Z^J\\ n_j\ne0}}
  \chi_{[L_n(\lambda_{\mathrm{b}}),L_n(0)]}(\Lambda)\,
  \theta\bigl(\kappa(n)\bigr)\,\theta\bigl(\rho - \lambda(n)\bigr)\,.
\end{equation}
The derivative of $\Omega_J(\rho,\Lambda)$ with respect to $\rho$ is
the density of states of the model
\begin{equation}
  \label{eq:126}
  \omega_J(\rho,\Lambda) =
  \frac{\partial\Omega_J(\rho,\Lambda)}{\partial\rho}
  \,.
\end{equation}
As a result of the discreteness of the landscape, $\lambda$ values are
drawn from a discrete set, and thus $\Omega_J(\rho,\Lambda)$ is a
stepwise-varying non-decreasing function of $\rho$, while
$\omega_J(\rho,\Lambda)$ has Dirac deltas at the values of $\rho$
coincident with actual $\lambda(n)$ of existing states at $n$.  The
amplitudes of the Dirac peaks are given by the degeneracies of the
corresponding states.  In this subsection we will obtain some analytic
approximations of the density of states in the regime of small
$\lambda$, and we will use the expressions thus obtained to study the
$\lambda$ spectrum.

We will denote by $\lambda_{\mathrm{max}}$ the maximum $\lambda$ value
a state can have.  Clearly, if $\rho \ge \lambda_{\mathrm{max}}$ then
\begin{equation}
  \label{eq:127}
  \Omega_J(\rho,\Lambda) = \mathcal{N}_J(\Lambda)\,.
\end{equation}
Analogously, we denote by $\lambda_{\mathrm{min}}$ the minimum
$\lambda$ value a state can have.  It is also clear that, if
$\rho\le\lambda_{\mathrm{min}}$, then
\begin{equation}
  \label{eq:128}
  \Omega_J(\rho,\Lambda) = 0\,.
\end{equation}
Thus, the interval $[\lambda_{\mathrm{min}},\lambda_{\mathrm{max}}]$
is the support of the density $\omega_J$.

The upper bound $\lambda_{\mathrm{max}}$ can be computed as follows.
Let us consider the gradient of the function $\lambda(n)$ computed as
if the components of $n$ were continuous variables.  We can derive
equation \eqref{eq:117} implicitly:
\begin{equation}
  \label{eq:129}
  J\partial_{n_j}\lambda + \sum_{i=1}^J\biggl\{
  \delta_{ij}\partial_{n_j}K_i + \partial_{n_j}\lambda\partial_\lambda K_i
  \biggr\} = 0
  \,.
\end{equation}
It follows
\begin{equation}
  \label{eq:130}
  \partial_{n_j}\lambda = -\frac{\partial_{n_j}K_j}{J +
    \sum_{i=1}^J \partial_\lambda K_i}
  = \frac{2}{n_j}\,
  \frac{K_j + \frac{\lambda}{\sqrt{1-2\lambda n_j^2}}}{J -
    \sum_{i=1}^J \frac{1}{\sqrt{1- 2\lambda n_i^2}}}
  \,.
\end{equation}
The denominator of \eqref{eq:130} is clearly negative, and thus the
gradient is always pointing from the branching surface to the limiting
surface.  Right at the branching surface $\lambda=0$ and the gradient
is always infinite.  Right at the limiting surface all components of
the gradient vanish except that of maximum $n_j$.  Thus, the gradient
is always pointing towards the diagonal, except right at the diagonal,
where it points towards the origin.  From this gradient configuration
we conclude that the maximum value of $\lambda$ is achieved at the
cusp of the limiting surface.  But this point is not in the stability
window, and thus the maximum $\lambda$ should be achieved at the onset
of stability along the diagonal.  But right on the diagonal the
stability matrix $H$ is permutation-invariant and we can compute
exactly its stability eigenvalue, which is
\begin{equation}
  \label{eq:131}
  \kappa_{\mathrm{diag}} = K-(J+2)\lambda\,,
\end{equation}
where $K$ is the common value of all curvatures on the diagonal of
flux space.  But then, equation \eqref{eq:117} reads
\begin{equation}
  \label{eq:132}
  \Lambda = \frac{J}{2}\bigl[\lambda + K\bigr]\,,
\end{equation}
which allows us to eliminate $K$ and gives the exact diagonal
stability condition:
\begin{equation}
  \label{eq:133}
  \lambda < \frac{2\Lambda}{J(J+3)} = \lambda_{\mathrm{max}}\,.
\end{equation}
Thus, we have exactly computed the maximum $\lambda$ value.

Things are far more difficult when we address
$\lambda_{\mathrm{min}}$.  We know that the minimum will be close to
the branching surface, but its exact position is unpredictable in
general.  The $J=1$ case is easier, because the landscape is a single
state chain.  In this case, equation \eqref{eq:132} is exact, so we
can obtain the dS spectrum as
\begin{equation}
  \label{eq:134}
  \lambda(n) = 2
  \Biggl(\frac{1}{|n|} - \sqrt{\frac{1}{n^2} - \Lambda}\Biggr)
  \sqrt{\frac{1}{n^2} - \Lambda}
  \,.
\end{equation}
The end state of this chain is at node
$n_{\mathrm{max}}=\lfloor\frac{1}{\sqrt{\Lambda}}\rfloor$, and this
implies
\begin{equation}
  \label{eq:135}
  \lambda_{\mathrm{min}} = 2
  \Biggl(\frac{1}{|\lfloor\frac{1}{\sqrt{\Lambda}}\rfloor|}
  - \sqrt{\frac{1}{\lfloor\frac{1}{\sqrt{\Lambda}}\rfloor^2} - \Lambda}\Biggr)
  \sqrt{\frac{1}{\lfloor\frac{1}{\sqrt{\Lambda}}\rfloor^2} - \Lambda}
  \,.
\end{equation}
In this case, it is possible to give an exact answer to the minimum
$\lambda$ value.  When $\frac{1}{\sqrt{\Lambda}}$ is an integer, then
$\lambda_{\mathrm{min}}=0$, but, as we know, this is not the generic
situation.

Generalizing the result \eqref{eq:135} is difficult.  We can argue as
in the previous subsection and approximate the $J=2$ case by the $J=1$
case just considered.  Then we can assume that the longest state
chains will host the minimum $\lambda$ states at its end nodes.  These
end nodes have an approximate $J=1$ spectrum which can be computed by
replacing $\Lambda$ in \eqref{eq:132} (with $J=1$) by $\Lambda -
\bigl(\lfloor\frac{1}{\sqrt{\Lambda}}\rfloor+1\bigr)^{-2}$.  Now the
same substitution can be performed in formula \eqref{eq:135}, giving a
fairly cumbersome expression of nested fractions, square roots and
floor functions, and hence it will be omitted.  This formula will
approximately give the minimum $\lambda$ provided the corresponding
state is located at the end of the longest chains.  The
minimum-$\lambda$ state can also be located among bulk states, but
this is non-generic, as we will see shortly, because long state chains
are generically low-lying states.  So such a formula can be trusted,
but it is neither exact, nor a bound, but rather it is an approximate
expression for the particular (if generic) case when the min-$\lambda$
state is located at the end of the longest state chain.

Similar arguments can give analogous (but much more complex)
expressions for higher $J$, with the same caveats as before.

Nevertheless, $\lambda_{\mathrm{min}}$ is not quite relevant for our
purposes, because a precise computation of it requires taking into
account even the finest details of the lattice, and thus no continuous
approximation can yield this value.  Instead, we will be interested in
an approximation of the density of states, whose expression will allow
us to estimate $\lambda_{\mathrm{min}}$ in more familiar terms.

\begin{figure}[htbp]
  \centering
  \includegraphics[width=0.75\textwidth]{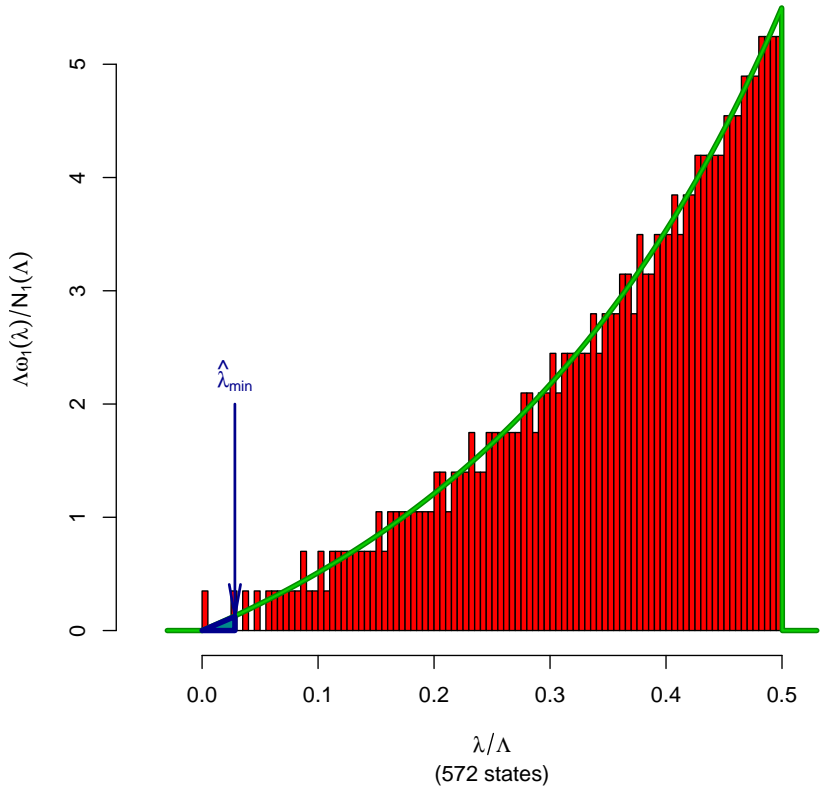}
  \caption{Density of states of the $J=1$ landscape (thick line)
    compared with the actual $\lambda$-spectrum for $\Lambda=10^{-8}$.
    This model has 572 stable states, and the density shown is
    normalized to unity.  The agreement between discrete data and
    continuous density is complete, in the sense that the histogram,
    which approximates a continuous curve when the spacing between
    neighboring states is much smaller than the bin width, accurately
    fits the approximation $\omega_1(\lambda,\Lambda)$.  The
    construction of the estimate $\widehat\lambda_{\mathrm{min}}$ is
    also shown: the small triangle located at the origin has area
    1/572 in this model, and its vertical side marks the position of
    the minimum-$\lambda$ estimate.  This value of $\Lambda$ allows
    for a $\lambda=0$ state, which is isolated by the bins used in the
    histogram.}
  \label{fig:lambda-spectrum-J1}
\end{figure}

Let us consider first the case $J=1$.  By inverting the relation
$\lambda(n)=\rho$, we obtain
\begin{equation}
  \label{eq:136}
  n_\rho = \frac{2\sqrt{\Lambda - \rho}}{2\Lambda - \rho}\,,
\end{equation}
so that the condition $\lambda(n) \le \rho$ can be rephrased as $n \ge
n_\rho$.  Thus, the distribution function $\Omega_1(\rho,\Lambda)$ is
simply the number of integers between $n_\rho$ and $n_{\mathrm{max}}$,
that is,
\begin{equation}
  \label{eq:137}
  \Omega_1(\rho,\Lambda) =
  \biggl\lceil\frac{1}{\sqrt{\Lambda}}\biggr\rceil
  - \biggl\lceil \frac{2\sqrt{\Lambda - \rho}}{2\Lambda - \rho}
  \biggr\rceil
  \,,
\end{equation}
supplemented with conditions \eqref{eq:128} and \eqref{eq:127}.  This
exact result can be approximated by a continuous function by simply
omitting the ceiling functions.  Doing this and taking the
$\rho$-derivative afterwards, we obtain
\begin{equation}
  \label{eq:138}
  \omega_1(\rho,\Lambda) =
  \frac{\rho\,\chi_{[0,\frac{\Lambda}{2}]}(\rho)}{
    (2\Lambda - \rho)^2\sqrt{\Lambda-\rho}}\,.
\end{equation}
Equation \eqref{eq:138} is the $J=1$ density of stable dS states.  It
has the following properties:
\begin{equation}
  \label{eq:139}
  \begin{split}
    \mathcal{N}_1(\Lambda) &\approx
    \int_{\R} \omega_1(\rho,\Lambda)\,\dif\rho 
    = \biggl(1 -
    \frac{2\sqrt{2}}{3}\biggr)\frac{1}{\sqrt{\Lambda}}
     \,, \\
    \langle\lambda\rangle_{\omega_1} &= \frac{1}{\mathcal{N}_1(\Lambda)}\,
    \int_{\R} \rho\,\omega_1(\rho,\Lambda)\,\dif\rho
    = \frac{12 + 12\tan^{-1}\frac{1}{\sqrt{2}} - 3\pi -
      7\sqrt{2}}{3-2\sqrt{2}}\,\Lambda
    \,,
  \end{split}
\end{equation}
That is, it is consistent with equation \eqref{eq:108}, and the mean
value of the density is around $0.7167\frac{\Lambda}{2}$.  This curve
has a jump discontinuity at $\rho=\frac{\Lambda}{2}$, which is the
upper limit of its support, and the position of its maximum, which is
$\frac{2\sqrt{2}}{9\Lambda^{\frac{3}{2}}}$.  Figure
\ref{fig:lambda-spectrum-J1} illustrates this density compared with
the actual spectrum of a $J=1$ model.

\begin{figure}[htbp]
  \centering
  \includegraphics[width=0.75\textwidth]{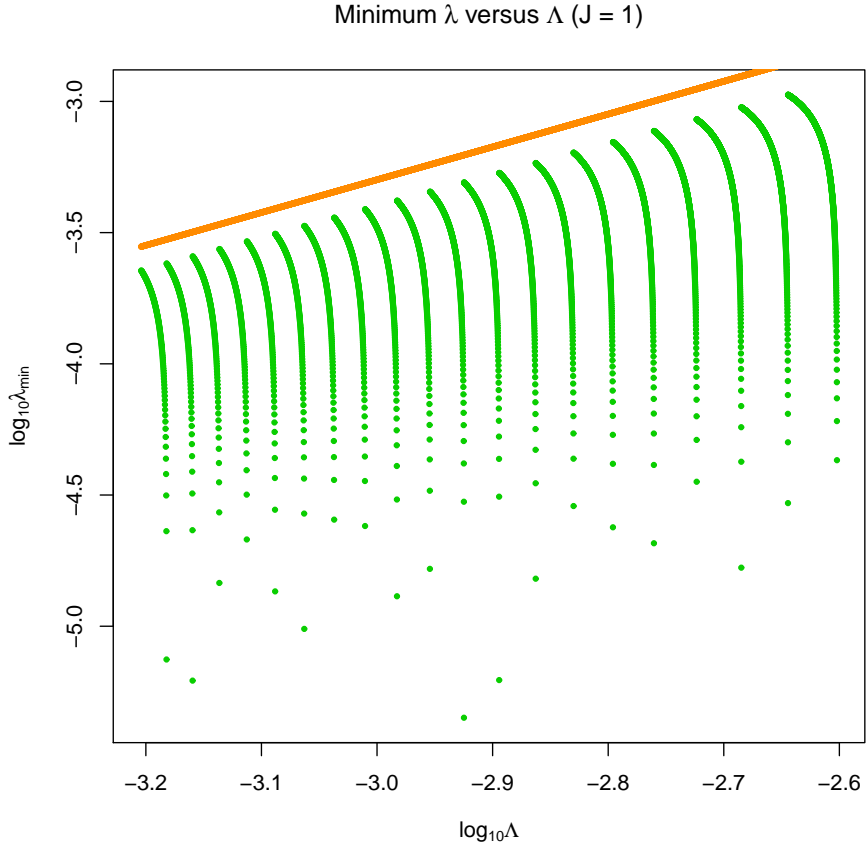}
  \caption{Comparison between the exactly computed minimum-$\lambda$
    (equation \eqref{eq:135}) and its estimate
    $\widehat\lambda_{\mathrm{min}}$ (equation \eqref{eq:141}).  The
    latter seems to be an upper bound of the former, closely following
    the decreasing of the worst-case minimum as $\Lambda$ decreases.
    The spikes shown by $\log\lambda_{\mathrm{min}}$ are the values
    for which $\frac{1}{\sqrt{\Lambda}}$ is an integer.  This
    continuous-density-based estimate is not accurate because the
    value of $\lambda_{\mathrm{min}}$ is dictated by the finest
    details of the lattice and not by the continuous density
    $\omega_1(\lambda,\Lambda)$.}
  \label{fig:lambda-min-J1}
\end{figure}

In figure \ref{fig:lambda-spectrum-J1} is also displayed a naive
estimate of $\lambda_{\mathrm{min}}$, defined as the abscissa
$\widehat\lambda_{\mathrm{min}}$ which encloses area 1 under the
density's graph:
\begin{equation}
  \label{eq:140}
  \int_0^{\widehat\lambda_{\mathrm{min}}}
  \omega_1(\rho,\Lambda)\,\dif\rho = 1\,.
\end{equation}
Assuming that $\widehat\lambda_{\mathrm{min}}$ is small enough, we can
approximate $\omega_1(\rho,\Lambda) =
\frac{\rho}{4\Lambda^{\frac{5}{2}}} + \mathcal{O}(\rho^2)$ and obtain
\begin{equation}
  \label{eq:141}
  \widehat\lambda_{\mathrm{min}} = 2\sqrt{2}\,\Lambda^{\frac{5}{4}}\,.
\end{equation}
Figure \ref{fig:lambda-min-J1} shows this estimate versus the exact
minimum.  Of course, this estimate does not provide the true minimum:
it has neither zeros nor peaks, but it grows at the same average rate.

We will now consider the $J=2$ case.  The recurrence relation
\eqref{eq:121} extends to distributions and densities as well, and
thus we have the approximation
\begin{equation}
  \label{eq:142}
  \omega_2(\rho,\Lambda) = 2
  \sum_{m = \bigl\lfloor\frac{1}{\sqrt{\Lambda}}\bigl\rfloor+1}
  ^{\bigl\lfloor\sqrt{\frac{2}{\Lambda}}\bigr\rfloor}
  \omega_1\biggl(\rho,\Lambda - \frac{1}{m^2}\biggr)
  \,.
\end{equation}
We interpret this equation as a decomposition of the $J=2$ landscape
in a superposition of several $J=1$ landscapes, which are state
chains.  Of course, the previous equation will be only valid for small
$\rho$.  The maximum value of $m$ in the sum \eqref{eq:142}, which is
$\bigl\lfloor\sqrt{\frac{2}{\Lambda}}\bigr\rfloor$, gives a maximum
effective cosmological constant $\Lambda_{\mathrm{eff,max}} = \Lambda -
\bigl\lfloor\sqrt{\frac{2}{\Lambda}}\bigr\rfloor^{-2}$, and the
maximum $\lambda$ of the corresponding state chain is
\begin{equation}
  \label{eq:143}
  \lambda_{\mathrm{max}} = \frac{\Lambda_{\mathrm{eff,max}}}{2}
  \approx \frac{\Lambda}{4}\,,
\end{equation}
in contrast with \eqref{eq:133}, which gives $\lambda_{\mathrm{max}} =
\frac{\Lambda}{5}$.  Thus, we see that this approximation gives a
wrong maximum $\lambda$ value.  The origin of this discrepancy is the
stability condition, because the superposition of state chains extends
the validity of the $J=1$ stability criterion to $J=2$, and this is
true only for small $\lambda$.

\begin{figure}[htbp]
  \centering
  \includegraphics[width=0.5\textwidth]{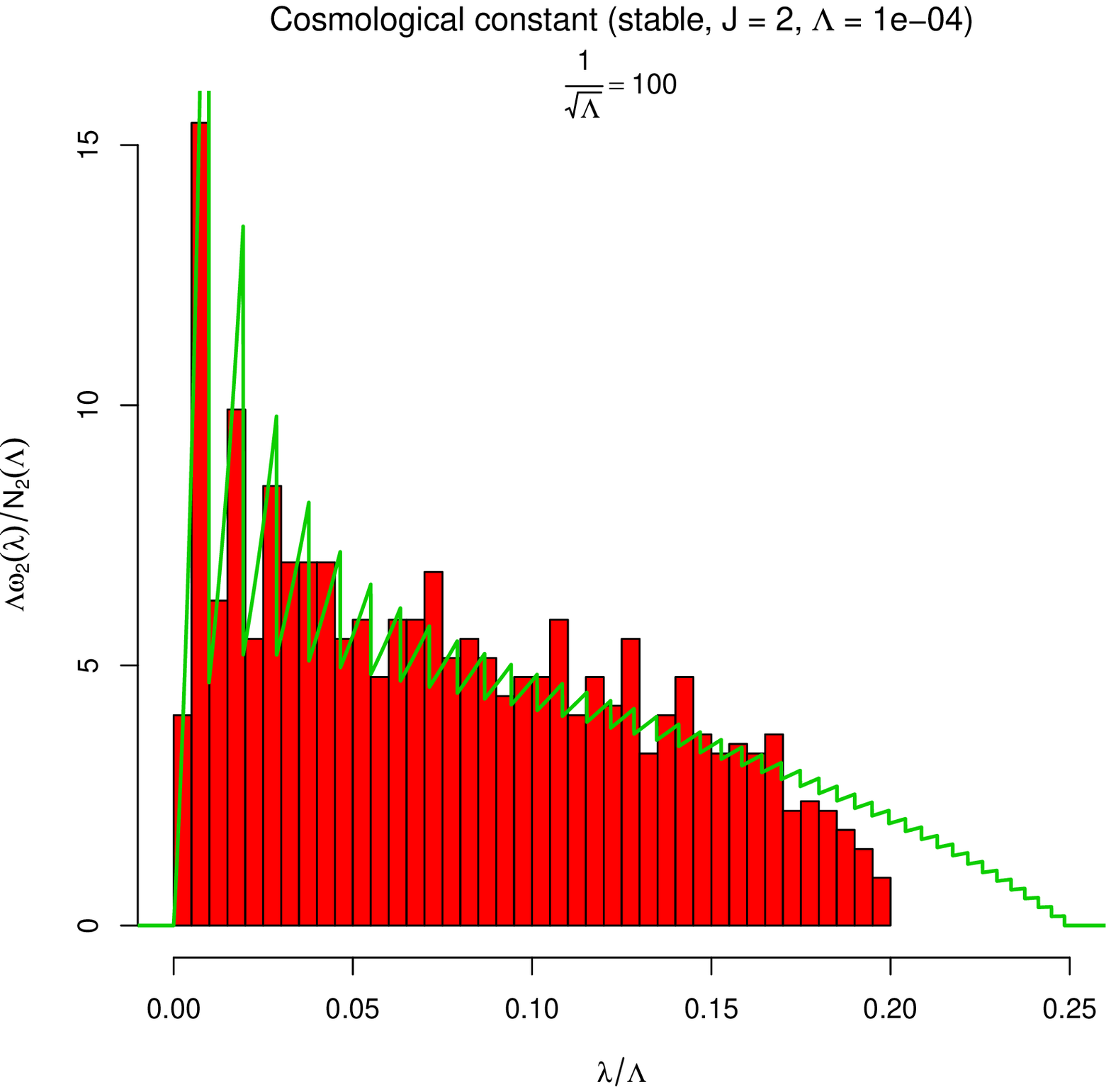}%
  \includegraphics[width=0.5\textwidth]{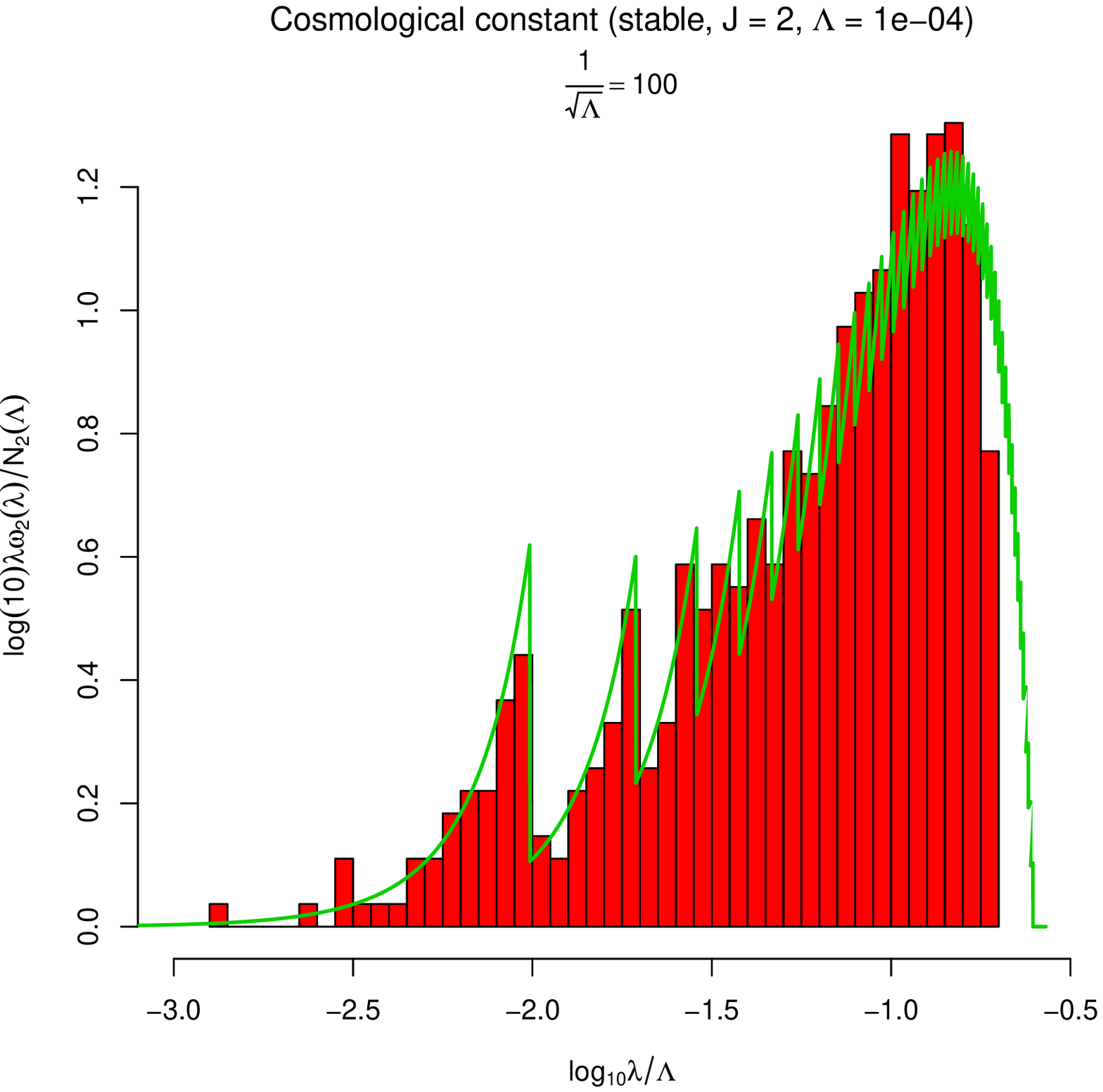}
  \caption{Comparison between a brute-force-computed
    $\lambda$-spectrum (histograms) and the approximated density of
    states (thick line) in a $J=2$ model of the Einstein-Maxwell
    landscape ($\Lambda=10^{-4}$).  Left panel: The ordinary histogram
    shows a narrow peak near the origin (most of them come from long
    state chains) and a tail of bulk states.  The jagged density of
    states accurately accounts for the first few peaks, but it fails
    to describe the high-$\lambda$ values.  Right panel: The
    logarithmic histogram shows the resolved structure of the first
    peaks, well described by the density of states.  This histogram
    also accumulates the bulk states in a single, broad peak.  The
    corresponding bulk part of the density of states is shifted to the
    right.}
  \label{fig:J2-lambda-spectrum}
\end{figure}

An example of this density of states compared with actual
$\lambda$-spectrum data is given in figure
\ref{fig:J2-lambda-spectrum}.  We can see that the histogram shows a
peak near the origin, and the density of states extends its support to
$\frac{\Lambda}{4}$ instead of the correct $\frac{\Lambda}{5}$ value.
In the logarithmic version of the histogram, we can see the first
peaks resolved enough, and the correctness of the state chain
approximation in the low-$\lambda$ region.  Only a few peaks get
resolved; the remaining peaks merge in a bulk distribution whose
approximation computed from \eqref{eq:142} is incorrect in the
high-$\lambda$ region.  We will not need this bulk distribution here.

\begin{figure}[htbp]
  \centering
  \includegraphics[width=0.5\textwidth]{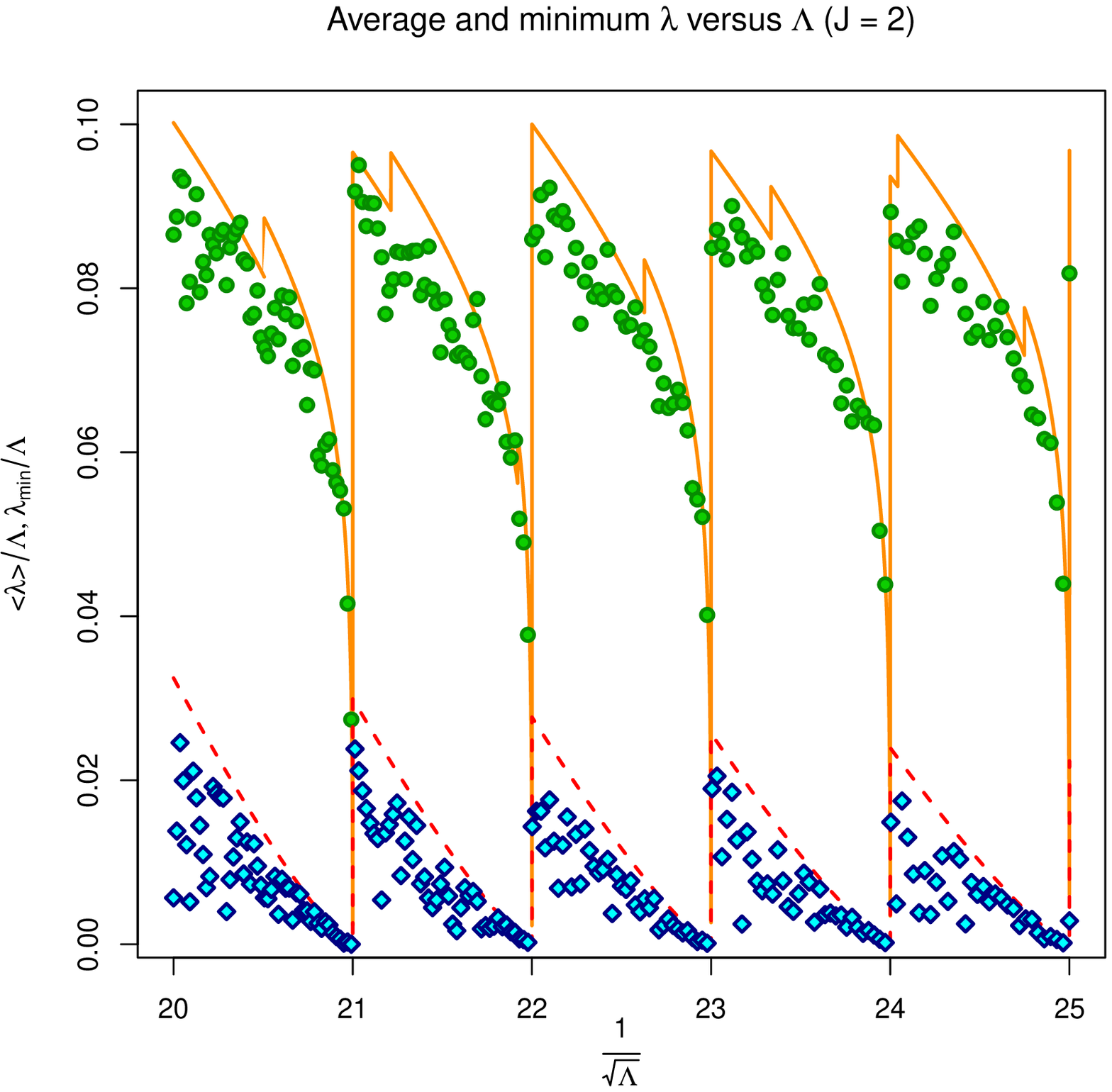}%
  \includegraphics[width=0.5\textwidth]{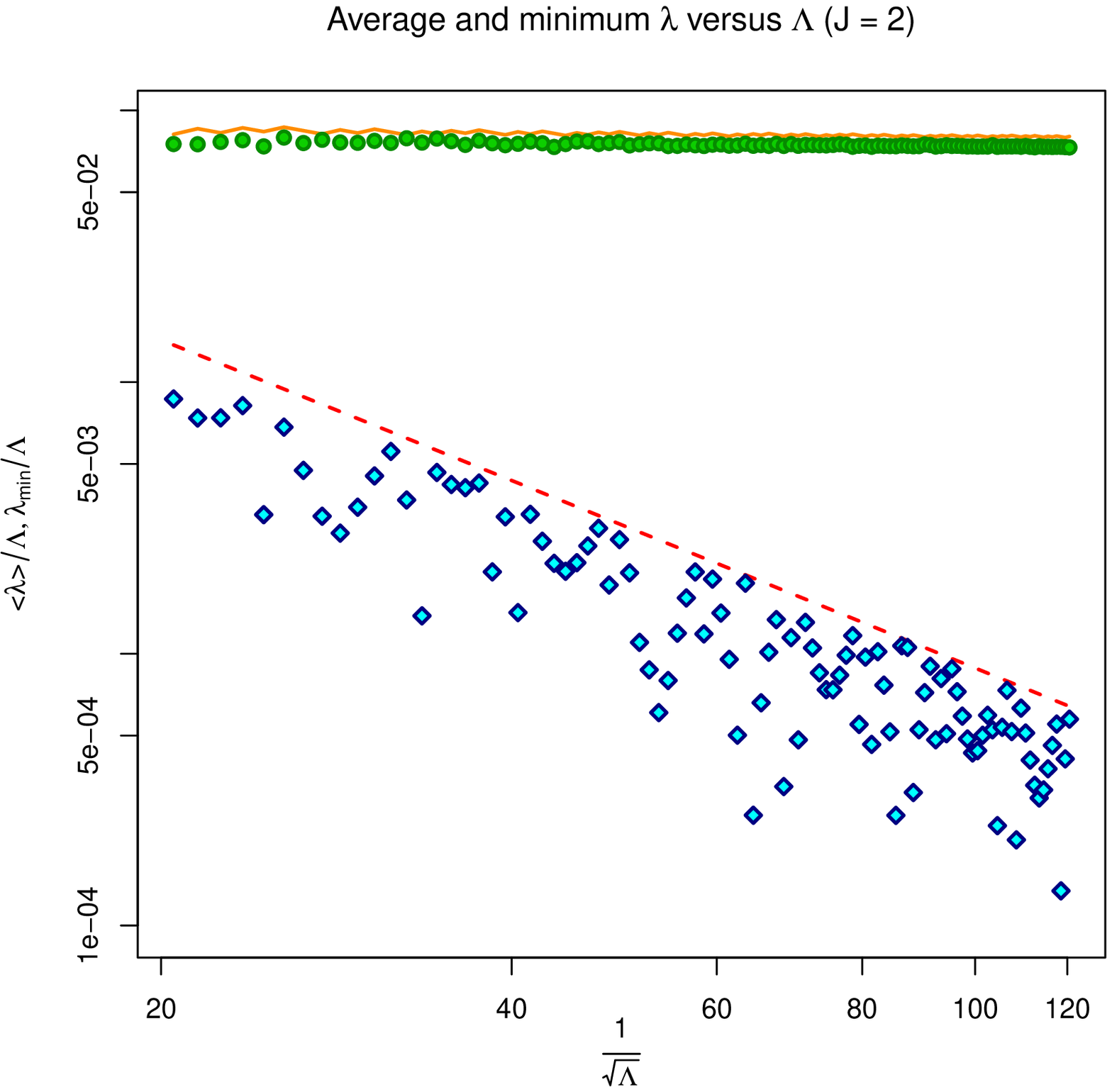}
  \caption{Comparison between brute-force data (bullets, diamonds) and
    approximate formulae (solid, dashed lines) for
    $\langle\lambda\rangle_{\omega_2}$ (bullets, solid lines) and
    $\lambda_{\mathrm{min}}$ (diamonds, dashed lines).  Left panel
    focuses on a small interval enclosing six integer values of
    $\frac{1}{\sqrt{\Lambda}}$.  Valleys of both magnitudes at those
    integer values are caused by long state chains.  The wild
    fluctuation of samples is a consequence of the lattice details.
    The small unevenly-spaced peaks of the solid line are there
    because formula \eqref{eq:145} lacks a diagonal correction as in
    figure \ref{fig:N2-vs-Lambda-J2}.  Right panel shows a much larger
    interval, but samples have been taken only at half-integer values
    of $\frac{1}{\sqrt{\Lambda}}$.  The apparently constant profile of
    $\langle\lambda\rangle_{\omega_2}$ is caused by the scale: it is
    actually decreasing at a rate ten times smaller than
    $\lambda_{\mathrm{min}}$.  Note how formula \eqref{eq:141} for
    $\widehat\lambda_{\mathrm{min}}$ (dashed line) works as an almost
    saturated upper bound for $\lambda_{\mathrm{min}}$.}
  \label{fig:J2-lambda-mean-min}
\end{figure}

The mean value of the $\omega_2$ distribution can be directly computed
from equations \eqref{eq:142} and \eqref{eq:139}.  We denote the
summation interval as $I(\Lambda)$:
\begin{equation}
  \label{eq:144}
  I(\Lambda) = \biggl[
  \Bigl\lfloor\frac{1}{\sqrt{\Lambda}}\Bigl\rfloor+1,
  \Bigl\lfloor\sqrt{\frac{2}{\Lambda}}\Bigr\rfloor\biggr]\,,
\end{equation}
and then we have
\begin{equation}
  \label{eq:145}
  \begin{split}
    \langle\lambda\rangle_{\omega_2} &= 
    \frac{1}{ \mathcal{N}_2(\Lambda) }
    \int_{\R} \rho\, \omega_2(\rho,\Lambda) \,\dif\rho \\
    &= \frac{2}{ \mathcal{N}_2(\Lambda) }
    \sum_{m \in I(\Lambda) }
    \mathcal{N}_1\Bigl(\Lambda - {\textstyle\frac{1}{m^2}}\Bigr)
    \langle\lambda\rangle_{ \omega_1\bigl(\Lambda\to\Lambda - \frac{1}{m^2}\bigr)} \\
    &= \xi\ \frac{\sum_{m\in I(\Lambda)}
      \sqrt{\Lambda - {\textstyle\frac{1}{m^2}}} 
    }{\sum_{m\in I(\Lambda)}
      \frac{1}{\sqrt{\Lambda - {\textstyle\frac{1}{m^2}}}}}
    \,.
  \end{split}
\end{equation}
In \eqref{eq:145}, the constant $\xi$ is the prefactor of $\Lambda$ in
the formula for $\langle\lambda\rangle_{\omega_1}$ appearing in
equation \eqref{eq:139}.  An illustration of the general behavior of
the mean cosmological constant is given in figure
\ref{fig:J2-lambda-mean-min}.  The average $\lambda$ diminishes
towards zero when $\frac{1}{\sqrt{\Lambda}}$ approaches integer values
from below, as a consequence of the development of large state chains.
Formula \eqref{eq:144} is not diagonal-corrected (see above), which
gives small unevenly-spaced jumps.  It is compared with
brute-force-computed averages, which fluctuate because of lattice
details.  The global decreasing of the mean value as $\Lambda$
decreases when $\frac{1}{\sqrt{\Lambda}}$ is half-integer is shown
also in figure \ref{fig:J2-lambda-mean-min} (right panel).

Figure \ref{fig:J2-lambda-mean-min} also shows the minimum $\lambda$,
computed using formula \eqref{eq:141} with the longest chain of $J=2$
models, compared with brute-force-computed minimum values.  The
fluctuation here is caused by the unpredictable nature of the minimum,
which can be located at any point near the branching curve.  When
$\frac{1}{\sqrt{\Lambda}}$ approaches an integer, the very long state
chains are mainly formed out of low-lying states, and thus the
approximate and exact minima approach zero.  We can see a weak
correlation between mean and minimum values: this happens because both
values are strongly influenced by the presence of large state chains,
but the minimum value depends on lattice details in an even stronger
way.

\begin{figure}[htbp]
  \centering
  \includegraphics[width=0.34\textwidth]{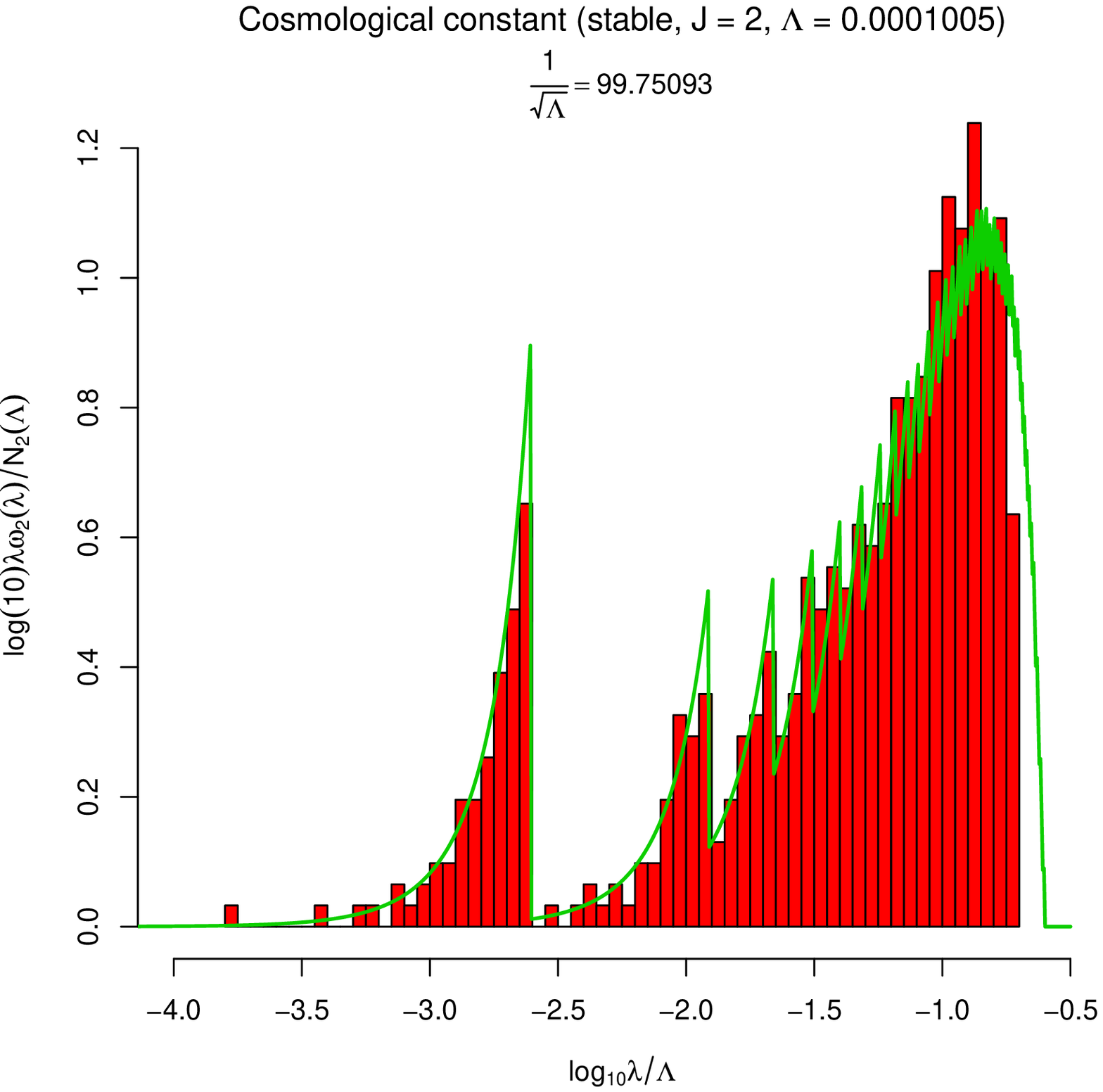}%
  \includegraphics[width=0.34\textwidth]{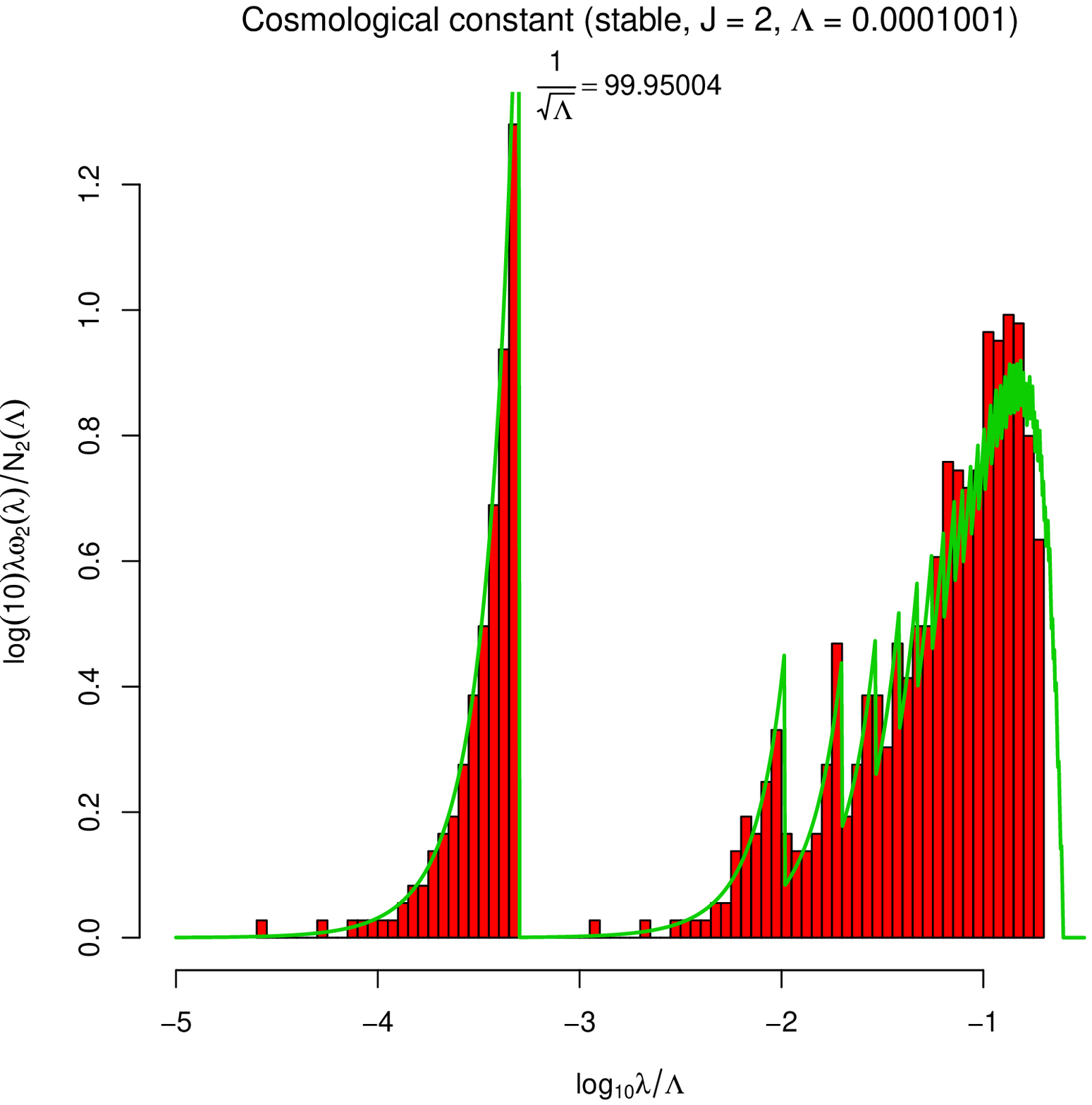}%
  \includegraphics[width=0.34\textwidth]{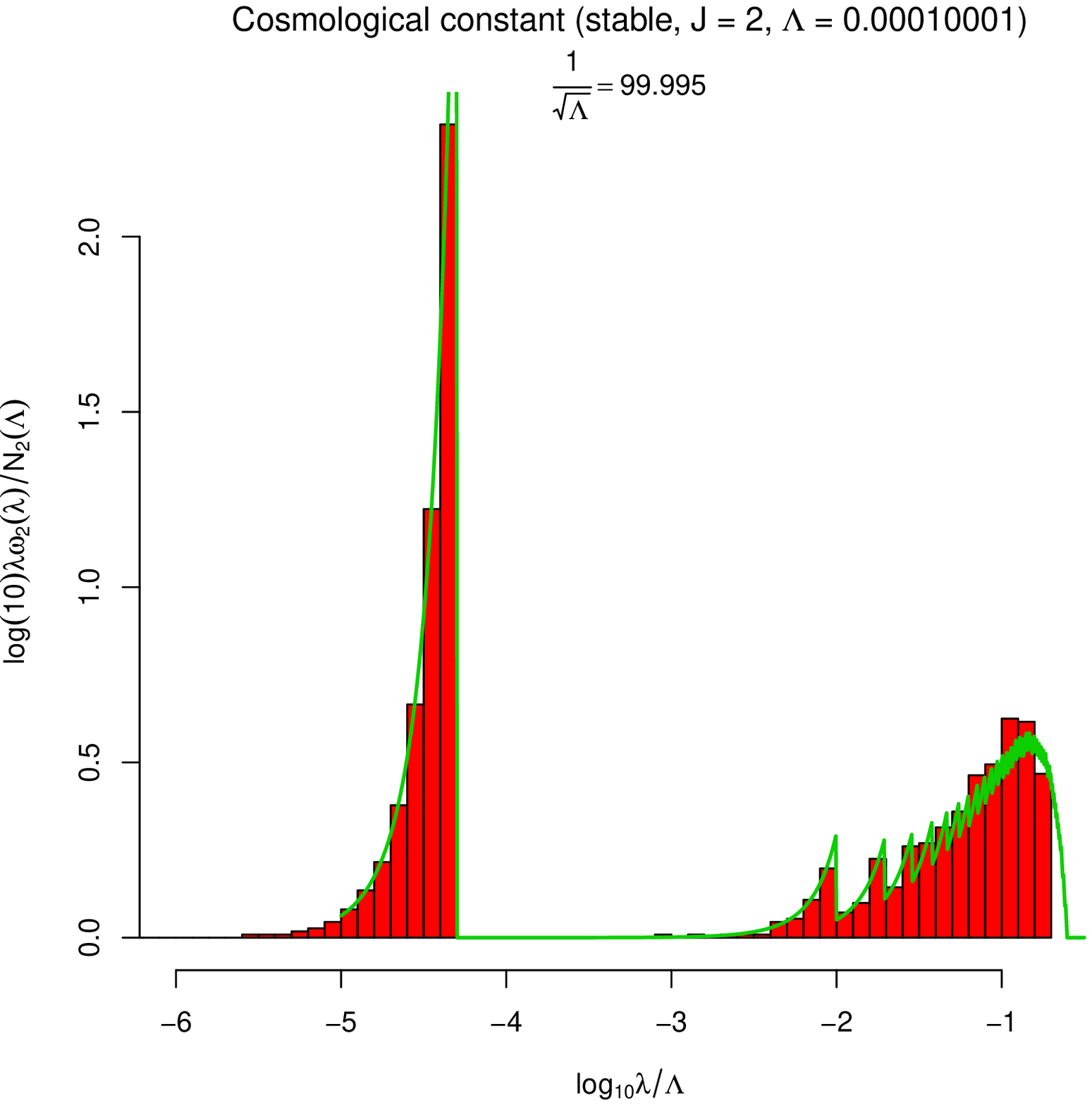}
  \caption{Logarithmic histograms and density of states (solid line)
    for three near values of $\Lambda$ characterizing three different
    examples of the $J=2$ Einstein-Maxwell landscape.  When
    $\frac{1}{\sqrt{\Lambda}}$ approaches an integer from below (100
    in this case), the first peak shifts to the left and becomes
    isolated, thus creating a gap in the $\lambda$-spectrum.}
  \label{fig:J2-gap-generation}
\end{figure}

Large state chains give rise also to a gap between the two
lowest-lying peaks.  This gap develops as $\frac{1}{\sqrt{\Lambda}}$
approaches an integer from below, giving the $\lambda$-spectrum a very
different aspect, as shown in figure \ref{fig:J2-gap-generation}.  As
the longest state chain grows, the peak near the origin becomes taller
and well separated from the second peak.  This separation is greater
than the first peak's width, so that it is effectively isolated from
the second peak.

\begin{figure}[htbp]
  \centering
  \includegraphics[width=0.75\textwidth]{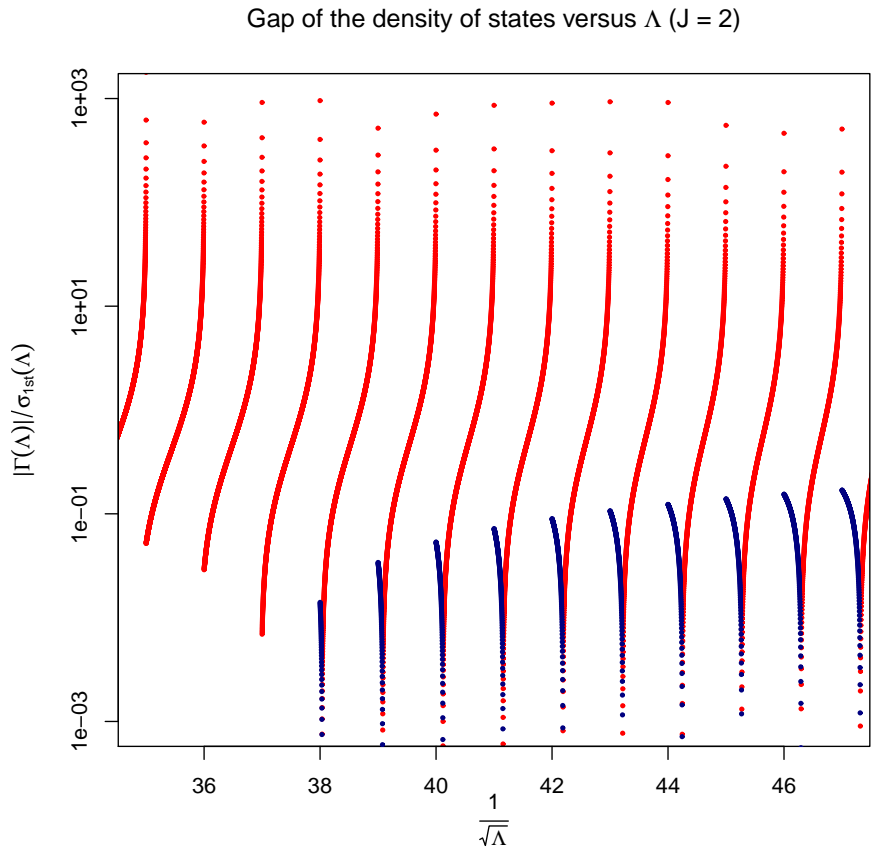}
  \caption{Gap of the $J=2$ density of states as a function of
    $\Lambda$ (equation \eqref{eq:146}).  It is shown the quotient
    between the (absolute value of the) gap and the first peak width.
    Gaps can be negative for sufficiently low $\Lambda$, and the
    negative-gap intervals become greater when $\Lambda$ becomes
    smaller.  Decreasing curves represent negative values of the gap,
    while increasing ones represent positive values.}
  \label{fig:J2-gap-vs-Lambda}
\end{figure}

We can compute an estimate of the gap using the minimum $\lambda$
estimate given in equation \eqref{eq:141}.  This estimate will be
reliable because the approximations leading to it are valid in the two
first peaks of the distribution.  We can define the gap $\Gamma$ as
the distance between the maximum $\lambda$ of the first peak and the
minimum $\lambda$ of the second.  Both of them are known (see
equations \eqref{eq:141} and \eqref{eq:133}), so we have the following
formula for the gap as a function of $\Lambda$:
\begin{equation}
  \label{eq:146}
  \Gamma(\Lambda) = \widehat\lambda^{(\mathrm{2nd})}_{\mathrm{min}} -
  \lambda^{\mathrm{1st}}_{\mathrm{max}} =
  2\sqrt{2}\,
  \biggl[
     \Lambda -
     \biggl(
        \Bigl\lfloor\frac{1}{\sqrt{\Lambda}}\Bigr\rfloor + 2
     \biggr)^{-2}
  \biggr]^{\frac{5}{4}} - \frac{1}{2}\,
  \biggl[
     \Lambda - 
     \biggl(
        \Bigl\lfloor\frac{1}{\sqrt{\Lambda}}\Bigr\rfloor + 1
     \biggr)^{-2}
  \biggr]
  \,.
\end{equation}
This gap is shown in figure \ref{fig:J2-gap-vs-Lambda}, where it is
shown with respect to the width of the first peak.  This width is
computed using the standard deviation $\sigma_1(\Lambda)$ of the
$\omega_1(\rho,\Lambda)$ distribution, which is
\begin{equation}
  \label{eq:147}
  \begin{split}
    \sigma_1(\Lambda)^2 &=
    \langle\bigl(\lambda -
    \langle\lambda\rangle_{\omega_1}\bigr)^2\rangle_{\omega_1} \\
    &=
    \frac{1}{\mathcal{N}_1(\Lambda)} \int_{\R}
    \bigl(\rho-\langle\lambda\rangle_{\omega_1}\bigr)^2
    \omega_1(\rho,\Lambda)\, \dif\rho \\
    &\approx \bigl(0.11\Lambda\bigr)^2
    \,.
  \end{split}
\end{equation}
The first peak of $\omega_2$ is a $\omega_1$ distribution with
$\Lambda$ replaced by $\Lambda -
\bigl(\lfloor\frac{1}{\sqrt{\Lambda}}\rfloor+1\bigr)^{-2}$, and thus
its width is given by
\begin{equation}
  \label{eq:148}
  \sigma_{\mathrm{1^{st}}}(\Lambda) \approx 0.11\biggl(\Lambda -
  \biggl(\biggl\lfloor\frac{1}{\sqrt{\Lambda}}\biggr\rfloor+1\biggr)^{-2}\biggr) 
  \,.
\end{equation}
or some multiple of it.  Immediately we can see that this width will
approach zero as $\frac{1}{\sqrt{\Lambda}}$ approaches an integer, and
therefore the relative gap will become enormous.  There is some values
of $\Lambda$ for which the gap becomes negative, that is, the two
first peaks of $\omega_2$ overlap.  This can happen if
$\frac{1}{\sqrt{\Lambda}}>37$, as can be seen in figure
\ref{fig:J2-gap-vs-Lambda}.  This gap will never disappear for large
values of $\frac{1}{\sqrt{\Lambda}}$ because the width of the first
peak will always vanish at integers, but the intervals of positive
gaps are smaller when $\Lambda$ decreases.  That is, the gap is
positive for $\frac{1}{\sqrt{\Lambda}}\le 36$, and for greater values
the gap changes sign between two consecutive integer values of
$\frac{1}{\sqrt{\Lambda}}$, with the zero being closer and closer to
$\bigl\lceil\frac{1}{\sqrt{\Lambda}}\bigr\rceil$ as $\Lambda$
decreases.

\section{Anthropic states in the multi-sphere Einstein-Maxwell landscape}
\label{sec:anthropic}

Stable dS states in the multi-sphere Einstein-Maxwell landscape can be
interpreted as inflating 1+1 cosmologies.  Such states are devoid of
matter, of course, and thus no real observers can live in such
universes.  Nevertheless, as a toy model of a multiverse, a natural
question one may ask is if anthropic states are present in this model,
that is, if states with very low, realistic effective cosmological
constant exist, and if they are generic, or some fine-tuning is needed
to obtain them.  We have seen above that special values of $\Lambda$
can yield particularly small values of $\lambda$, but a huge amount of
fine-tuning is needed to obtain a realistic value.  Surprisingly
enough, anthropic states \emph{do} exist indeed, and this section is
devoted to describe how to find and count them.  We also draw some
conclusions regarding the multiverse prediction of the cosmological
constant with these anthropic states in mind.

\subsection{Step-by-step construction of anthropic states}
\label{sec:recurrence-relation}

We start by considering equation (\ref{eq:101}) for the branching
surface, which is the $\lambda=0$ locus.  We can try to solve this
equation by successive approximations, looking for the best choice of
an integer at each step in a greedy fashion:
\begin{equation}
  \label{eq:167}
  \sum_{j=1}^J \frac{1}{n_j^2} = \Lambda\equiv\Lambda_1
  \quad\longrightarrow\quad
  \sum_{j=2}^J \frac{1}{n_j^2} = \Lambda_1 -
  \frac{1}{n_1^2}\equiv\Lambda_2 > 0
  \quad\Rightarrow\quad
  n_1 = \left\lceil\frac{1}{\sqrt{\Lambda_1}}\right\rceil\,.
\end{equation}
We have called $\Lambda\equiv\Lambda_1$ for the start of a recurrence
relation replicating the previous step:
\begin{equation}
  \label{eq:168}
  \Lambda_{j+1} = \Lambda_{j} - \frac{1}{n_j^2}\,,\qquad
  n_j = \left\lceil\frac{1}{\sqrt{\Lambda_j}}\right\rceil\,.
\end{equation}
The recurrence relation (\ref{eq:168}) gives the best integer choice
at each step for getting the smallest possible difference between the
two sides of the formula
\begin{equation}
  \label{eq:169}
  \sum_{i=j}^J \frac{1}{n_i^2} = \Lambda_{j}\,.
\end{equation}
The last step of the approximation is
\begin{equation}
  \label{eq:170}
  \Lambda_J - \frac{1}{n_J^2} = \Lambda - \sum^J_{j=1}\frac{1}{n_j^2}
  \equiv \Lambda_{J+1} < 0\,,
\end{equation}
that is, the last remainder should be negative, so that the existence
condition (\ref{eq:103}) can be satisfied.  This gives the last
integer as
\begin{equation}
  \label{eq:171}
  n_J = \left\lfloor\frac{1}{\sqrt{\Lambda_J}}\right\rfloor\,,
\end{equation}
where the floor function is taken instead of the ceiling to guarantee
that the last remainder is negative.  Thus, we can run the recurrence
relation (\ref{eq:168}) starting from any positive value of
$\Lambda_1$ until some desired number of steps $J$ is reached, and
then finish it with the last step (\ref{eq:171}).

\begin{figure}[htbp]
  \centering
  \includegraphics[width=0.75\textwidth]{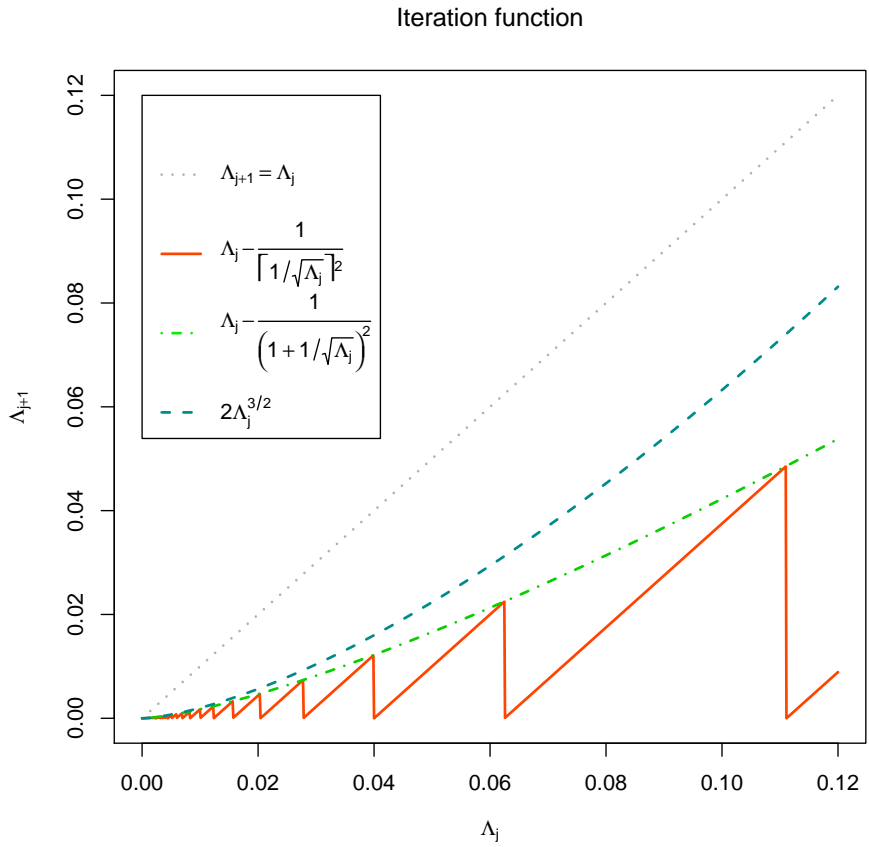}
  \caption{Iteration function of recurrence relation (\ref{eq:172}).
    It is shown along with its envelope and its first-order Taylor
    approximation, which is used as upper bound in equation
    (\ref{eq:174}).  The first quadrant diagonal is also drawn,
    showing that the only fixed point of the iteration is at
    $\Lambda=0$.}
  \label{fig:iteration-function}
\end{figure}

Before the final step closes the algorithm, we can rewrite the
recurrence relation as a fixed-point iteration:
\begin{equation}
  \label{eq:172}
  \Lambda_{j+1} = f(\Lambda_j)\,,\quad\text{with}\quad
  f(x) = x - \frac{1}{\left\lceil\frac{1}{\sqrt{x}}\right\rceil^2}\,.
\end{equation}
The iteration function just defined $f(\Lambda)$ has jump
discontinuities when $\frac{1}{\sqrt{\Lambda}}$ is an integer, and it
is simply $\Lambda-1$ if $\Lambda>1$.  Its continuous envelope, which
is easily obtained replacing
$\left\lceil\frac{1}{\sqrt{x}}\right\rceil$ with
$\frac{1}{\sqrt{x}}+1$, gives the magnitude of the jumps, and it has a
particularly attractive behaviour when $x\to0$:
\begin{equation}
  \label{eq:173}
  f(x) = x - \frac{1}{\left\lceil\frac{1}{\sqrt{x}}\right\rceil^2}
  \le x - \frac{1}{\left(\frac{1}{\sqrt{x}} + 1\right)^2}
  \xrightarrow{\quad x\to 0\quad} 2x^{\frac{3}{2}}\,.
\end{equation}
The iteration function, its envelope and its first-order term are
plotted in figure \ref{fig:iteration-function}.  The figure also shows
the first-quadrant diagonal, thereby proving that the only fixed point
of the recurrence is at $x=0$.

The super-linear behaviour of $f(x)$ near $x\to0$ has the nice
consequence of giving the recurrence relation a very fast convergence
rate.  This can be seen by replacing $f(x)$ by its first-order
approximation, which is an upper bound:
\begin{equation}
  \label{eq:174}
  \Lambda_{j+1} = f(\Lambda_j) < 2\Lambda_j^{\frac{3}{2}}\,.
\end{equation}
The resulting approximate recurrence relation is exactly solvable, and
its solution starting from $\Lambda_1$ is
\begin{equation}
  \label{eq:175}
  \Lambda_j = 2^{\sum^{j-2}_{k=0}(\frac{3}{2})^k}
  \Lambda_1^{(\frac{3}{2})^{j-1}}
  \qquad(j\ge2)\,.
\end{equation}
Thus, as long as $\Lambda_1 < 1$, the previous upper bound decreases
at a double-exponential rate, thus providing very small values of the
negative remainder $\Lambda_{J+1}$ when the last step is taken for
moderate values of $J$.

We have thus a recipe for obtaining a node $\{n_1,\cdots,n_J\}$ with
the property of being an approximate solution of the branching surface
equation with a very small negative remainder.  Nevertheless, it
should be shown that both inequalities of the existence condition
(\ref{eq:103}) are satisfied, because only one of them is guaranteed
by the last step (\ref{eq:171}).  The solution of the existence
equation (\ref{eq:102}) should be smaller that the branching point
$\lambda_\mathrm{b}$, which in this case is
\begin{equation}
  \label{eq:176}
  \lambda_\mathrm{b} = \frac{1}{2n_J^2} \approx \frac{\Lambda_J}{2}\,.
\end{equation}
Equation (\ref{eq:171}) guarantees that $L_n(0) > \Lambda$, thus it
remains to show that $L_n(\lambda_\mathrm{b}) < \Lambda$.  The $L_n$
function evaluated at $\lambda_\mathrm{b}$ is
\begin{equation}
  \label{eq:177}
  L_n(\lambda_\mathrm{b}) = \frac{1}{2}\biggl(
  J\lambda_b + \sum_{j=1}^J \frac{1 +
    \sqrt{1-2\lambda_\mathrm{b}n_j^2}}{n_j^2}
  \biggr) =
  \frac{J}{4n_J^2} + \frac{1}{2}\sum_{j=1}^J \frac{1 +
    \sqrt{1-\frac{n_j^2}{n_J^2}}}{n_j^2}\,.
\end{equation}
The convergence rate of the approximate recurrence relation
(\ref{eq:174}) is so fast that the integers $n_j$ grow in such a way
that $n_J$ is overwhelmingly larger than the rest, and thus all square
roots in equation (\ref{eq:177}) can be approximated by first-order
Taylor expansions, except for the last, which is zero:
\begin{equation}
  \label{eq:178}
  L_n(\lambda_\mathrm{b}) \approx \frac{J}{4n_J^2} +
  \frac{1}{2}\sum_{j=1}^{J-1} \Bigl[\frac{2}{n_j^2} - \frac{1}{2n_J^2}\Bigr]
  + \frac{1}{2n_J^2}
  = \sum_{j=1}^{J}\frac{1}{n_j^2} - \frac{1}{4n_J^2}\,.
\end{equation}
The difference with $\Lambda$ is
\begin{equation}
  \label{eq:179}
  L_n(\lambda_\mathrm{b}) - \Lambda = \sum_{j=1}^{J}\frac{1}{n_j^2} - \Lambda -
  \frac{1}{4n_J^2} \approx |\Lambda_{J+1}| - \frac{\Lambda_J}{4} < 0\,,
\end{equation}
where the last inequality follows from the recurrence relation at its
final step, because $|\Lambda_{J+1}|$ is much smaller than
$\Lambda_J$.  This proves that a state exists at the node provided by
the recurrence relation.

We now estimate the corresponding solution $\lambda$ of the existence
equation $L_n(\lambda)=\Lambda$, and prove its stability.  We can
obtain a solution by using again that the integers
$\{n_1,\cdots,n_J\}$ grow very fast, so that we can replace all
curvatures by its first-order expansions in $\lambda$ (we know that
$\lambda<\lambda_\mathrm{b}$) except for the last, which remains
untouched, thus respecting the location of the branching point:
\begin{equation}
  \label{eq:180}
  \Lambda = L_n(\lambda) = \frac{1}{2}\biggl(
  J\lambda + \sum^J_{j=1} K_j
  \biggr)
  \approx
  \frac{1}{2}\biggl[
  J\lambda + \sum^J_{j=1} \biggl(
  \frac{2}{n_j^2} - \lambda
  \biggr) + K_J
  \biggr]
  = \sum^{J-1}_{j=1} \frac{1}{n_j^2}
  + \frac{1}{2} \bigl( \lambda + K_J \bigr)
  \,.
\end{equation}
We can rewrite the previous equation as
\begin{equation}
  \label{eq:181}
  \Lambda_J = 
  \Lambda - \sum^{J-1}_{j=1} \frac{1}{n_j^2}
  = \frac{1}{2} \bigl( \lambda + K_J \bigr)\,,
\end{equation}
which is exactly the existence equation for a EM landscape with a
single curvature $K_J$ and an effective four-dimensional cosmological
constant $\Lambda_J$.  This $J=1$ EM landscape has been obtained by
fixing the integers $\{n_1,\cdots,n_{J-1}\}$ by means of the
recurrence relation (\ref{eq:168}).  The last integer $n_J$, if chosen
as in (\ref{eq:171}), gives the last node verifying the existence
equation.  We know that no greater value of $n_J$ will satisfy the
existence equation, but smaller values can also give valid solutions.
Thus, varying $n_J$ downwards from (\ref{eq:171}) provides us with a
state chain embedded in the $J$-sphere EM landscape: This state chain
is simply the single-sphere EM landscape described by equation
(\ref{eq:181}).

The analysis of the $J=1$ EM landscape performed in subsections
\ref{sec:one-flux}, \ref{sec:modulus-stab} and
\ref{sec:small-cc-distribution} is now entirely applicable to
(\ref{eq:181}).  In particular, the exact minimum two-dimensional
cosmological constant of this chain is given by equation
(\ref{eq:135}) (with $\Lambda$ replaced by $\Lambda_J$) or by its
continuum approximation given in equation (\ref{eq:141}) (with the
same replacement):
\begin{equation}
  \label{eq:182}
  \widehat{\lambda}_{\mathrm{min}} \approx 2\sqrt{2}
  \bigl(\Lambda_J\bigr)^{\frac{5}{4}}\,.
\end{equation}
We know that $\Lambda_J$ is very small, and we now see that
$\widehat{\lambda}_{\mathrm{min}}$ is even smaller.

The stability condition for the $J=1$ landscape (\ref{eq:133}) reads
$\lambda < \frac{\Lambda_J}{2}$.  We can see that
$\widehat{\lambda}_{\mathrm{min}} =
4\sqrt{2}\Lambda_{J}^{\frac{1}{4}}\,\frac{\Lambda_J}{2} \lll
\frac{\Lambda_J}{2}$, and thus this minimum-$\lambda$ state is always
stable.  Moreover, we can let $n_J$ decrease until it reaches the
stability limit.  This generates all dS stable states in the chain,
whose number is given by (\ref{eq:108}), which is
\begin{equation}
  \label{eq:183}
  \mathcal{N}_{1}(\Lambda_J;n_1,\cdots,n_{J-1})
  \approx \biggl(1 - \frac{2\sqrt{2}}{3}\biggr)\frac{1}{\sqrt{\Lambda_J}}
  \approx 0.05719\cdot n_J\,.
\end{equation}
This is an enormous number, as we now see.  We will choose a reference
value $\lambda_A$, and we wish $\widehat{\lambda}_{\mathrm{min}}$ to
reach it.  We can compute the value of $J$ we need for this to happen
by inserting the worst-case approximate formula (\ref{eq:175}) in
equation (\ref{eq:182}) for the minimum $\lambda$ value:
\begin{equation}
  \label{eq:184}
  \lambda_A = 2\sqrt{2} \bigl(\Lambda_J\bigr)^{\frac{5}{4}} =
  2^{\frac{3}{2} + \frac{5}{4}\bigl[\frac{(\frac{3}{2})^{J-1} - 1}{\frac{3}{2}-1}\bigr]}
  \Lambda^{\frac{5}{4}(\frac{3}{2})^{J-1}}\,.
\end{equation}
Solving for $J$, we obtain
\begin{equation}
  \label{eq:185}
  J = 1 + \log_{\frac{3}{2}}\biggl(\frac{4}{5}\, 
  \frac{\log(2\lambda_A)}{\log(4\Lambda)}\biggr)\,.
\end{equation}
We can also demand a much more restrictive condition, that the whole
chain is inside the anthropic range.  The peak of the density is
located at $\frac{\Lambda_J}{2}$, and thus the relation $\lambda_A =
\frac{\Lambda_J}{2}$ together with (\ref{eq:175}) leads to a value of
$J$ given by
\begin{equation}
  \label{eq:187}
  J = 1 + \log_{\frac{3}{2}}\biggl(\frac{\log(\lambda_A/2)}{\log(4\Lambda)}\biggr) \,.
\end{equation}
Using the emblematic number $\lambda_A=10^{-120}$ and $\Lambda=0.1$,
we obtain a non-integer $J=14.5$ with the first formula and $15.08$
with the second; using $J=15$ we find $2.43\cdot10^{58}$ states in the
chain with a minimum of order $10^{-146}$.  In this case, the
stability limit is around $10^{-117}$, well inside the anthropic
range.  As another example, starting from $\Lambda=0.0008$, we obtain
$J=10$ almost exactly with the first formula and $10^{47}$ states.
The second formula provides $J=10.55$, and with $J=11$ we obtain
$10^{72}$ states.

Therefore, we can see that moderate values of $J$ and $\Lambda$ can
yield an enormous number of anthropic states in the multi-sphere EM
landscape.

We may ask if the states just found are generic inside the
$J$-dimensional landscape, because the recurrence relation
(\ref{eq:168}), (\ref{eq:171}) leading to them gives very precise
values for the integers $\{n_1,\cdots,n_J\}$, and therefore they seem
to be located at a very special place in flux space.  We will now see
that, despite being very numerous, these anthropic states are not
generic.

We have just obtained a very long state chain by fixing
$n_1,\cdots,n_{J-1}$ and letting $n_J$ to vary from (\ref{eq:171})
downwards.  This state chain is a one-dimensional landscape embedded
in $J$-dimensional flux space.  We can let $n_{J-1}$ vary downwards as
well, thus generating a two-dimensional landscape embedded in
$J$-dimensional flux space.  The effective high-dimensional
cosmological constant of this landscape is $\Lambda_{J-1}$, and
it is very small, which allows us to use formula (\ref{eq:123}) with
the approximation (\ref{eq:124}) to give the number of states of this
two-dimensional landscape as
\begin{equation}
  \label{eq:186}
  \mathcal{N}_2(\Lambda_{J-1};n_1,\cdots,n_{J-2}) \approx
  2\biggl(1 - \frac{2\sqrt{2}}{3}\biggr)
  \bigl(n_J + n_{J-1}^2\bigr)\,,
\end{equation}
where we have used that
$n_{J-1}\approx\frac{1}{\sqrt{\Lambda_{J-1}}}$.  The first
contribution in formula (\ref{eq:186}), $n_J$, comes from the longest
state chain, while the second, $n_{J-1}^2$, comes from the bulk.  The
simplified recurrence (\ref{eq:174}) shows that $n_J\approx
\frac{1}{\sqrt{2}}n_{J-1}^{3/2}$, that is, the number of states in the
chain scales as $n_{J-1}^{3/2}$ while the number of states in the bulk
scales as $n_{J-1}^2$.  Thus, the fraction of states in this
two-dimensional landscape belonging to the chain scales as
$n_{J-1}^{-1/2}$, and therefore they are non-generic.

For example, choosing $\Lambda = 0.0008$ and $J=10$ we obtain $n_J
\approx 10^{48}$, but $n_{J-1}^2 \approx 10^{64}$.  Thus, states in
the chain are in a proportion $1 : 10^{16}$.

We may as well let the remainder of the integers $n_1,\cdots,n_{J-2}$
vary downwards from (\ref{eq:168}), thus generating the entire
$J$-dimensional landscape.  In this complete landscape the proportion
will be much smaller than $n_{J-1}^{-1/2}$, and thus we see that
anthropic states are very rare, despite being very numerous.  We
cannot exclude the possibility that other corners of flux space may
contain low-$\lambda$ states, either as isolated, randomly close
nodes, or as very long chains obtained in a different way, but they
will be non-generic also.

Summarizing, we have seen that for any value of $\Lambda$ (say,
between $10^{-4}$ and $10^{-1}$) moderate values of $J$ (between 10
and 15 respectively) lead to the existence of a huge chain of
anthropic states, that is, states having a two-dimensional
cosmological constant of order $10^{-120}$. Those states represent a
tiny fraction of the total number of states, and thus they are
non-generic.  But they are very numerous, and they can be found with
no fine tuning at all, which is a very remarkable feature of the
multi-sphere EM landscape.

\subsection{Implications for the multiverse prediction of the cosmological constant}
\label{sec:lambda-prediction}

The very long chains of anthropic states found in the previous
subsection are another form of the discretuum introduced by Bousso and
Polchinski \cite{BP} as part of the solution of the cosmological
constant problem.  Moderate values of $J$ can yield a 1+1 effective
cosmological constant of the order of the observed value in our
universe.  The only parameter of the model, $\Lambda$, can be chosen
as any positive real number to achieve that.  Thus, the multi-sphere
EM landscape do not need fine-tuning $\Lambda$ to contain anthropic
states in the discretuum.

There are another possibilities to produce a discretuum.  A very small
charge produces a finely spaced tower of states as in the
Brown-Teitelboim mechanism \cite{BT-1,BT-2}, or a number of different,
incommensurable elemental charges can yield a BP-like discretuum.  In
the first case, a single, very small parameter is needed, while in the
multi-sphere EM model the parameter is not restricted at all.  In the
second case, as commented above, a number of parameters are given from
the start, thus bypassing the need for a stabilization mechanism.
This mechanism works only for large $J$, which are easily obtained in
some Calabi-Yau compactification scenarios, but large values of $J$
give rise to the $\alpha^*$-problem discussed in section
\ref{sec:intro}.  Thus, no fine-tuning is needed in such cases, but
the counting of states becomes tricky, because most of them might be
unstable if its stability were correctly addressed, as is demonstrated
in the multi-sphere Einstein-Maxwell model.  Thus, we conclude that,
as a discretuum-generating method, state chains circumvent some
previously encountered problems.

Anthropic state chains have further implications in the prediction of
the cosmological constant distribution in realistic landscapes.  As
stated in section \ref{sec:intro}, a multiverse prediction of the
cosmological constant requires a prior probability distribution
counting the states present in the model, a cosmological measure to
weigh relative probabilities, and an anthropic factor taking into
account the existence of observers \cite{Multiv-1}.  Authors in
\cite{Multiv-1} state that the prediction is very sensitive to changes
in the prior distribution, so we may wonder how state chains can
change the prediction.  

The current multiverse prediction of the cosmological constant assumes
that the prior distribution has a scale of variation of order the
Planck scale, which is enormous when compared with the anthropic
range.  Thus it is safe to consider that the prior distribution is
almost constant in the anthropic range, and the cosmological constant
prediction is dominated by the anthropic factor.

The anthropic range, also called ``Weinberg window'', is an interval
of values of the cosmological constant which allow the formation of
structures, such as galaxies, which may contain observers like us.
The order of magnitude of such an interval is large when compared with
the observed value of the cosmological constant \cite{WW}.  Thus, if
the prior probability has a very narrow peak inside the anthropic
range of width comparable to the observed value
$\lambda_{\mathrm{obs}}$, then the anthropic factor, varying on a much
larger scale, can be considered as almost constant.  Therefore, the
prediction of the cosmological constant would be dominated by the
prior distribution.  This is precisely the case with anthropic chains
in the multi-sphere EM model.

Obviously, an anthropic factor is entirely out of question in the
context of a 1+1 cosmology.  Thus, the prior distribution will
dominate the prediction if anthropic state chains can be shown to
exist in a multi-sphere EM model with a 3+1 cosmology.  The
construction of such a model is left as future work.

\section{Comparison between the Bousso-Polchinski and multi-sphere
  Einstein-Maxwell landscapes}
\label{sec:EM-vs-BP}

Obviously, the multi-sphere Einstein-Maxwell landscape cannot be
considered as a model of the string theory landscape, because it
belongs to a completely different family of theories.  Nevertheless,
the features we have described in the previous sections are not
excluded from the string theory landscape, and they are qualitatively
different in other simplified models, such as the Bousso-Polchinski
(BP) landscape \cite{BP}.  We will now provide a brief summary of the
main features of the BP landscape, and then we will stress the
differences with the multi-sphere EM landscape.

\subsection{The Bousso-Polchinski landscape}
\label{sec:BP-model}

The BP landscape is a simplified model which provides an elegant
method for solving the cosmological constant problem\footnote{Good
  reviews of the cosmological constant problem can be found in
  references \cite{WW2,B-CC}}.  The starting point is M-theory, which
is formulated in 10+1 dimensions, compactified down to 3+1.  One of
the main ingredients of this theory is a seven-form, which is used to
introduce the Brown-Teitelboim cosmological constant neutralization
mechanism \cite{BT-1,BT-2}, which is a generalization of the Schwinger
pair creation process responsible for the spontaneous lowering of a
strong electric field.

In the presence of a compactification manifold having three-cycles,
the seven-form is expanded in a basis of harmonic three-forms, whose
coefficients are four-forms.  After dimensional reduction, the
four-dimensional duals of the four-forms are zero-forms, that is,
scalars, which are quantized by virtue of generalized Dirac
quantization conditions.  The total value of the flux of a four-form
in the $j^{\mathrm{th}}$ three-cycle is an integer multiple of a
fundamental charge $q_j$ which is proportional to the volume of the
three-cycle.  These charges are moduli of the theory, whose
stabilization is given \emph{a priori} in the BP model.  

A vacuum state of this model is given by specifying the integers
representing the value of the four-form flux stored in each
three-cycle.  Transitions between the states are mediated by
instantons, which can be viewed as M5-brane bubbles with two ``legs''
enclosing a three-dimensional interior of a different vacuum energy
density, while having three remaining ``legs'' wrapping the flux in a
three-cycle.

Thus, the vacuum states of the model are arranged in the nodes of a
lattice in flux space.  A given state is specified by $J$ integers
$n_1,\cdots,n_J$, whose effective cosmological constant $\lambda$ is
given by
\begin{equation}
  \label{eq:149}
  \lambda = \Lambda + \frac{1}{2}\sum_{j=1}^J q_j^2n_j^2 \,.
\end{equation}
In equation \eqref{eq:149}, $J$ represents the number of three-cycles
inside the compactification manifold; $\Lambda$ is the bare
cosmological constant of the theory, which should be negative, so that
$\lambda$ can reach a small value; and $q_j$ are the moduli, that is,
the elementary charges of the fluxes.

There is a Minkowski surface in flux space separating AdS and dS
states, which is obtained by setting $\lambda=0$ in \eqref{eq:149}.  A
node of the lattice can be located very close to this surface, and the
number of such nodes can be huge by choosing a large enough $J$.  The
existence of these nodes, randomly close to the $\lambda=0$ surface,
is essentially the BP mechanism solving the cosmological constant
problem.

In a BP landscape with a large amount of fluxes, the vast majority of
the nodes are located far away from the origin.  Some criterion is
needed to limit the value of the integers $n_j$ and render the
landscape finite.  Usually, this is accomplished by introducing a
cut-off $\Lambda_{\mathrm{cutoff}}$ in flux space which characterizes
the maximum value of $\lambda$ to be possibly reached.  The
computation of the probability of a given state among all available
states based on abundance of states gives very small values for a
large $J$, and a large $J$ is needed to reach a value of $\lambda$ as
low as the observed value $10^{-120}$ \cite{SN-1,SN-2}.  So this model
has the necessary states, but a very low probability for them to be
occupied, which leads to anthropic arguments.

The anthropic window is an interval of cosmological constant values
which allow the formation of observed structures (like galaxies, stars
and planets) \cite{WW}.  Even inside this anthropic window, the
number of states is so huge that the probability of a state having
$\lambda=10^{-120}$ is tiny.  Dynamical relaxation inside the BP
landscape reduces the states to a shell wider than the anthropic
window \cite{BY}, and thus do not solve this problem.  This requires
less convincing anthropic arguments to explain the observed value of
the cosmological constant.

We can rephrase this problem by saying that the distribution of
$\lambda$ values near $\lambda=0$ is flat \cite{SV,RHM}, and thus
specially small values do not get rewarded.  The observed value of
$\lambda$ lies in a very thin shell, very small when compared with the
anthropic or dynamically relaxed shells.  Thus, the flatness of the
distribution gives rise to such small probabilities.

In addition, there is another complication with large values of $J$.
When the dimension of flux space is large, the vast majority of states
in \emph{any} spherical shell are confined to coordinate hyperplanes
with a dimension of near $J\alpha^*$ with $\alpha^* < 1$
\cite{Alpha-star}.  The bulk of the spherical shell\footnote{That is,
  the region of the spherical shell surrounding the diagonals of flux
  space, where states are located far away from the coordinate
  hyperplanes.}  is almost devoid of states, and the number of
non-vanishing fluxes is generically less than $J$.  Nevertheless,
stability arguments often force the integers $n_j$ to be nonzero, even
large ones; this would dramatically lower the number of states in the
BP landscape, resulting in an empty anthropic shell.  This
$\alpha^*$-problem of the BP landscape is not restricted to sets with
spherical symmetry; secant states are not spherically distributed, and
share the same problem.

\subsection{Comparison between BP and ms-EM landscapes}
\label{sec:BP-vs-EM}

The previously described features of the BP landscape contrast with
their counterparts in the multi-sphere Einstein-Maxwell landscape.
First of all, this landscape is derived from a $2J+2$-dimensional
theory, after its dimensional reduction to 1+1 dimensions.  Therefore,
the resulting cosmologies are not comparable.  Nevertheless, we will
focus in the distribution of states and qualitative features of the
landscape.

The ``bare'' cosmological constant $\Lambda$ is negative in the BP
case, allowing cancellation in the effective cosmological constant
$\lambda$; if $\Lambda$ were positive, no AdS nor low-lying dS states
would longer exist.  In the EM case, $\Lambda$ should be positive;
otherwise, dS states would not exist at all.  Thus, both landscapes
have twin versions with reversed $\Lambda$ which are not physically
interesting.

The BP model assumes that its moduli are frozen by some external,
unspecified mechanism.  Therefore, the elementary charges are
parameters of the model, as well as $\Lambda$.  On the other hand, the
moduli of the EM theory, which are the radii of the internal spheres,
are fixed (at least at a linear level) by an effective potential built
from the magnetic field, the curvatures and the vacuum energy density.
Thus, this theory needs only one parameter, $\Lambda$.  It is
generally believed that the same stabilization mechanism should work
in the BP model, but as far as we know it has not been implemented
yet.

The simplicity of the formula for $\lambda$ in the BP model,
\eqref{eq:149}, is to be compared with the equation determining
$\lambda$ in the EM model, \eqref{eq:68}, \eqref{eq:98} or
\eqref{eq:102}.  In this equation, $\lambda$ cannot be isolated in
general, and there are several branches for each node.  Nevertheless,
only the principal branch has solutions with positive curvatures, and
these are the only ones with a chance of being stable.  Moreover, this
equation can have zero, one or two solutions, depending on $\Lambda$,
giving zero, one or two states per node in flux space.  In contrast,
\eqref{eq:149} always has one solution, and no more, per node.  Thus,
the correspondence between nodes and states is one-to-one in the BP
model, but this is not the case in the EM model.

The finiteness of the BP landscape is a consequence of a cutoff
introduced in flux space.  As commented above, were this cutoff
absent, the theory would have an infinite family of states with
infinitely high-$\Lambda$, which would raise the problem of choosing
initial conditions.  The EM model has a finite amount of stable dS
states because of the presence of a branching point in the equation
determining $\lambda$.  The number of unstable dS states is much
greater, but these states are excluded from the landscape.  Thus, the
stability analysis gets rid of the majority of dS states, and so we
expect this situation to be analogous in the BP model completed with a
stability analysis.  This ingredient can thus significantly change a
lot the general properties of the BP model, because it would exclude a
huge amount of states from the landscape.  This might be a feature,
though, because it might raise the probability for the system to be in
an anthropic state, which is currently very small because of the
enormous amount of dS states present.  But it is impossible to
establish this claim or the opposite without a well-defined model to
work with.

AdS states are finite in number in the BP model, because they are
located inside a sphere in flux space.  In the EM model with $J > 1$
there is an infinite number of them and they are always stable.
Therefore, the probabilistic arguments based on the number of states
cannot be applied here, because the probability of dS states would
always be zero.  This argument might be interpreted as indicating that
the method of computing probabilities using simply amounts of states
could be completely wrong in both models.  As a consequence, the
probability measure used in these landscapes should be revised from
scratch.

Both models have a $\lambda=0$ surface separating dS from AdS states,
which in the BP model is
\begin{equation}
  \label{eq:150}
  2|\Lambda| = \sum_{j=1}^J q_j^2 n_j^2\,,
\end{equation}
that is, a sphere in flux space (parametrized in $q_jn_j$
coordinates), while in the EM model it is
\begin{equation}
  \label{eq:151}
  \Lambda = \sum_{j=1}^J \frac{1}{n_j^2}\,,
\end{equation}
which is a sphere after performing a coordinate inversion.  These
surfaces provide the BP mechanism for solving the cosmological
constant problem: if the landscape contains a state randomly close to
this surface then this state can have a realistic value of $\lambda$.
Both models have this property.  Nevertheless, the surface
\eqref{eq:151} is not compact, and it allows for long state chains
whose cosmological constant can approach very small values.  This
phenomenon is absent in the BP model, and constitutes a basic
difference because it increases the amount of states in the anthropic
shell.  As stated above, we don't know the correct way of computing
probabilities, but state chains provide a new source of low-lying
states which is absent in the BP model.

State chains are also responsible for a crucial modification in the
distribution of $\lambda$ values, which is flat near $\lambda=0$ in
the BP case, as stated above.  In the EM case, the $\omega(\lambda)$
density vanishes at $\lambda=0$, but it has a huge peak of small
values, corresponding precisely to those lying in the state chains.
Thus, this distribution is not flat, which means that the
randomly-close-state mechanism is less important in the EM model than
in the BP case, because only states near the diagonal in flux space
contribute to it, while all states near the sphere contribute in the
BP case, accounting for the difference.  State chains provide a
dominant peak of small $\lambda$ values, which is another different
mechanism for solving the cosmological constant problem.  This
mechanism can provide a peak very near $\lambda=0$ for very specific
values of $\Lambda$ when $J$ is small (namely, when
$\frac{1}{\sqrt{\Lambda}}$ is very close to an integer from below), or
for generic $\Lambda$ values when $J$ is moderate $J\approx10,15$,
leading even to anthropic states.  Thus, both mechanisms are
different, and both have states with very small values of the
cosmological constant, but they differ deeply in the form of the
$\omega(\lambda)$ distribution.

Finally, the $\alpha^*$-problem is absent in the EM model, because the
$\lambda=0$ surface never approaches the coordinate hyperplanes where
one or more $n_j=0$.  Thus, if the stability results found in the EM
model translate to the BP model completed with a stability analysis,
then we are forced to conclude that the vast majority of dS states,
which are near the hyperplanes, would be unstable, and thus there
would be excluded from the BP landscape.  This would change all
reasoning based on number of states, if it were to be of any use.

Table \ref{tab:diff-BP-msEM} summarizes all the issues we have
addressed while comparing the BP and EM landscapes.

\begin{table}
  \centering
  \includegraphics[width=\textwidth]{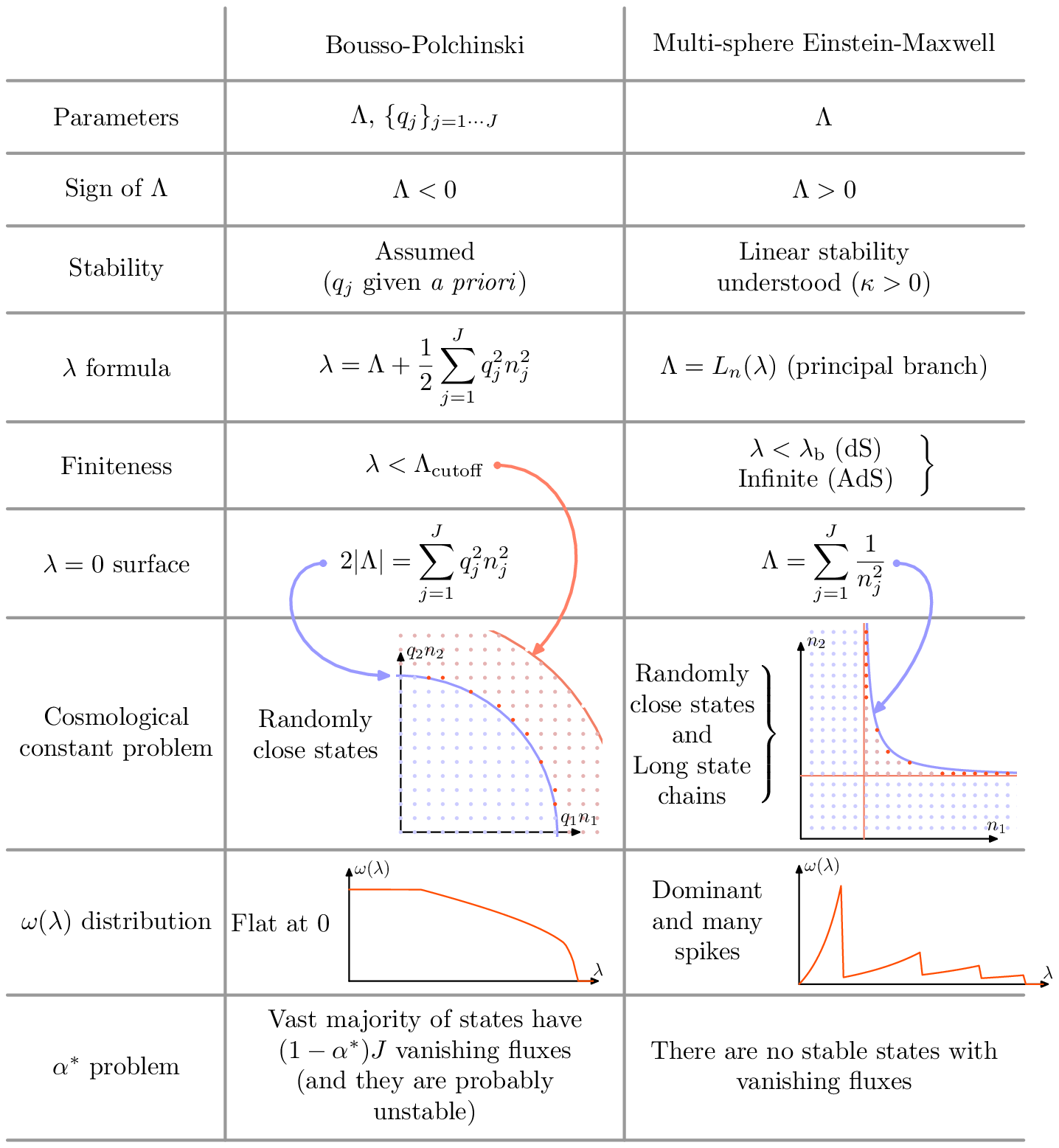}
  \caption{Summary of the differences between the Bousso-Polchinski and
    multi-sphere Einstein-Maxwell landscapes.  The first two graphics
    emphasize the different distribution of states in flux space,
    which is the origin of the state chains.  These state chains are
    responsible of the dominant spike in the 
    $\lambda$-density, as shown in the two last graphics.}
  \label{tab:diff-BP-msEM}
\end{table}

\section{Possible implications for the string theory landscape }
\label{sec:implications}

The comparison carried out in the previous section leads to some
features that a complete treatment of the BP landscape interpreted as
a toy model of the true string theory landscape would bear when
compared to what is currently believed.  This section is devoted to
envision what the BP landscape would look like if some of the main
features of the EM landscape were found to hold.

Two key points should be stressed:
\begin{itemize}
\item Theories with four-form fluxes have duals which are
  gauge-gravity theories.  The main difference between them is that
  gauge-gravity theories have a built-in cut-off mechanism which
  limits the maximum values of the integers characterizing the
  landscape.  Thus, those landscapes have natural finiteness
  conditions, such as the branching point in the multi-sphere
  Einstein-Maxwell landscape.  This would avoid the neccessity of a
  cut-off scale put by hand in the model.  Nevertheless, the KKTL
  model has some natural cut-off mechanisms built-in
  \cite{Frey:2003dm}, \cite{Conlon:2008cj}.
\item Stability conditions are important not only beacuse they
  complete the model, but also because they exclude a huge amount of
  states from the model.  If the same exclusion were to take place in
  the BP landscape, the vast majority of its nodes would not be true
  states of the landscape, and the predictions of existence of
  anthropic states would dramatically change.  Thus, stability
  conditions have a two-fold purpose: on the one hand they fix the
  values of the elementary charges to be used in the model, on the
  other hand they limit which nodes have physically relevant states.
  A priori frozen moduli fulfill the first purpose, but do not help in
  deciding which nodes have states.  This causes a huge proliferation
  of states, which may be spurious ones.  The EM model shows that the
  vast majority of dS states are unstable.  Thus, we can expect the
  same to be true in a completed BP landscape.
\end{itemize}

Other features of the EM landscape may not have a direct translation
to a completed BP model, such as state chains.  They are a consequence
of the asymptotes found in the null-$\lambda$ surface, which is
non-compact.  This is an indication that the details of the
null-$\lambda$ surface provide different sources of low-lying states
which change the density of states $\omega(\lambda)$.  This
distribution is needed when one has to compute probabilities in a
given landscape model; but a completely clear, unambiguous, quantum
prescription for computing probabilities with a general model is still
lacking, and therefore the implications of the details of
$\omega(\lambda)$ in the computation of probabilities cannot go beyond
the naive arguments based on state abundances.  At this simple level,
details of the null-$\lambda$ surface translate in peaks in the
$\omega(\lambda)$ density, thus producing very different probabilities
for the states in the anthropic shell.  The BP model has a spherical
null-$\lambda$ surface, and thus $\omega(\lambda)$ has no peaks; if
the null-$\lambda$ surface of more realistic flux compactifications of
M-theory had other nontrivial shapes, this would be reflected in the
$\omega(\lambda)$ distribution and in the final computation of the
probabilities.  So this is the last point that the EM landscape brings
in: the details of the null-$\lambda$ surface are very important for
probability computations.

\section{Conclusions}
\label{sec:conc}

We have addressed a simple sector of the Einstein-Maxwell theory as an
exactly solvable model of a landscape.  The theory, formulated in
$2J+2$ spacetime dimensions, has a single parameter in the Lagrangian,
namely, the ``bare'' cosmological constant $\Lambda>0$.  The
compactification has the form
$\mathrm{(A)dS}_2\times(\mathrm{S}^2)^J$, which is referred to as
\emph{multi-sphere Einstein-Maxwell} compactification.  Equations of
motion for the corresponding metric ansatz are algebraic equations for
the values of the curvatures of the inner spheres and the effective
cosmological constant $\lambda$ of the cosmological part.  In the
presence of a magnetic monopole, the magnetic flux in each sphere,
which is quantized by a Dirac condition, stabilizes the configuration
which spontaneously would decompactify.  The cosmological constant
$\Lambda$ helps to evade the Maldacena-Nu\~nez no-go theorem in this
case \cite{Maldacena-no-go,Denef-2}.  The different combinations of
the flux quanta stored in the spheres give rise to a complicated
landscape, in which each configuration of integers (called a
\emph{node}) can host a true stable state of the model, two stable
states, an unstable state and a stable one, or no state, giving rise
to two branches of (AdS and dS) states.  It is found that for $J>1$ an
infinite family of stable AdS states exist, but stable dS states exist
only near the \emph{branching surface}, which is the locus at which
both branches meet, that is, the null-$\lambda$ surface.  The
structure of the null-$\lambda$ surface gives rise to the state
chains, which provide a different source of low-lying states besides
the randomly close states which help to solve the cosmological
constant problem in the Bousso-Polchinski landscape.  State chains
also help in counting states approximately, and they translate in
peaks in the density of states $\omega(\lambda)$, providing anthropic
states for moderately large values of $J$.

All the previous features of the model are qualitatively different
from its counterparts in the Bousso-Polchinski landscape.  We think
that in a completed BP model, all these differences would render a
very different picture with respect to the number of states,
probabilities and anthropic reasoning.  Thus, despite not being a
realistic landscape model, the multi-sphere Einstein-Maxwell model has
very appealing features that might propagate in more realistic models
of the true string theory landscape.

The account of the multi-sphere Einstein-Maxwell model given in this
paper has three main limitations: firstly, it is difficult to
extrapolate the stability conditions found from 1+1 spacetime
dimensions to a more realistic 3+1 cosmology.  Secondly, we have
considered a restricted class of linear perturbations; the inclusion
of fully general linear perturbations could render unstable some
states which are stable.  The combination of the two ingredients, that
is, 3+1 cosmology and a full set of linear perturbations, can lead to
a qualitatively very different sector of the Einstein-Maxwell
landscape.  Finally, a fundamental missing piece is the cosmological
measure.  This problem and the construction of the corresponding model
will be addressed in future papers.

\section*{Acknowledgments}
\label{ack}

We would like to thank Concha Orna for carefully reading this
manuscript, and the Pedro Pascual Benasque Center of Science.  We also
thank Frederik Denef, Roberto Emparan, Jaume Garriga, Bert Janssen,
Donald Marolf and Jorge Zanelli for useful discussions and
encouragement.  This work has been supported by CICYT (grant
FPA-2009-09638) and DGIID-DGA (grant 2011-E24/2).  We thank also the
support by grant A9335/10 (F\'{\i}sica de alta energ\'{\i}a:
Part\'{\i}culas, cuerdas y cosmolog\'{\i}a).

\appendix

\section{Effect of derivative couplings in the multi-radion evolution equations}
\label{sec:deriv-couplings}

In this appendix we give a heuristic argument leading to the
conclusion that the linear stability analysis of the multi-radion
field evolution equations, equation~\eqref{eq:85}, which is achieved
by neglecting the derivative couplings, can be promoted to a
non-linear stability analysis in which the linear stability is
preserved as long as perturbation amplitudes are sufficiently small.

To begin with, we consider again equation \eqref{eq:85}:
\begin{equation}
  \label{eq:152}
  -e^{-2\phi}\eta^{\alpha\beta}
  \biggl[
  (\xi_j)_{\alpha\beta}
  + 2(\xi_j)_\alpha\Bigl(\sum_k(\xi_k)_\beta\Bigr)
  - \sum_k(\xi_k)_\alpha (\xi_k)_\beta
  \biggr]
  = 
  \lambda - e^{-2\sum_k\xi_k} U'_j(\xi_j)\,.
\end{equation}
The derivative couplings appear in a quadratic form.  We will use the
symbol $\boldsymbol{\xi}$ to denote the $J$-component column vector of
the perturbations $\xi_j$, and then we will write the derivative
couplings in matrix form as
\begin{equation}
  \label{eq:153}
  -e^{-2\phi}\eta^{\alpha\beta}
  \biggl[
  (\xi_j)_{\alpha\beta}
  + \langle\boldsymbol{\xi}_\alpha, M_j\boldsymbol{\xi}_\beta\rangle
  \biggr]
  = 
  \lambda - e^{-2\sum_k\xi_k} U'_j(\xi_j)\,,
\end{equation}
where the constant $J\times J$ matrix $M_j$ has the number $-1$ along
the diagonal except for 1 along the $j^{\text{th}}$ row and
$j^{\text{th}}$ column and zeroes elsewhere:
\begin{equation}
  \label{eq:154}
  M_j =
  \begin{pmatrix}
    -1     & \cdots &  0 &      1 &     0  & \cdots &       0 \\
    \vdots & \ddots &    & \vdots & \vdots &        & \vdots  \\
    0      & \cdots & -1 &      1 &     0  & \cdots &       0 \\
    1      & \cdots &  1 &      1 &     1  & \cdots &       1 \\
    0      & \cdots &  0 &      1 &    -1  & \cdots &       0 \\
    \vdots &        &    & \vdots & \vdots & \ddots & \vdots  \\
    0      & \cdots &  0 &      1 &     0  & \cdots &      -1
  \end{pmatrix}
  \begin{matrix}
    \phantom{0} \\ \text{($j^{\text{th}}$ row)} \\ \phantom{0}
  \end{matrix}
  \,.
\end{equation}
The matrix $M_j$ has a spectrum with two simple eigenvalues
$\pm\sqrt{J}$ and a $(J-2)$-degenerate $-1$ eigenvalue.  It is
therefore an ``almost negative-definite'' quadratic form, which
represents a non-linear force.

We are interested in the effect of the derivative couplings;
therefore, we approximate the system of equations by Taylor-expanding
to first order the right-hand side of equation \eqref{eq:153}, thus
considering the much simpler system (written in matrix form)
\begin{equation}
  \label{eq:155}
  -e^{-2\phi}\eta^{\alpha\beta}
  \biggl[
  (\xi_j)_{\alpha\beta}
  + \langle\boldsymbol{\xi}_\alpha, M_j\boldsymbol{\xi}_\beta\rangle
  \biggr]
  = 
  -\sum_{k} H_{jk}\xi_k\,,
\end{equation}
where the frequency matrix $H$ is given by equation \eqref{eq:88} in
section \ref{sec:moduli-stab}.

As said above, a complete analysis of the system \eqref{eq:155} is not
possible, but we can form a single equation out of it by projecting
the system of equations along a constant direction $\mathbf{w}=(w_j)$:
\begin{equation}
  \label{eq:156}
  -e^{-2\phi}\eta^{\alpha\beta}
  \biggl[
  \langle\mathbf{w},\boldsymbol{\xi}\rangle_{\alpha\beta}
  + \langle\boldsymbol{\xi}_\alpha, M_\mathbf{w}\boldsymbol{\xi}_\beta\rangle
  \biggr]
  = -\langle\mathbf{w},H\boldsymbol{\xi}\rangle
  \,.
\end{equation}
The matrix $M_\mathbf{w}$ is
\begin{equation}
  \label{eq:157}
  M_\mathbf{w} = \langle\mathbf{w},\mathbf{M}\rangle = \sum_jw_jM_j\,.
\end{equation}
Any solution of the system \eqref{eq:155} is a solution of equation
\eqref{eq:156}, but the reciprocal is false.  Nevertheless, we can look
for normal modes of the form $\boldsymbol{\xi}(t,x) =
\chi(t,x)\boldsymbol{\upsilon}$ for constant $\boldsymbol{\upsilon}$
and a single scalar function $\chi(t,x)$:
\begin{equation}
  \label{eq:158}
  -e^{-2\phi}\eta^{\alpha\beta}
  \biggl[
  \langle\mathbf{w},\boldsymbol{\upsilon}\rangle\chi_{\alpha\beta}
  + \langle\boldsymbol{\upsilon}, M_\mathbf{w}\boldsymbol{\upsilon}\rangle
  \chi_\alpha\chi_\beta
  \biggr]
  = -\langle\mathbf{w},H\boldsymbol{\upsilon}\rangle\chi
  \,.
\end{equation}
Taking $\chi = \chi(t)$ and $\mathbf{w}=\boldsymbol{\upsilon}$ for
simplicity, we obtain
\begin{equation}
  \label{eq:159}
  e^{-2\phi}
  \biggl[
  \ddot\chi
  + \frac{\langle\boldsymbol{\upsilon},
    M_{\boldsymbol{\upsilon}}\boldsymbol{\upsilon}\rangle}
  {\langle\boldsymbol{\upsilon},\boldsymbol{\upsilon}\rangle}
  \dot\chi^2
  \biggr]
  = -\frac{\langle\boldsymbol{\upsilon},H\boldsymbol{\upsilon}\rangle}
  {\langle\boldsymbol{\upsilon},\boldsymbol{\upsilon}\rangle}\chi
  \,.
\end{equation}
If the non-linear term were absent, we would have a simple oscillator
equation with a frequency given by the Rayleigh quotient of the matrix
$H$.  The solution of this equation would be a solution of the linear
system if $\boldsymbol{\upsilon}$ were chosen as an eigenmode of $H$.
In this sense, the projected equation \eqref{eq:159} is an
\emph{average} equation, and its solution (a \emph{weak} solution
henceforth) can indicate the behavior of the true solutions we are
inspecting.  Of course, this is a heuristic argument, but we can argue
that true solutions provide weak solutions; thus, an unstable true
solution should be reflected by an unstable weak solution.  This
argument has the obstacle of the existence of the normal modes we are
using as ansatz; thus, as long as the normal modes constitute a
reasonable description of the system \eqref{eq:158}, the projected
equation will reflect accurately the character of its non-linear
counterpart.

Nevertheless, we can use the projected equation \eqref{eq:159} to see
if the presence of the non-linear term can render unstable a linearly
stable solution.

\begin{figure}
  \centering
  \includegraphics[width=0.5\textwidth]{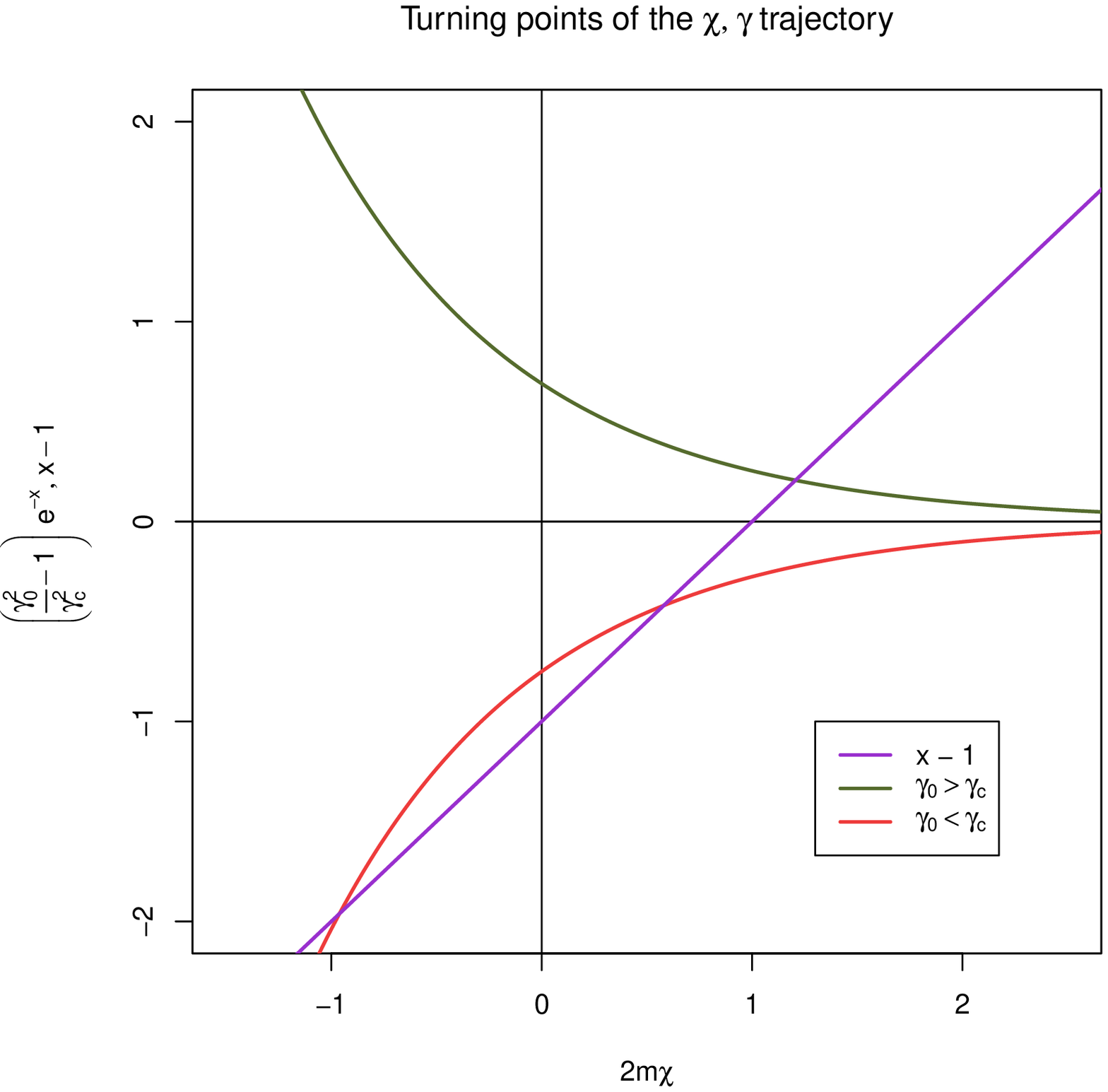}%
  \includegraphics[width=0.5\textwidth]{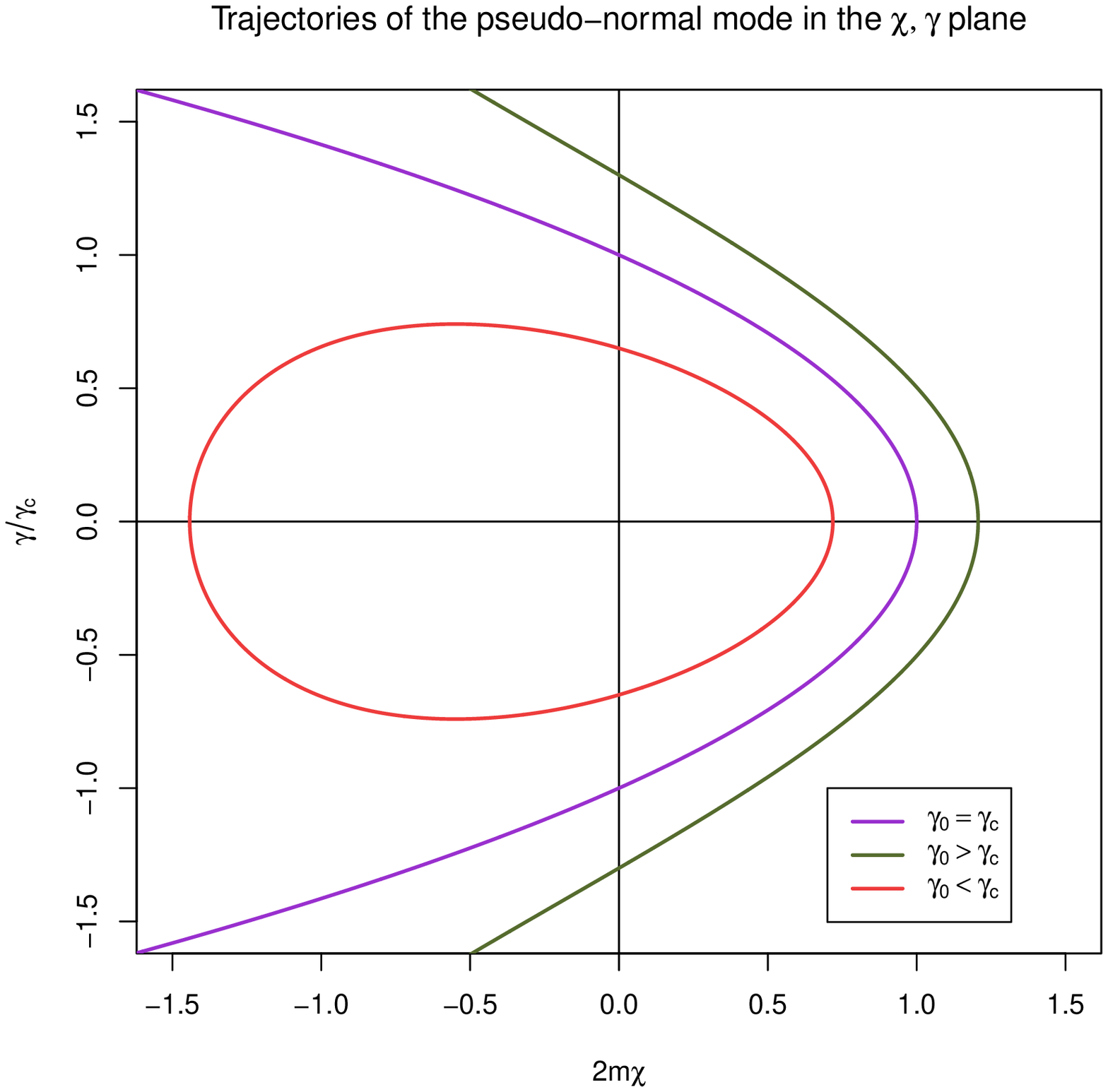}
  \caption{Turning points (left) and actual trajectories (right) of the
    normal modes of the projected multi-radion evolution equation.}
\label{fig:traj}
\end{figure}

We will rewrite equation \eqref{eq:159} as
\begin{equation}
  \label{eq:160}
  \ddot\chi + m\dot\chi^2 = -h\chi\,,
  \quad \text{with}\quad
  m = \frac{\langle\boldsymbol{\upsilon},
    M_{\boldsymbol{\upsilon}}\boldsymbol{\upsilon}\rangle}
  {\langle\boldsymbol{\upsilon},\boldsymbol{\upsilon}\rangle}
  \,,\quad\text{and}\quad
  h = e^{2\phi}\,
  \frac{\langle\boldsymbol{\upsilon},H\boldsymbol{\upsilon}\rangle}
  {\langle\boldsymbol{\upsilon},\boldsymbol{\upsilon}\rangle}\,.
\end{equation}
The parameter $m$ depends on the projection direction
$\boldsymbol{\upsilon}$ but not on time.  In contrast, $h$ depends
also on time by the presence of the $e^{2\phi}$ factor.  We will now
discuss the expected domain of both parameters in the following.
\begin{itemize}
\item The parameter $m$ is the Rayleigh quotient of the matrix
  $M_{\boldsymbol{\upsilon}}$ on the projection direction
  $\boldsymbol{\upsilon}$.  The eigenvalues of
  $M_{\boldsymbol{\upsilon}}$ are $\pm\sqrt{J}\sqrt{\sum_j
    \upsilon_j^2}$, both of them nondegenerate, and
  $-\sum_j\upsilon_j$ with $J-2$ degeneracy.  Thus, depending on the
  projection direction $\boldsymbol{\upsilon}$, $m$ can have both
  signs.  If $\boldsymbol{\upsilon}$ is taken to have unit norm, then
  $m$ will be some value in the interval $[-\sqrt{J},\sqrt{J}]$.
\item The 1+1 cosmological solution $\phi(t)$ has a characteristic
  evolution time $t_\phi = \frac{1}{\sqrt{|\lambda|}}$.  On the other
  hand, the longest characteristic evolution time of the oscillator
  term is $t_H=\frac{1}{\sqrt{\kappa}}$, where $\kappa$ is the minimum
  eigenvalue of the frequency matrix $H$.  Of course, we are
  considering a linearly stable equation, so that $\kappa>0$.  We can
  consider $\phi$ as slowly-varying if its characteristic time is much
  greater than the oscillator characteristic time:
  \begin{equation}
    \label{eq:161}
    t_\phi \gg t_H
    \quad\Rightarrow\quad
    \lambda \ll \kappa
  \end{equation}
  Thus, for times $t\approx t_H$, we may consider $\phi$ as constant,
  and thus $h$ will be a positive number.
\end{itemize}
Equation \eqref{eq:160} is integrable.  We can show the form of its
trajectories by writing $\dot\chi = \gamma$:
\begin{equation}
  \label{eq:162}
  \left.
    \begin{aligned}
      \dot\chi &= \gamma \\
      \dot\gamma &= -m\gamma^2 - h\chi
    \end{aligned}
  \right\}
  \quad\Rightarrow\quad
  \frac{\dif\gamma}{\dif\chi} = -m\gamma - h\frac{\chi}{\gamma}\,.
\end{equation}
The last equation has the exact solution
\begin{equation}
  \label{eq:163}
  \gamma(\chi)^2 = \Bigl(\gamma_0^2 - \frac{h}{2m^2}\Bigr)\,e^{-2m\chi}
  + \frac{h}{m}\Bigl(\frac{1}{2m} - \chi\Bigr)\,.
\end{equation}
In the previous equation, $\gamma_0 = \gamma(0)$.  Specializing $m=0$
(which removes the nonlinear term) we obtain the oscillator trajectory
$\gamma^2 + h\chi^2 = \gamma_0^2$.  The points where $\gamma=0$ are
called \emph{turning points} of the trajectory, and they mark its
domain because of the square in \eqref{eq:163}.  The trajectory has one
or two turning points given by the equation
\begin{equation}
  \label{eq:164}
  \Bigl(\gamma_0^2 - \frac{h}{2m^2}\Bigr)\,e^{-2m\chi}
  = \frac{h}{m}\Bigl(\chi - \frac{1}{2m}\Bigr)
\end{equation}
If $m$ is a fixed positive value, then the previous equation has a
single solution if $\gamma_0^2 - \frac{h}{2m^2}>0$ but it has two
solutions if $\gamma_0^2 - \frac{h}{2m^2} < 0$.  A single turning
point describes an open trajectory, while two turning points describe
a closed one, see figure \ref{fig:traj}.
Thus, the oscillator trajectory remains closed when we turn on the
non-linearity if
\begin{equation}
  \label{eq:165}
  \gamma_0^2 < \frac{h}{2m^2}
\end{equation}
Therefore, a linearly stable trajectory remains non-linearly stable if
the amplitude $\gamma_0$ does not exceed the critical value
$\gamma_c^2=\frac{h}{2m^2}$.  Beyond this value, the trajectory is
open and therefore the linearly stable solution becomes non-linearly
unstable.

The lowest value of the critical amplitude is reached when $m$ is
largest; for a unit-norm projection direction, the largest value of
$m$ is $\sqrt{J}$, as discussed above.  Thus, the lowest value of the
critical amplitude is
\begin{equation}
  \label{eq:166}
  \gamma_{c,\text{min}}^2 = \frac{e^{2\phi}\kappa}{2J}\,.
\end{equation}
As long as amplitudes are smaller than this value, the non-linear
derivative couplings cannot spoil linear stability.  Nevertheless,
when $\kappa$ is small, perturbations have a chance of trigger a
non-linear instability and destabilize a linearly stable state.

Summarizing, the non-linear terms in the multi-radion evolution
equations respect the linear stability criterion except in the regime
of large amplitudes, which is most easily accessible when the minimum
eigenvalue of the frequency matrix becomes small, that is, in the
onset of instability.

\end{document}